\pdfoutput=1

\documentclass[ASNA]{USG}

\graphicspath{{./figs/}}
\hypersetup{hidelinks}

\makeatletter
\def\oddfoot@titlepage@info{%
  \vbox{%
    \hsize\textwidth
    \hrule height0.5pt
    \vskip3pt
    \hbox to \textwidth{{\pagenumfont\itshape\@journal}\hfil{\pagenumfont\thepage\ of \@LastPg}}%
  }%
}
\makeatother

\usepackage{amssymb}
\usepackage{amsmath}
\usepackage{booktabs}
\usepackage{array}
\usepackage{algorithm}
\usepackage{algpseudocode}
\usepackage[section]{placeins}  %

\definecolor{darkBrown}{rgb}{0.55,0.27,0.07}
\definecolor{darkYellow}{rgb}{0.9,0.7,0.1}
\definecolor{darkRed}{rgb}{0.75,0.1,0.1}
\definecolor{deepPink}{rgb}{0.65,0.2,0.5}
\definecolor{gtBlue}{rgb}{0.1,0.45,0.8}

\newcommand{\colorlabel}[1]{%
    \begin{tcolorbox}[
        on line,
        baseline=1.5pt,
        colback=#1!30,
        colframe=black,
        sharp corners,
        size=small,
        width=0.6cm,
        height=0.3cm,
        left=0pt, right=0pt,
        boxsep=0pt,
        boxrule=0.5pt
    ]
    \end{tcolorbox}%
}

\articletype{RESEARCH ARTICLE}
\journal{Int. J. Numer. Meth. Engng}

\begin{document}
\title{A Level Set Method on Particle Flow Maps}
\author[1]{Jinjin He}[https://orcid.org/0009-0000-4319-1191]
\author[2]{Taiyuan Zhang}
\author[1]{Zhiqi Li}
\author[3]{Junwei Zhou}
\author[1]{Duowen Chen}
\author[1]{Bo Zhu}

\authormark{HE \textsc{et al.}}
\titlemark{A Level Set Method on Particle Flow Maps}

\address[1]{\orgdiv{School of Interactive Computing, }\orgname{Georgia Institute of Technology, }\orgaddress{\state{Atlanta, GA 30332, }\country{USA}}}
\address[2]{\orgdiv{Department of Computer Science, }\orgname{Dartmouth College, }\orgaddress{\state{Hanover, NH 03755, }\country{USA}}}
\address[3]{\orgdiv{Computer Science and Engineering, }\orgname{University of Michigan, }\orgaddress{\state{Ann Arbor, MI 48109, }\country{USA}}}

\corres{Jinjin He (\email{jhe433@gatech.edu})}

\fundingInfo{National Science Foundation (NSF): CAREER \#2420319, IIS \#2433307, OISE \#2433313, IIS \#2433322, ECCS \#2318814, and CNS \#2450401; NVIDIA Academic Grant (computational resources).}


\abstract[ABSTRACT]{Accurate interface tracking is central to numerical simulations involving moving boundaries, including multiphase flows, free-surface dynamics, phase-change problems, and other interface-driven processes in computational mechanics. Although level set methods provide a flexible implicit representation of evolving interfaces, their grid-based advection and reinitialization procedures often introduce numerical diffusion, volume loss, and degradation of thin or highly curved structures. This paper introduces a Particle Flow Map Level Set (PFM-LS) method that addresses these limitations by transporting interface geometry through particle-carried bidirectional flow maps. In contrast to particle level set methods that use particles mainly as corrective markers, PFM-LS uses particles as primary carriers of level-set values, gradients, and Hessians within a narrow band around the interface. The method combines long-range flow-map transport, high-order particle-grid reconstruction, adaptive particle control, and a hybrid particle-grid redistancing scheme to preserve both signed-distance consistency and sub-grid geometric features. Numerical benchmarks in two and three dimensions, including topology-change tests, chamfer and Hausdorff shape-error metrics, and coupling to an incompressible free-surface flow solver, show that PFM-LS achieves strong volume preservation, high shape fidelity, and near fourth-order convergence, outperforming a broad range of existing level set and reference-map-based methods.}

\maketitle

\section{Introduction}
\label{sec-intro}

Interface tracking is a central task in computational mechanics problems involving moving material boundaries, including free-surface flows~\cite{sussman1994level,bargteil2006semi,Largetimestep}, multiphase flows~\cite{sussman2000coupled,tryggvason2001front,gibou2018review}, phase-change processes~\cite{anumolu2018gradient}, fluid--structure interaction~\cite{cottet2008eulerian}, and curvature-driven interface motion~\cite{osher1988fronts,smereka2003semi,kolahdouz2013semi}. In these applications, the accuracy of the computed interface directly affects volume conservation, curvature evaluation, surface-force calculation, and the preservation of small-scale geometric features~\cite{gibou2019sharp,shin2005accurate}. A persistent challenge is to maintain these geometric quantities during long-time advection, especially when the interface develops thin filaments, sharp corners, or strongly deformed structures~\cite{enright2002hybrid,ferstl2016narrow,heo2010detail,chentanez2013mass}.

The level set method, introduced by Osher and Sethian~\cite{osher1988fronts, sethian1999level, osher2004level},
represents interfaces implicitly as the zero isocontour of a signed distance
function and has since become a foundational tool for interface tracking in
free-surface simulations~\cite{sussman1994level, bargteil2006semi, Largetimestep,
fedkiw2001visual, chentanez2013mass,sussman1999improved}, multiphase
flows~\cite{sussman2000coupled, lyras2020coupled, shin2002modeling, shin2005accurate, smereka2003semi, tryggvason2001front},
and other interface-driven dynamics~\cite{osher2001level, gibou2018review, gibou2019sharp, cottet2008eulerian, chen2025neural, he2024multi}.
The central idea of the level set method is to represent an interface
$\partial\Omega$ implicitly as the zero isocontour of a signed distance
function $\varphi(\boldsymbol{x}, t)$:
\begin{equation}
    \partial\Omega(t) = \{\boldsymbol{x} \in \mathbb{R}^d : \varphi(\boldsymbol{x}, t) = 0\},
\end{equation}
where $\varphi > 0$ in the air domain $\Omega_0$ and $\varphi < 0$ in the liquid
domain $\Omega_1$. The evolution of the interface under a velocity field
$\boldsymbol{u}(\boldsymbol{x}, t)$ is governed by the level set advection equation:
\begin{equation}
    \frac{\partial \varphi}{\partial t} + \boldsymbol{u} \cdot \nabla \varphi = 0.
    \label{eq:ls_advection}
\end{equation}
This formulation is well suited to interface-evolution problems because it provides a compact implicit description of geometry while retaining convenient access to normals, curvature, and signed-distance information through finite differences of $\varphi$; these quantities are estimated from the discrete field rather than stored directly.
This implicit representation naturally accommodates complex topological
changes such as merging and splitting without explicit
remeshing~\cite{adalsteinsson1995fast, peng1999pde, sethian1996fast}, making
level set methods indispensable for a wide range of computational physics
applications~\cite{osher2001level,sethian2003level, gibou2018review, gibou2019sharp}.
Despite their versatility, classical level set methods suffer from two fundamental limitations when solving Eq.~\eqref{eq:ls_advection} numerically. First, numerical diffusion during advection causes substantial volume loss, particularly for thin structures and small-scale features undergoing large deformation~\cite{enright2002hybrid, ferstl2016narrow, chentanez2013mass}. Even with high-order spatial discretizations such as ENO~\cite{shu1988efficient} or WENO~\cite{liu1994weighted, jiang2000weighted} schemes, the repeated interpolation and reinitialization steps introduce cumulative errors that erode interface fidelity over time. Second, the discrete grid representation lacks the resolution to capture sub-grid geometric details, leading to the gradual erosion of sharp corners and filaments that are crucial for visual and physical fidelity~\cite{heo2010detail, shin2005accurate}.

\noindent\textbf{Alternative interface representations} We note that level set methods are not the only approach to interface tracking. Volume-of-fluid (VOF) methods~\cite{hirt1981volume, rider1998reconstructing, owkes2014computational} represent interfaces through volume fractions and achieve exact mass conservation by construction, with modern high-order variants capable of capturing thin filaments within a single cell. Moment-of-fluid (MOF) methods~\cite{dyadechko2008reconstruction, jemison2013coupled} further augment this representation with material centroids for improved reconstruction. Conservative level set methods~\cite{olsson2005conservative} reformulate the level set advection in conservative form to reduce mass loss. Edge-based interface tracking methods~\cite{glimm2001conservative} explicitly track interface elements for geometric fidelity. Conservative level set formulations and modern geometric VOF schemes thus recover volume conservation by construction, but trade it for interface smearing or reconstruction complexity; front-tracking methods~\cite{tryggvason2001front} likewise maintain an explicit marker representation of the interface, providing sharp resolution at the cost of more demanding topology handling. These approaches use fundamentally different interface representations (volume fractions, material moments, or explicit meshes) compared to our signed-distance-based formulation. While they offer advantages in mass conservation, level set methods naturally provide continuous gradient and curvature information essential for surface-tension-driven flows and other curvature-dependent physics.

\noindent\textbf{Particle Level Set}
The numerical-diffusion and sub-grid-feature limitations of classical level set methods described above motivate the development of numerical methods assisted by Lagrangian structures that can better preserve interface geometry over long time horizons. The hybrid particle level set (PLS) method~\cite{enright2002hybrid} employs marker particles distributed near the interface within a narrow band to correct the signed distance function $\varphi$. Each particle $p$ carries its own signed distance value $\varphi_p$ and is advected along the flow:
\begin{equation}
    \frac{d\boldsymbol{x}_p}{dt} = \boldsymbol{u}(\boldsymbol{x}_p, t).
\end{equation}
When discrepancies arise between the particle values $\varphi_p$ and the grid values $\varphi(\boldsymbol{x}_p)$, the particles provide correction information to reduce volume loss and maintain sharper interfaces~\cite{osher2004level, hieber2005lagrangian, ianniello2010self, leung2009grid, mercier2013numerical, fu2016gradient}. Subsequent extensions have further improved accuracy and robustness: for instance, oriented and gradient-augmented variants~\cite{vartdal2013oriented, fu2016gradient} embed additional geometric information such as the interface normal $\mathbf{n} = \nabla\varphi / |\nabla\varphi|$ or gradient $\boldsymbol{g} = \nabla\varphi$ within particles to better preserve curvature and local detail. In addition, the one-layer PLS formulation~\cite{ianniello2010self,vartdal2013oriented,zhao2018one,zhao2020three} simplifies particle management by confining particles to a single layer near the interface without compromising reinitialization accuracy. More recently, diffusion-driven characteristic mapping~\cite{yin2021diffusion} has enhanced particle redistribution by maintaining uniform and stable marker distributions during large deformations, and neural extensions~\cite{chen2025neural} leverage learned corrections for adaptive and data-driven interface tracking.

Despite these advances, existing particle-based level set methods still face fundamental limitations in long-range advection. First, the level set function $\varphi$ on the grid is advected using conventional schemes (e.g., semi-Lagrangian), which introduces numerical dissipation that particles can only partially correct. Second, periodic particle reseeding necessary to maintain adequate particle coverage discards old particles and initializes new ones by interpolating $\varphi$ from the current grid, thereby introducing additional errors and erasing the trajectory history carried by the original particles. Most critically, particles in these methods store only their current position $\boldsymbol{x}_p$ and level set value $\varphi_p$, without retaining their initial positions $\boldsymbol{X}_p$. As a result, the natural flow map encoded by particle trajectories is never exploited, and the long-range characteristic structure of the advection equation remains unutilized.

\noindent\textbf{Gradient Augmented Level Set}
Complementary grid-based strategies have also been developed to improve advection accuracy and volume preservation. Among these, gradient-augmented level set (GALS) methods~\cite{nave2010gradient} are most closely related to the present work. GALS methods extend the formulation by evolving both the level set $\varphi$ and its gradient $\boldsymbol{g} = \nabla\varphi$ as independent variables, satisfying
\begin{equation}
    \frac{\partial \boldsymbol{g}}{\partial t} + (\boldsymbol{u} \cdot \nabla)\boldsymbol{g} + (\nabla\boldsymbol{u})^\top \boldsymbol{g} = 0,
\end{equation}
and using high-order Hermite interpolation to reconstruct $\varphi$ from both values and gradients~\cite{bockmann2014gradient, kolomenskiy2016adaptive}. This enhances local geometric expressiveness and enables sub-grid accuracy, with extensions to phase-change simulations~\cite{anumolu2018gradient} and curvature-driven flows~\cite{kolahdouz2013semi}. However, since GALS methods store and evolve information on fixed Eulerian grids, they still rely on grid-based interpolation during advection, which introduces cumulative errors over long time horizons. The above two classes of improvements---particle-based correction and gradient-augmented grid methods---address the limitations of classical level set methods from different angles, but neither fundamentally resolves the issue of accumulated interpolation errors inherent in local temporal updates.

\subsection{Characteristic Map Methods}
Rather than correcting errors locally at each time step, characteristic mapping methods~\cite{mercier2020characteristic, yin2021characteristic, tessendorf2011characteristic} take a fundamentally different approach by maintaining long-range mappings that trace back to initial conditions. The key observation is that the advection equation~\eqref{eq:ls_advection} admits a characteristic solution: along particle trajectories, the level set value is conserved. Specifically, if we define the forward flow map $\boldsymbol{\phi}_{t_0 \to t}(\boldsymbol{X})$ as the mapping from the initial position $\boldsymbol{X}$ at time $t_0$ to the current position $\boldsymbol{x}$ at time $t$, satisfying
\begin{equation}
    \frac{\partial \boldsymbol{\phi}_{t_0 \to t}}{\partial t} = \boldsymbol{u}(\boldsymbol{\phi}_{t_0 \to t}, t), \quad \boldsymbol{\phi}_{t_0 \to t_0} = \boldsymbol{X},
    \label{eq:forward_map}
\end{equation}
then the solution to the advection equation can be written as
\begin{equation}
    \varphi(\boldsymbol{x}, t) = \varphi_0(\boldsymbol{\psi}_{t \to t_0}(\boldsymbol{x})),
    \label{eq:characteristic_solution}
\end{equation}
where $\boldsymbol{\psi}_{t \to t_0}$ denotes the backward (inverse) flow map and $\varphi_0$ is the initial level set function. This formulation shows that, in principle, the value of the advected scalar field at any point in space and time can be recovered exactly by tracing back to the initial configuration, free of the interpolation diffusion introduced by repeated grid resampling. We note an important distinction: the recovered value is the exact transport of the initial field $\varphi_0$, so every isolevel, and in particular the interface, is advected exactly; away from the zero isocontour, however, the transported field is in general no longer the signed distance function of the deformed interface. Exact value transport therefore does not preserve the signed-distance property, and redistancing remains part of any practical scheme.

The concept of characteristic mapping builds upon semi-Lagrangian advection schemes~\cite{bargteil2006semi, stam1999stable}, which approximate the backward map over a single time step:
\begin{equation}
    \varphi(\boldsymbol{x}, t^{n+1}) \approx \varphi(\boldsymbol{x} - \boldsymbol{u}\Delta t, t^n).
\end{equation}
However, classical semi-Lagrangian methods construct this mapping only within a single time step, and repeated interpolation still introduces numerical dissipation. Long-range characteristic mapping methods~\cite{mercier2020characteristic, yin2021characteristic, yin2023characteristic} extend this concept by maintaining mappings over multiple time steps, storing the reference coordinates $\boldsymbol{X}$ on grid nodes so that each grid point at time $t$ records its original coordinates at time $t_0$. Similarly, reference-map techniques~\cite{li2023garmls} track material deformation through an Eulerian reference map $\boldsymbol{\xi}(\boldsymbol{x}, t)$ and have demonstrated strong volume preservation for level set methods. Quantities can then be transported directly from the initial frame via Eq.~\eqref{eq:characteristic_solution}, avoiding the accumulation of interpolation errors. This framework has also been combined with gradient-augmented formulations~\cite{taylor2023projection} for transport on spheres and other manifolds. However, since these methods store the mapping on fixed Eulerian grids, they still incur interpolation errors when evaluating the map at non-grid locations.

\noindent\textbf{Particle Flow Map}
To overcome the interpolation limitations of grid-based flow maps, the Particle Flow Map (PFM) method~\cite{zhou2024eulerian} builds upon the observation that Lagrangian particles inherently encode bidirectional flow maps without grid interpolation. Each particle stores both its current position $\boldsymbol{x}_p(t)$ and its initial position $\boldsymbol{X}_p = \boldsymbol{x}_p(t_0)$, thereby providing direct samples of both the forward map $\boldsymbol{\phi}$ and the backward map $\boldsymbol{\psi}$:
\begin{equation}
    \boldsymbol{x}_p(t) = \boldsymbol{\phi}_{t_0 \to t}(\boldsymbol{X}_p), \quad \boldsymbol{X}_p = \boldsymbol{\psi}_{t \to t_0}(\boldsymbol{x}_p(t)).
\end{equation}
To achieve higher-order accuracy, PFM also evolves the forward Jacobian $\mathcal{F}_p = \partial \boldsymbol{\phi}_{t_0 \to t} / \partial \boldsymbol{X}$ along each particle trajectory:
\begin{equation}
    \frac{d\mathcal{F}_p}{dt} = \nabla \boldsymbol{u}(\boldsymbol{x}_p, t) \cdot \mathcal{F}_p, \quad \mathcal{F}_p(t_0) = \mathbf{I}.
    \label{eq:deformation_gradient}
\end{equation}
By discretizing the flow map into a set of Lagrangian particles that advect both position and Jacobian information, PFM obtains accurate flow maps implicitly without auxiliary computation or historical velocity buffers. The transfer of information between particles and the Eulerian grid follows the Affine Particle-In-Cell (APIC)  style formulation~\cite{jiang2015affine}, which enables particles to carry higher-order information transferred to grid nodes via weighted Taylor expansion $q_i = \sum_{p} [ q_p + \nabla q_p \cdot (\boldsymbol{x}_i - \boldsymbol{x}_p) ] \omega_{ip} / \sum_{p} \omega_{ip}$, reducing numerical dissipation and achieving high-order accuracy in long-range transport without global map reconstruction.

Flow-map methods have been developed for impulse transport in incompressible fluids~\cite{zhou2024eulerian,nabizadeh2022covector,deng2023neural} and for vorticity transport~\cite{wang2025fluid,wang2024eulerian}, and have been extended to further settings and formulations~\cite{chen2025fluid,chen2024solid,li2024particle,li2025clebsch,li2025edge,wang2025cirrus,li2025adjoint}. These methods motivate the use of long-range flow-map transport for interface geometry.
However, the application of particle flow maps to interface tracking problems, particularly level set methods with narrow-band confinement and high-order geometric preservation, remains unexplored.

Although particle level set, gradient-augmented level set, and reference-map methods have improved interface tracking from different directions, they still leave a gap between Lagrangian geometric correction and long-range characteristic transport.
Particle flow map methods achieve long-range transport accuracy through bidirectional flow maps, but have not been applied to level set interface tracking. In contrast to ``virtual-particle'' approaches~\cite{hieber2005lagrangian, enright2002hybrid, ferstl2016narrow, fedkiw2001visual} that reconstruct flow trajectories through backward tracing and incur significant temporal buffering and substepping overhead, PFM establishes a compact, physically consistent bridge between Lagrangian advection and Eulerian projection. This motivates our central question: can we combine the high-order geometric representation of GALS with the long-range flow map accuracy of PFM to achieve high-fidelity level set advection while maintaining the efficiency of narrow-band formulations?

\subsection{Particle Flow Map Level Set}

Building upon these insights, we present a Particle Flow Map Level Set (PFM-LS) method that uses particle-carried flow maps as the primary mechanism for transporting level-set geometry. By transporting
$\varphi$, $\nabla\varphi$, and $\nabla^2\varphi$ on particles according to
their respective differential forms, we achieve a high-order geometric
representation that remains accurate under complex flow deformations.
The proposed framework confines particles to a narrow band
$\Omega_\epsilon = \{\boldsymbol{x} : |\varphi(\boldsymbol{x})| < \epsilon\}$
around the interface, significantly reducing computational cost while focusing
resources where accuracy is most critical.
Each particle $p$ stores the level-set value $\varphi_p$
(transported as a 3-form), its gradient $\boldsymbol{g}_p = \nabla\varphi_p$
(transported as a 1-form), and Hessian $\boldsymbol{H}_p = \nabla^2\varphi_p$.
These quantities are transported along particle trajectories according to their
respective differential structure:
\begin{equation}
    \varphi_p(t) = \varphi_p(0), \qquad
    \boldsymbol{g}_p(t) = \mathcal{T}_p^\top \boldsymbol{g}_p(0), \qquad
    \boldsymbol{H}_p(t) = \mathcal{T}_p^\top \boldsymbol{H}_p(0)\,\mathcal{T}_p + \nabla\mathcal{T}_p^\top\,\boldsymbol{g}_p(0),
    \label{eq:ls_transport}
\end{equation}
where the level set value is preserved along characteristics, and the gradient
transforms via the transposed backward Jacobian $\mathcal{T}_p^\top$.
The Hessian evolution, given here for completeness, additionally involves the spatial derivative $\nabla\mathcal{T}_p$ of the flow-map Jacobian; the treatment of this term is detailed in Section~\ref{subsec-advection}.

We reconstruct the level set field using APIC-style~\cite{jiang2015affine}
interpolation with second-order Taylor expansion, leveraging the gradient and
Hessian carried by each particle to capture sub-grid features such as thin
filaments and small droplets. To control error accumulation over time, we adopt
a dual-timescale flow map strategy: a long-range map preserves $\varphi$ and
$\nabla\varphi$ over extended periods, while a frequently reinitialized
short-range map handles the distortion-sensitive Hessian. Since particles may
disperse during advection, adaptive particle control dynamically generates new
particles when local density falls below a threshold, ensuring sufficient
coverage near evolving interfaces. Finally, to maintain the signed-distance
property essential for curvature computation, we employ a hybrid particle-grid
redistancing scheme that combines quasi-Newton projection with
gradient-consistency blending, preserving fine-scale features captured by
particles. Together, these components enable accurate reconstruction of complex
interface geometries that are typically lost in conventional grid-based schemes.

The main advancement of PFM-LS is to change the role of particles from auxiliary error-correction samples to carriers of the evolving interface geometry.
Our method differs from existing approaches in several key aspects. Unlike GALS and reference-map methods that store and evolve geometric information on fixed Eulerian grids, our method leverages particles to carry high-order geometric information through accurate long-range flow maps, avoiding the interpolation errors inherent in grid-based characteristic mapping. Unlike previous particle level set methods that use particles primarily to correct the level set function post hoc, our particles directly encode the interface geometry through their flow maps and higher-order derivative information, serving as the primary carriers of interface evolution rather than auxiliary correction tools.
Compared to the original PFM framework~\cite{zhou2024eulerian}, which transports impulse across the entire domain, PFM-LS adapts particle flow maps to level set interface tracking by transporting $\varphi$ (conserved along characteristics), $\nabla\varphi$ (covector via $\mathcal{T}^\top$), and $\nabla^2\varphi$ (via $\mathcal{T}^\top (\nabla^2\varphi)\,\mathcal{T}$). This adaptation further introduces narrow-band confinement, adaptive particle control, hybrid particle-grid quasi-Newton redistancing, and a dual-timescale flow map tailored to the distinct sensitivity of level set quantities.

The key contributions of our work include:
\begin{enumerate}
    \item A framework that integrates particle-based tracking with flow maps to achieve fourth-order accurate interface representation and advection, where particles carry level set values $\varphi$, gradients $\boldsymbol{g}$, and Hessians $\boldsymbol{H}$ that are transported exactly along characteristics.

    \item A high-order particle-to-grid transfer scheme that preserves geometric information within a narrow band around the interface, using the transported derivatives to enable accurate Hermite interpolation during reconstruction.

    \item An adaptive particle control mechanism that maintains consistent particle density near the interface during advection, including seeding, deletion, and redistribution strategies that preserve flow map accuracy.

    \item A novel particle-based quasi-Newton projection method for redistancing that preserves fine-scale features during interface reconstruction, avoiding the smoothing typically introduced by PDE-based reinitialization.
\end{enumerate}

The remainder of this paper is organized as follows. In Section~\ref{sec-FM}, we present the mathematical background of our method, including level sets, gradient-augmented level sets, flow maps, and particle flow maps. Section~\ref{sec-ls} details our particle flow map approach for gradient-augmented level sets, covering advection, reconstruction, and narrow band implementation. Section~\ref{sec-re} focuses on reinitialization procedures, including flow map reinitialization and our particle Newton redistancing method. Section~\ref{sec:numerical} presents numerical results comparing our method against a wide range of existing approaches. Finally, Section~\ref{conclusion} concludes with a discussion of our findings and directions for future work.
\section{Mathematical Background}\label{sec-FM}
In this section, we present the mathematical foundations of our method, including level sets, gradient-augmented level sets, flow maps, and particle flow maps. We first establish the notation used throughout the paper in Table~\ref{tab:notations}, then provide the necessary background on each concept.

\begin{table}[t]
\centering
\caption{Summary of important notations used in the paper.}
\label{tab:notations}
\begin{tabular}{cll}
\toprule
Notation & Type & Definition \\
\midrule
$\varphi$ & scalar & level set function \\
$\Omega_0$ & region & air domain ($\varphi > 0$) \\
$\Omega_1$ & region & liquid domain ($\varphi < 0$) \\
$\partial\Omega$ & manifold & interface between materials ($\varphi = 0$) \\
$\nabla\varphi$, $\boldsymbol{g}$ & vector & gradient of level set function \\
$\nabla^2\varphi$, $\boldsymbol{H}$ & matrix & Hessian of level set function \\
$\mathbf{n}$ & vector & unit normal vector at interface ($\frac{\nabla\varphi}{|\nabla\varphi|}$) \\
$\mathbf{v}$, $\boldsymbol{u}$ & vector & velocity field $(u, v, w)^T$ \\
$\boldsymbol{X}$ & vector & initial position of material point \\
$\boldsymbol{x}$ & vector & current position of material point \\
$\boldsymbol{\phi}$ & function & forward flow map from initial to current position \\
$\boldsymbol{\psi}$ & function & backward flow map from current to initial position \\
$\mathcal{F}$ & matrix & Jacobian of forward flow map $\boldsymbol{\phi}$ \\
$\mathcal{T}$ & matrix & Jacobian of backward flow map $\boldsymbol{\psi}$ \\
$\mathcal{F}_{[c,a]}$ & matrix & forward Jacobian from $c$ to $a$ \\
$\mathcal{T}_{[a,c]}$ & matrix & backward Jacobian from $a$ to $c$ \\
$\omega$ & function & interpolation weight function \\

\bottomrule
\end{tabular}
\end{table}
\subsection{Level Sets}
In this paper, we focus on level set methods \cite{osher1988fronts} for interface tracking between two materials in $\mathbb{R}^2$. Without loss of generality, we denote materials as 0 (air) and 1 (liquid). The level set function $\varphi(\boldsymbol{x}, t)$ is defined in the entire domain and satisfies:
\begin{equation}
\varphi(\boldsymbol{x}, t) =
\begin{cases}
< 0, & \boldsymbol{x} \in \Omega_1, \\
= 0, & \boldsymbol{x} \in \partial\Omega, \\
> 0, & \boldsymbol{x} \in \Omega_0,
\end{cases}
\end{equation}
where $\Omega_1$ represents the liquid region, $\Omega_0$ represents the air region, and $\partial\Omega$ denotes the liquid-air interface, which is a codimension-1 geometric structure (i.e., a curve in $\mathbb{R}^2$, or a surface in $\mathbb{R}^3$). This implicit representation enables direct computation of key geometric properties from the level set function. The unit normal vector at any point on the interface is given by $\mathbf{n} = \frac{\nabla\varphi}{|\nabla\varphi|}$, pointing from the liquid region toward the air region. The mean curvature is computed as $\kappa = \nabla \cdot \mathbf{n} = \nabla \cdot \left(\frac{\nabla\varphi}{|\nabla\varphi|}\right)$, which quantifies the local geometry of the interface. The evolution of the interface under a velocity field $\mathbf{v} = (u, v, w)^T$ is governed by the advection equation $\frac{\partial\varphi}{\partial t} + \mathbf{v} \cdot \nabla\varphi = 0$. For numerical robustness $\varphi$ should remain close to a signed distance function ($|\nabla\varphi|=1$); the classical Eulerian discretization and reinitialization choices reviewed in the introduction address these two requirements, and our treatment of both is presented in Sections~\ref{sec-ls} and~\ref{sec-re}.

The standard level set method often suffers from numerical dissipation during advection. \citet{enright2002hybrid} addressed this by introducing marker particles near the interface $\varphi=0$. These particles, with positions $\mathbf{x}_p$, radii $r_p$, and a sign marking the side of the interface each particle belongs to, are advected alongside the level set and help detect and correct interface errors when they cross the zero isocontour incorrectly. This hybrid approach demonstrates that incorporating particles into the level set framework is both natural and significantly improves accuracy for complex interface dynamics.

\noindent\textbf{Level Set Gradient}
\citet{nave2010gradient} introduced a method that maintains the level set's gradient $\nabla\varphi$ as an auxiliary field and uses it to enhance the level-set function's local expressiveness through high-order interpolation. This approach significantly improves the accuracy of the level set method and related interface property calculations, such as curvature. \citet{li2023garmls} and \citet{nave2010gradient} demonstrated that, with the aid of a fourth-order Hermite interpolation scheme on an Eulerian grid, they could better preserve droplets or thin films much smaller than the grid resolution. To reinitialize the gradient-augmented level set, quasi-Newton methods are typically employed to reinitialize both $\varphi$ and $\nabla\varphi$ around the interface by iteratively solving for the nearest point on $\partial\Omega$.

\subsection{Flow Map}
Consider interfacial dynamics governed by a spatiotemporal velocity field
$\boldsymbol{u}(\boldsymbol{q}, \tau)$ evolving from time $0$ to time $t$. We denote the initial position
of a material point as $\boldsymbol{X}$ and its current position at time $t$ as $\boldsymbol{x}$. The fluid motion over this time interval can be
characterized by a forward flow map $\boldsymbol{\phi}$, defined as:

\begin{equation}
\begin{cases}
\frac{\partial \boldsymbol{\phi}(\boldsymbol{X}, \tau)}{\partial \tau} = \boldsymbol{u}[\boldsymbol{\phi}(\boldsymbol{X}, \tau), \tau], \\
\boldsymbol{\phi}(\boldsymbol{X}, 0) = \boldsymbol{X}, \\
\boldsymbol{\phi}(\boldsymbol{X}, t) = \boldsymbol{x}.
\end{cases}
\end{equation} The backward flow map $\boldsymbol{\psi}$ represents the inverse mapping of the
forward flow map and is defined as:

\begin{equation}
\begin{cases}
\boldsymbol{\phi}(\boldsymbol{\psi}(\boldsymbol{x}, \tau), \tau) = \boldsymbol{x}. \\
\boldsymbol{\psi}(\boldsymbol{x}, 0) = \boldsymbol{x}, \\
\boldsymbol{\psi}(\boldsymbol{x}, t) = \boldsymbol{X}.
\end{cases}
\end{equation}We denote the Jacobian matrices of $\boldsymbol{\phi}$ and $\boldsymbol{\psi}$ by $\mathcal{F}$ and $\mathcal{T}$, respectively:

\begin{equation}
\mathcal{F}(\boldsymbol{X}, \tau) = \frac{\partial \boldsymbol{\phi}(\boldsymbol{X}, \tau)}{\partial \boldsymbol{X}}, \quad \mathcal{T}(\boldsymbol{x}, \tau) = \frac{\partial \boldsymbol{\psi}(\boldsymbol{x}, \tau)}{\partial \boldsymbol{x}}.
\end{equation} As proven in \cite{CambridgeBook}, the temporal evolution of
$\mathcal{F}$ and $\mathcal{T}$ is described by:
\begin{equation}\label{eq:evolution}
\frac{\mathrm{\partial}\mathcal{F}}{\mathrm{\partial}t} = \nabla\boldsymbol{u}\mathcal{F}, \quad \frac{\mathrm{D}\mathcal{T}}{\mathrm{D}t} = -\mathcal{T}\nabla\boldsymbol{u}.
\end{equation}
The operator $\frac{D(\cdot)}{Dt}$ represents the material derivative, which describes the rate of change of the Jacobians moving with the particle along its flow map's trajectory.

\subsection{Particle Flow Map}
\begin{figure}[!htb]
    \centering
    \includegraphics[width=0.8\linewidth]{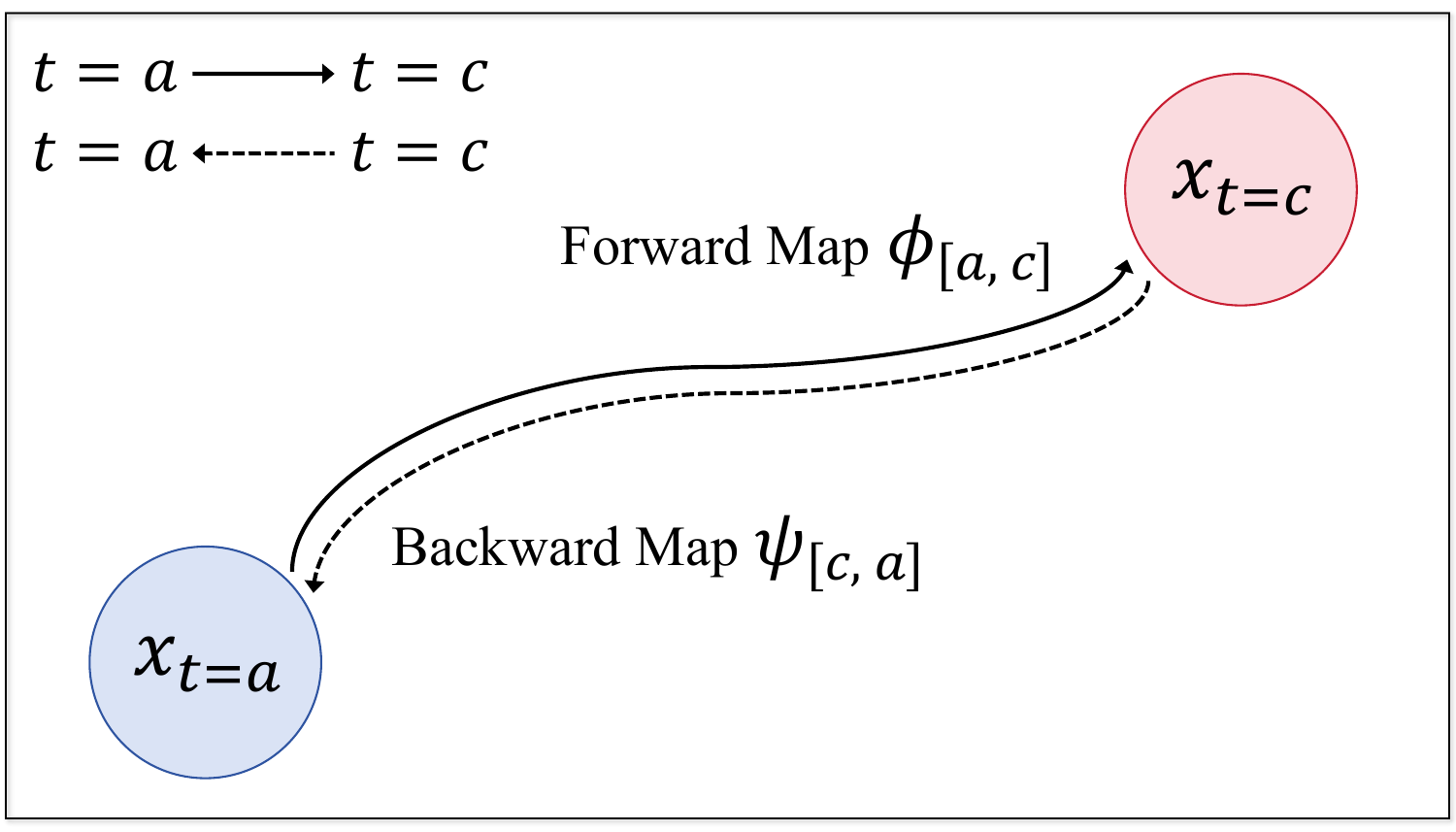}
    \caption{Particle trajectory from time $a$ to $c$ naturally defines both forward map $\boldsymbol{\phi}_{[a,c]}$ and backward map $\boldsymbol{\psi}_{[c,a]}$, forming a perfect bidirectional flow map.}
    \label{fig:pfm_fig}
\end{figure}
Building upon the flow map concept, \citet{zhou2024eulerian} observed that particle trajectories naturally embody bidirectional flow maps. When a particle moves according to velocity field $\boldsymbol{u}(\boldsymbol{x}, t)$, its path from time $a$ to time $c$ defines both the forward map $\boldsymbol{\phi}_{[a,c]}$ and the backward map $\boldsymbol{\psi}_{[c,a]}$ as in Figure~\ref{fig:pfm_fig}. A particle flow map constitutes a perfect flow map by naturally satisfying bidirectionality requirements. A particle moving from $\boldsymbol{X}$ at time $0$ to $\boldsymbol{x}$ at time $t$ via $\boldsymbol{\phi}(\boldsymbol{X}, t)$ will follow the same trajectory in reverse when moving according to $\boldsymbol{\psi}(\boldsymbol{x}, t)$. The Jacobians defined in Equation \eqref{eq:evolution} can be integrated numerically along these trajectories: in practice we advance $\mathcal{T}$ by fourth-order Runge--Kutta integration of $\mathrm{D}\mathcal{T}/\mathrm{D}t=-\mathcal{T}\nabla\boldsymbol{u}$, synchronously with the particle position update (Section~\ref{subsec-advection}), with $\nabla\boldsymbol{u}$ evaluated analytically for the benchmark velocity fields or by grid interpolation in coupled settings. The dominant error sources are the interpolation of $\nabla\boldsymbol{u}$, the accumulated drift of $\mathcal{T}$ over a flow-map window, the long--short map composition, and the particle-to-grid reconstruction. Carrying the flow map and its Jacobians directly on particles in this way provides an efficient means of tracking deformations in complex systems.

\begin{figure}[!htb]
    \centering
    \includegraphics[width=0.99\linewidth]{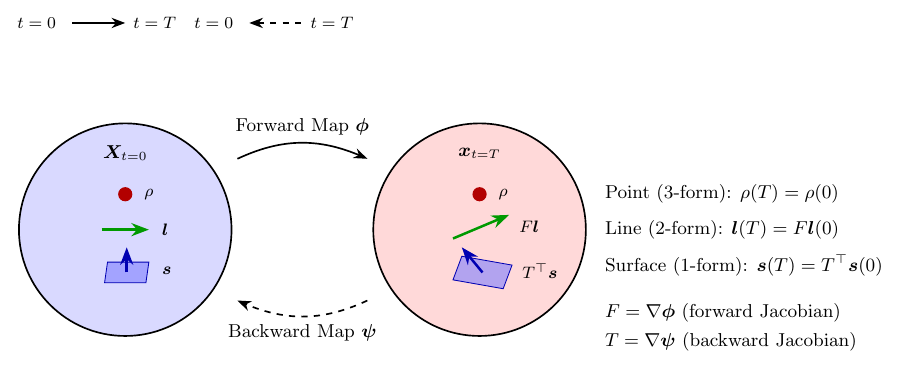}
    \caption{Transport of differential forms under the flow map. Point elements (3-forms, e.g., $\rho$) are preserved; line elements (2-forms, e.g., vorticity) transform via $F = \nabla\boldsymbol{\phi}$; surface elements (1-forms, e.g., impulse) transform via $T^\top$.}
    \label{fig:differential_forms}
\end{figure}

\noindent\textbf{Geometric Insight: Transport of Differential Forms on Particles}
The success of the Particle Flow Map methods stems from a unified geometric principle:
different physical quantities transform according to their differential
structure under flow map transport (see Figure~\ref{fig:differential_forms} and Table~\ref{tab:differential_forms}).
We summarize the three canonical cases:
\begin{table}[!htb]
\centering
\renewcommand{\arraystretch}{1.6}
\resizebox{0.99\linewidth}{!}{
\begin{tabular}{>{\bfseries}l c c c}
\toprule
 & \textbf{Surface Elements (1-forms)} & \textbf{Line Elements (2-forms)} & \textbf{Point Elements (3-forms)} \\
\midrule
Evolution
&
$\displaystyle \frac{D\boldsymbol{s}}{Dt} = -(\nabla \boldsymbol{u})^{\top}\boldsymbol{s}$
&
$\displaystyle \frac{D\boldsymbol{l}}{Dt} = (\nabla \boldsymbol{u})\,\boldsymbol{l}$
&
$\displaystyle \frac{D\rho}{Dt} = 0$
\\
\midrule
Flow Map
&
$\boldsymbol{s}(t) = T^{\top}\boldsymbol{s}(0)$
&
$\boldsymbol{l}(t) = F\,\boldsymbol{l}(0)$
&
$\rho(t) = \rho(0)$
\\
\midrule
Example
&
impulse~\cite{cortez1995impulse,nabizadeh2022covector,feng2022impulse,deng2023neural,zhou2024eulerian}
&
vorticity~\cite{wang2024eulerian, wang2025fluid}
&
density~\cite{stam1999stable, fedkiw2001visual}
\\
\bottomrule
\end{tabular}}
\caption{Differential forms and their transformation under flow maps. The level set value $\varphi$ is transported as a 3-form (point element), while the gradient $\nabla\varphi$ is transported as a 1-form (surface element).}
\label{tab:differential_forms}
\end{table}
\textit{(1) Point elements (3-forms):}
Scalar quantities such as density~\cite{stam1999stable, fedkiw2001visual} are preserved along trajectories:
\begin{equation}
    \frac{D \rho}{Dt} = 0, \qquad \rho(t) = \rho(0).
\end{equation}

\textit{(2) Line elements (2-forms):}
Quantities such as vorticity~\cite{wang2024eulerian, wang2025fluid}
stretch and rotate via the velocity gradient:
\begin{equation}
    \frac{D \boldsymbol{l}}{Dt} = (\nabla \boldsymbol{u}) \boldsymbol{l}, \qquad
    \boldsymbol{l}(t) = \mathcal{F} \, \boldsymbol{l}(0).
\end{equation}

\textit{(3) Surface elements (1-forms):}
Quantities such as impulse~\cite{cortez1995impulse,nabizadeh2022covector,feng2022impulse,deng2023neural,zhou2024eulerian}
transform via the transposed Jacobian:
\begin{equation}
    \frac{D \boldsymbol{s}}{Dt} = -(\nabla \boldsymbol{u})^\top \boldsymbol{s}, \qquad
    \boldsymbol{s}(t) = \mathcal{T}^\top \boldsymbol{s}(0).
\end{equation}

A key observation is that the gradient of a 3-form yields a 1-form, a classical identity in the transport of differential forms (see, e.g., \cite{marsden1983mathematical,abraham2012manifolds}), restated here in the flow-map setting. This principle directly applies to level set methods: the level set value $\varphi$ is a scalar (3-form) preserved along characteristics, and its gradient $\nabla\varphi$ transforms as a 1-form---precisely analogous to the relationship between density and impulse. This observation is consistent with GALS methods~\cite{nave2010gradient}, which evolve both $\varphi$ and $\nabla\varphi$ as independent variables, but interprets them through the lens of differential forms. Understanding level set advection in this way allows us to unify interface tracking within the PFM framework, with the value, gradient, and Hessian evolving as:
\begin{equation}
    \varphi(t) = \varphi(0), \qquad
    \nabla\varphi(t) = \mathcal{T}^\top \nabla\varphi(0), \qquad
    \nabla^2\varphi(t) = \mathcal{T}^\top \nabla^2\varphi(0)\,\mathcal{T} + \nabla\mathcal{T}^\top\, \nabla\varphi(0).
\end{equation}

\section{Level Set on Particle Flow Maps}\label{sec-ls}
\begin{figure}[!htb]
    \centering
    \includegraphics[width=0.99\linewidth]{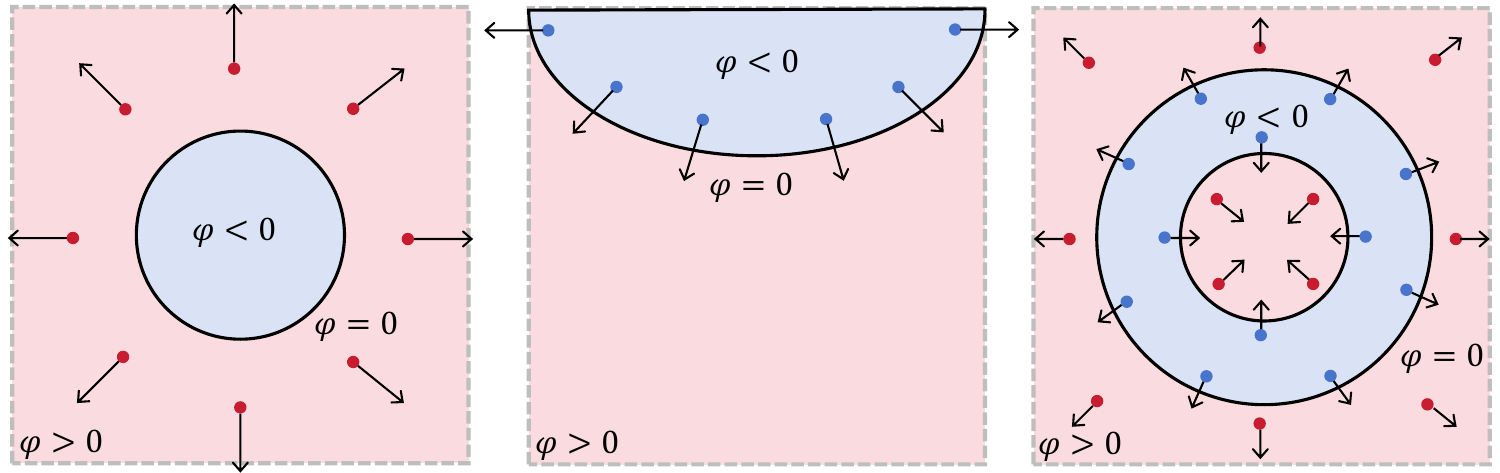}
    \caption{Illustration of gradient-augmented level sets with particles. Red and blue particles carry level set values and gradient information (arrows) near the interface $\varphi = 0$. This higher-order representation enables capturing sub-grid features such as small droplets (left), thin films (middle), and annular structures (right) that traditional methods cannot resolve.}
    \label{fig:particle_with_grad}
\end{figure}

We use the particle flow map to track the level set $\varphi$ and its higher-order information gradient $\nabla \varphi$ and hessian $\nabla^2\varphi$. As mentioned in \cite{zhou2024eulerian}, a perfect flow map can maintain exceptionally high accuracy during advection, and particles naturally follow a perfect flow map.
As illustrated in Figure~\ref{fig:particle_with_grad}, particles carrying both level set values and higher gradient information provide an APIC-style~\cite{jiang2015affine} high-order interpolation that significantly enhances the representation of sub-grid features. This approach allows our method to preserve fine structures such as small droplets and thin films during advection, even when these features are smaller than the grid resolution.

\subsection{Particle Advection}\label{subsec-advection}
We now describe how to advect the level set function $\varphi$, its gradient $\nabla\varphi$, and its Hessian $\nabla^2\varphi$ using particle flow maps.

We emphasize that particles do not represent the interface directly. Instead, particles serve as carriers of the level-set field ($\varphi$, $\nabla\varphi$, $\nabla^2\varphi$) and are advected by the same velocity field $\boldsymbol{u}$ that governs the level-set equation~\eqref{eq:ls_advection}. The interface is always extracted as the zero isocontour of the reconstructed $\varphi$ field (Section~\ref{subsec-recon}). There is no separate ``particle interface''---particles provide a high-order interpolant of the continuous level-set field, ensuring consistency between Lagrangian transport and Eulerian reconstruction.

The level set value is directly conserved along particle trajectories. Given the backward flow map $\boldsymbol{\psi}(\boldsymbol{x}, t)$, the advection follows:
\begin{equation}
\varphi(\boldsymbol{x}, t) = \varphi(\boldsymbol{\psi}(\boldsymbol{x}, t), 0)
\end{equation} For the gradient field, let $\boldsymbol{g} = \nabla\varphi$ denote the gradient. Using the transpose of the backward deformation gradient:
\begin{equation}
\boldsymbol{g}(\boldsymbol{x}, t) = \mathcal{T}_t^T(\boldsymbol{x}) \boldsymbol{g}(\boldsymbol{\psi}(\boldsymbol{x}, t), 0)
\end{equation}
where $\mathcal{T}_t(\boldsymbol{x}) = \frac{\partial \boldsymbol{\psi}(\boldsymbol{x}, t)}{\partial \boldsymbol{x}}$ is the Jacobian of the backward flow map. We emphasize that exact transport of $\varphi$ along characteristics preserves its values but not the signed-distance property $|\nabla\varphi|=1$ (except under rigid motion); redistancing (Section~\ref{subsec-renewton}) therefore remains necessary. For the Hessian matrix, let $\boldsymbol{H} = \nabla^2\varphi$ denote the Hessian. The transformation incorporates both deformation and its spatial variation:
\begin{equation}
\boldsymbol{H}(\boldsymbol{x}, t) = \mathcal{T}_t^T \nabla_{\psi}\boldsymbol{g}(\boldsymbol{\psi}(\boldsymbol{x}, t), 0) \mathcal{T}_t + \nabla\mathcal{T}_t^T \boldsymbol{g}(\boldsymbol{\psi}(\boldsymbol{x}, t), 0)
\end{equation}
where $\nabla_{\psi}$ denotes the gradient with respect to the variable $\boldsymbol{\psi}$. In practice, we use an approximate form
\begin{equation}
\boldsymbol{H}(\boldsymbol{x}, t) \approx \mathcal{T}_t^T \nabla_{\psi}\boldsymbol{g}(\boldsymbol{\psi}(\boldsymbol{x}, t), 0) \mathcal{T}_t,
\end{equation}
following \citet{zhou2024eulerian}, who found that the second term $\nabla\mathcal{T}_t^T \boldsymbol{g}$ is negligible in practice. This is because the short-range flow map $\mathcal{T}_{[b,c]}$ is reinitialized every $f_S$ steps (typically 5), keeping $\mathcal{T}_{[b,c]}$ close to the identity $\mathbf{I}$ and thus $\nabla\mathcal{T}_{[b,c]} \approx 0$. Moreover, computing $\nabla\mathcal{T}_t$ requires third-order spatial derivatives of the velocity field, which are expensive and numerically noisy. The frequent short-range reinitialization ensures that the approximation error remains small. We validate this approximation directly with an ablation study (Section~\ref{subsec:hessian-ablation}): across the LeVeque and distorted-circle benchmarks, transporting the Hessian with the approximate form matches the alternatives to within $0.1$ percentage points of volume preservation, confirming that the effect of the neglected term is negligible in these tests, while retaining the Hessian preserves the sub-grid reconstruction of Section~\ref{subsec-recon} and provides curvature access for the free-surface coupling of Section~\ref{subsec:freesurface}.

By tracking these three quantities using particle flow maps, we achieve a higher-order representation that maintains geometric fidelity during complex interface deformations. While the perfect flow map approach provides high accuracy, numerical errors can accumulate over time. To balance accuracy and efficiency, we implement a dual-timescale flow map approach. We store two backward Jacobians: $\mathcal{T}_{[a,b]}$ and $\mathcal{T}_{[b,c]}$, where $a$ is the initial time, $c$ is the current time, and $b$ is an intermediate time closer to $c$. The long backward Jacobian $\mathcal{T}_{[a,c]}$ can be computed as $\mathcal{T}_{[a,c]} = \mathcal{T}_{[a,b]}\mathcal{T}_{[b,c]}$. At each time step, we update the short-range flow map by marching $\mathcal{T}_{[b,c]}$ by one step in parallel with the particle's advection, using a fourth-order Runge--Kutta (RK4) scheme; ``custom'' refers to integrating the particle position and the Jacobian $\mathcal{T}_{[b,c]}$ within one synchronized RK4 loop (Algorithm~\ref{alg:nbpfm}). Subsequently, the long-range flow map is updated according to $\mathcal{T}_{[a,c]} = \mathcal{T}_{[a,b]}\mathcal{T}_{[b,c]}$, with $\mathcal{T}_{[a,b]}$ which has been reinitialized at time $b$. The long flow map is used for advecting the level set value $\varphi$ and its gradient $\nabla\varphi$, while the short flow map is used for the Hessian $\nabla^2\varphi$, which is more sensitive to flow distortion. Details of the flow map reinitialization process are presented in Section~\ref{subsec-pfmreinit}. In summary, RK4 integration appears in three places: (i) particle trajectories, (ii) the short-range Jacobian $\mathcal{T}_{[b,c]}$ along each trajectory, and (iii) the semi-Lagrangian backtracing used for the background grid outside the narrow band.

Next, we will describe how to utilize the level set value $\varphi$, gradient $\boldsymbol{g}$, and Hessian $\boldsymbol{H}$ stored on particles to interpolate and reconstruct a highly accurate, feature-preserving level set function throughout the computational domain.

\subsection{Grid Reconstruction}\label{subsec-recon}
Having tracked the level set values, gradients, and Hessians on particles through advection, we now need to reconstruct the continuous level set field throughout the domain. To leverage the higher-order information carried by particles, we adopt an APIC-style \cite{jiang2015affine} interpolation approach. For any query point $\boldsymbol{x}_\alpha$ in the domain, we reconstruct the level set value using a second-order Taylor series expansion:
\begin{equation}
\varphi_\alpha = \frac{\sum_{p \in N_\alpha} \left[ \varphi_p + \boldsymbol{g}_p \cdot (\boldsymbol{x}_\alpha - \boldsymbol{x}_p) + \frac{1}{2}(\boldsymbol{x}_\alpha - \boldsymbol{x}_p)^T \boldsymbol{H}_p (\boldsymbol{x}_\alpha - \boldsymbol{x}_p) \right] \omega(\boldsymbol{x}_p, \boldsymbol{x}_\alpha)}{\sum_{p \in N_\alpha} \omega(\boldsymbol{x}_p, \boldsymbol{x}_\alpha)}
\end{equation}
where $N_\alpha = \{p \in P \mid \omega(\boldsymbol{x}_p, \boldsymbol{x}_\alpha) > 0\}$ represents the set of particles in the neighborhood of query point $\alpha$, and $\omega(\boldsymbol{x}_p, \boldsymbol{x}_\alpha)$ is the interpolation weight. We use the tensor-product quadratic B-spline kernel $\omega(\boldsymbol{x}_p, \boldsymbol{x}_\alpha) = \prod_{d} N_2\!\left(\frac{|x_p^d - x_\alpha^d|}{\Delta x}\right)$, where $N_2(s) = \frac{3}{4} - s^2$ for $|s| \leq \frac{1}{2}$, $N_2(s) = \frac{1}{2}({\frac{3}{2} - |s|})^2$ for $\frac{1}{2} < |s| \leq \frac{3}{2}$, and $0$ otherwise. This kernel has a support radius of $1.5\Delta x$. Similarly, for the gradient field $\boldsymbol{g}_\alpha$ at the query point, we use a first-order Taylor series expansion of the gradient:
\begin{equation}
\boldsymbol{g}_\alpha = \frac{\sum_{p \in N_\alpha} \left[ \boldsymbol{g}_p + \boldsymbol{H}_p (\boldsymbol{x}_\alpha - \boldsymbol{x}_p) \right] \omega(\boldsymbol{x}_p, \boldsymbol{x}_\alpha)}{\sum_{p \in N_\alpha} \omega(\boldsymbol{x}_p, \boldsymbol{x}_\alpha)}
\end{equation}

These formulations naturally extend our particle-based flow map approach by incorporating both gradient and second-order derivative information in the reconstruction process. The first-order term $\boldsymbol{g}_p \cdot (\boldsymbol{x}_\alpha - \boldsymbol{x}_p)$ captures local directional changes, while the second-order term $\frac{1}{2}(\boldsymbol{x}_\alpha - \boldsymbol{x}_p)^T \boldsymbol{H}_p (\boldsymbol{x}_\alpha - \boldsymbol{x}_p)$ accounts for the local variations in the gradient of the level set field. This higher-order reconstruction preserves sharp features and complex interface geometries that would otherwise be lost with standard interpolation methods, aligning with our goal of maintaining the high accuracy promised by the perfect flow map described in Section \ref{subsec-advection}.

\subsection{Narrow Band}\label{subsec-nbpfm}
While the level set method provides an elegant framework for interface tracking, only the level set values near the interface are critical for capturing the interface geometry. Inspired by \cite{enright2002hybrid} and \cite{ferstl2016narrow}, we enhance our PFM-LS approach by adopting a narrow band strategy, which focuses computational resources where they matter most.
\begin{figure}
    \centering
    \includegraphics[width=0.99\linewidth]{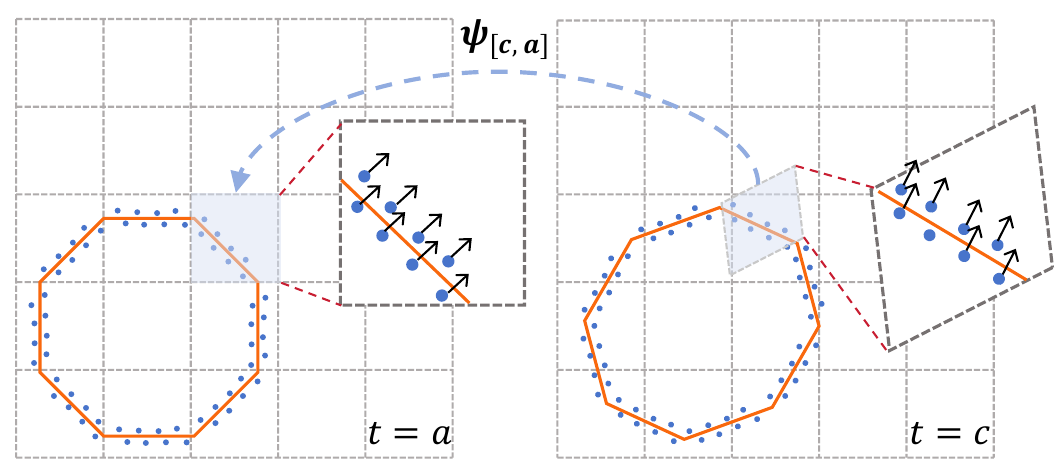}
    \caption{Narrow band particle flow map level set representation. The octagonal interface (orange) is tracked by particles (blue dots) concentrated only in a narrow band around the interface. The zoomed regions show particles carrying gradient information (arrows) at two time points, with the backward map $\boldsymbol{\psi}_{[c, a]}$ connecting their positions. This narrow-band approach maintains high accuracy near the interface while reducing computational cost elsewhere.}
    \label{fig:narrowband_fig}
\end{figure}
We define the narrow band as a region of width $n\Delta x$ surrounding the interface, where $n=4$ is used in all experiments unless otherwise noted and $\Delta x$ denotes the background grid cell size. As shown in Figure~\ref{fig:narrowband_fig}, our method concentrates particles only near the interface, tracking their level set values, gradients, and Hessians through the backward flow map $\boldsymbol{\psi}_{[c,a]}$. This narrow band approach significantly reduces computational cost while maintaining high accuracy near the interface. This focused approach offers significant advantages over traditional flow maps or reference maps \cite{deng2023neural, li2023garmls}. Specifically, the particle flow map naturally aligns with this narrow band concept, as we only need to maintain particles within this region of interest.

The narrow band particle flow map substantially reduces memory requirements and computational overhead compared to a full-domain approach. Outside the narrow band, where high interface accuracy is not needed, we use a standard grid-based representation of the level set with conventional semi-Lagrangian advection. To ensure continuity of the level set field across the entire domain, we apply a smooth transition within a small region ($\Delta x$ width) at the boundary between the narrow band and background grid. In this transition region, we blend the values as $\varphi_{\text{cross}} = w \cdot \varphi_{\text{particle}} + (1-w) \cdot \varphi_{\text{grid}}$, where $w$ is a blending weight that linearly decreases from 1 at the inner boundary to 0 at the outer boundary of this transition region. Since the transition region lies at a distance of $n\Delta x$ from the interface (the outer boundary of the narrow band), and the interface geometry is determined entirely by the particle-based representation within the band, the blending has negligible impact on the zero isocontour.

A challenge with particle flow maps is that particles may accumulate or disperse near the interface during advection if not reinitialized frequently, leading to non-uniform particle distribution and potential accuracy degradation. While \citet{zhou2024eulerian} use particle flow maps for impulse advection, which inherently requires more frequent reinitialization than level sets, PFM for level sets can maintain longer flow periods with fewer reinitializations to better preserve interface features. Additionally, unlike \cite{zhou2024eulerian}, which needs particles spread across the entire domain, our narrow band PFM only maintains particles within the narrow band region. This concentration of particles makes Narrow band PFM-LS more susceptible to particle dispersion near the interface, precisely where adequate particle density is most critical for accuracy. To address this issue, we implement adaptive particle control similar to the approach in NBFLIP \cite{ferstl2016narrow}, which dynamically manages particle density to maintain consistent accuracy throughout the simulation. The details of this adaptive particle control mechanism will be presented in the following section.

\subsection{Adaptive Particle Control}\label{subsec-adaptive}

A key challenge in the narrow-band PFM-LS framework is maintaining sufficient particle density near the interface during advection.
Standard PFM implementations \cite{zhou2024eulerian, chen2024solid, li2024particle} use periodic flow-map reinitialization (Section~\ref{sec-re}) to restore uniform coverage, but particle loss within the narrow band between reinitializations can degrade accuracy.
To address this, we introduce an adaptive particle control mechanism.

A background grid, matching the simulation resolution, monitors particle counts in each narrow-band cell.
We define a minimum threshold $n_{\min}$---typically half of the reinitialization density---and dynamically generate new particles when the count drops below $n_{\min}$.
New particles are seeded at jittered positions within the deficient cell: a regular sub-grid is constructed with spacing $\Delta x / \sqrt{n_{\text{ppc}}}$ in 2D (or $\Delta x / \sqrt[3]{n_{\text{ppc}}}$ in 3D), and each position is perturbed by a small random offset to avoid grid-alignment artifacts.
Particles are not deleted, as overpopulation is naturally corrected during the next reinitialization. For each newly generated particle, the level-set value $\varphi$ and gradient $\boldsymbol{g}$ are interpolated from nearby particles using the APIC-style scheme (Section~\ref{subsec-recon}).
The Hessian $\boldsymbol{H}$ is estimated via finite differences of the reconstructed gradient field, improving accuracy over direct interpolation. The finite-difference estimate is preferred at seeding time because newly created particles carry no transported Hessian history, and differencing the reconstructed gradient field aggregates information from all neighboring particles; in the Hessian-handling ablation on a representative scenario (Section~\ref{subsec:hessian-ablation}), the deviation between such estimation variants was below $0.1$ percentage points of volume error.
The flow-map Jacobians $\mathcal{T}_p$ and $\mathcal{F}_p$ are initialized as identity matrices since the particles are created within the current time step.

This adaptive control maintains adequate particle coverage near evolving interfaces and focuses computation where accuracy is most critical, achieving a robust balance between precision and efficiency.

\section{Reinitialization} \label{sec-re}

Our PFM-LS framework employs two complementary reinitialization procedures to maintain numerical stability and signed-distance accuracy:
(1) \textit{flow-map reinitialization}, which regenerates particles and resets flow maps; and
(2) \textit{level-set redistancing}, which enforces the signed-distance property of $\varphi$.
We detail both processes and their execution criteria below.

\subsection{Flow-Map Reinitialization}\label{subsec-pfmreinit}

Flow-map methods accumulate numerical errors over time (from the sources enumerated at the end of Section~\ref{sec-FM}) and require periodic reinitialization to preserve accuracy.
Following the single-sample long--short flow map scheme of \citet{zhou2024eulerian}, we maintain two coupled flow maps between initial time $a$ and current time $c$ with an intermediate timestamp $b$ ($a<b<c$).
This produces a long flow map $[a,c]$ and a short flow map $[b,c]$, with corresponding backward Jacobians $\mathcal{T}_{[a,b]}$ and $\mathcal{T}_{[b,c]}$.
The full backward map is obtained as $\mathcal{T}_{[a,c]} = \mathcal{T}_{[a,b]}\mathcal{T}_{[b,c]}$.
We use the long map for advecting $\varphi$ and $\nabla\varphi$, while the short map---reinitialized more frequently---is used for $\nabla^2\varphi$, which is more sensitive to flow distortion.

\noindent\textbf{Long-range reinitialization.}
Performed every $f_L$ time steps (typically 50), it:
\begin{enumerate}
    \item Uniformly redistributes particles within the narrow band (e.g., $4\times4$ per cell).
    \item Resets all Jacobians $\mathcal{T}_{[a,b]},\mathcal{T}_{[b,c]},\mathcal{F}_{[a,b]},\mathcal{F}_{[b,c]}$ to identity.
    \item Updates the initial time reference to the current frame.
    \item Interpolates $\varphi$, $\nabla\varphi$, and $\nabla^2\varphi$ to new particle positions using the adaptive interpolation in Section~\ref{subsec-adaptive}; if combined with redistancing, blending is applied as in Section~\ref{subsec:hybrid-blend}.
\end{enumerate}

\noindent\textbf{Short-range reinitialization.}
Executed every $f_S$ steps (typically 5), it:
\begin{enumerate}
    \item Resets $\mathcal{T}_{[b,c]}$ and $\mathcal{F}_{[b,c]}$ to identity.
    \item Recomputes Hessians $\nabla^2\varphi$ at current particle positions.
\end{enumerate}

This hierarchical reinitialization strategy effectively balances accuracy and efficiency by stabilizing long-term flow-map evolution while preserving fine-scale curvature information.

\subsection{Level-set Redistancing}\label{subsec-renewton}
During advection, the level-set function gradually deviates from the signed-distance property ($|\nabla\varphi|=1$), which is essential for accurate computation of interface normals and curvature. Existing redistancing techniques include fast marching methods \cite{sethian1996fast}, PDE-based WENO schemes \cite{liu1994weighted, jiang2000weighted}, Newton-type closest-point projections \cite{chopp2001some, anumolu2018gradient}, and particle-based corrections \cite{enright2002hybrid}. We adopt a modified quasi-Newton redistancing scheme that cooperates between particles and the background grid. Instead of redistancing every particle---which is computationally expensive due to their high density near the interface---we redistance grid nodes within the narrow band and transfer this information to particles through a subsequent flow-map reinitialization.

The redistancing frequency follows the same schedule as the long-range flow-map reinitialization; an adaptive trigger based on the gradient-consistency measure of Section~\ref{subsec:hybrid-blend} may additionally request reinitialization between scheduled steps. The concrete schedules used for each experiment are summarized in Table~\ref{tab:protocols}. For applications such as liquid simulation that can tolerate small deviations from the signed-distance condition, a lower frequency or adaptive strategy \cite{li2023garmls} is recommended.

Our procedure consists of four stages:
\begin{enumerate}
    \item \textbf{Initial guess:} Apply WENO iterations to obtain an initial $\varphi$ field for grid points within the narrow band.
    \item \textbf{Quasi-Newton projection:} For each grid node, perform a quasi-Newton projection onto the closest interface point following \cite{li2023garmls}. Each iteration includes (i) a gradient-descent step along $\nabla\varphi$, (ii) a perpendicular correction to align $(\boldsymbol{x}-\boldsymbol{x}_i)$ with $\nabla\varphi$, and (iii) a back-and-forth midpoint update to suppress oscillations. We replace grid-based Hermite interpolation with particle-based APIC-style interpolation for evaluating $\varphi$ and $\nabla\varphi$, significantly improving sub-grid feature preservation and projection accuracy.
    \item \textbf{Background propagation:} Extend the redistanced $\varphi$ values from the narrow band to the background domain $\Omega_0$ using a semi-Lagrangian Eikonal solver. Cells outside the band are processed in order of increasing $|\varphi|$; each cell traces one grid spacing back along the interface normal $\mathbf{n}=\nabla\varphi/|\nabla\varphi|$ and sets $\varphi(\boldsymbol{x})=\varphi(\boldsymbol{x}-\Delta x\,\mathbf{n})+\Delta x$, with the right-hand side evaluated by interpolation; sweeping repeats until all exterior cells are updated. This propagation dominates the cost of a redistancing pass (Section~\ref{subsec:cost}).
    \item \textbf{Particle update:} Reinitialize particle flow maps and redistribute particles, transferring grid-based $\varphi$ and $\nabla\varphi$ values through the hybrid blending strategy (Section~\ref{subsec:hybrid-blend}).
\end{enumerate}

This hybrid particle--grid formulation preserves the signed-distance property while maintaining fine-scale interface features. By performing redistancing on grid nodes rather than individual particles, it achieves substantial computational savings without compromising accuracy.
\begin{algorithm}[!htb]
\caption{Particle Newton Redistancing}
\label{alg:reinit}
\begin{algorithmic}[1]
\Require Level set field $\varphi$, gradient field $\nabla\varphi$, narrow band region
\Ensure Redistanced level set with $|\nabla\varphi| = 1$ while preserving features

\State Initialize grid points with WENO iterations for initial guess
\For{each grid point $\mathbf{x}_p$ in narrow band}
    \State Apply quasi-Newton projection to find the closest interface point:
    \State \quad Initial guess: $\mathbf{x}_0 = \mathbf{x}_p - \varphi(\mathbf{x}_p)\frac{\nabla\varphi(\mathbf{x}_p)}{|\nabla\varphi(\mathbf{x}_p)|}$
    \State \quad Gradient-descent: $\mathbf{x}_{k+1/2} = \mathbf{x}_k - \frac{\varphi(\mathbf{x}_k)\nabla\varphi(\mathbf{x}_k)}{|\nabla\varphi(\mathbf{x}_k)|^2}$
    \State \quad Perpendicular adjustment: $\mathbf{x}_{k+1} = \mathbf{x}_{k+1/2} + \alpha(\mathbf{I} - \mathbf{n}\mathbf{n}^T)(\mathbf{x}_p - \mathbf{x}_{k+1/2})$
    \State \quad If $|\mathbf{x}_{k+1} - \mathbf{x}_{k-1}| < |\mathbf{x}_{k+1} - \mathbf{x}_k|$: $\mathbf{x}_{k+1} = \frac{1}{2}(\mathbf{x}_k + \mathbf{x}_{k+1})$
    \State Update: $\varphi(\mathbf{x}_p) = \text{sgn}(\varphi_0)|\mathbf{x}_p - \mathbf{x}_k|$, $\nabla\varphi(\mathbf{x}_p) = \frac{\mathbf{x}_p - \mathbf{x}_k}{|\mathbf{x}_p - \mathbf{x}_k|}$
\EndFor

\State Propagate values from narrow band to background regions
\State Apply hybrid particle-grid blending: \hfill $\triangleright$ Section~\ref{subsec:hybrid-blend}
\State \quad Compute $\gamma = \frac{\nabla\varphi_{\text{particle}} \cdot \nabla\varphi_{\text{grid}}}{|\nabla\varphi_{\text{particle}}||\nabla\varphi_{\text{grid}}|}$ for each particle
\State \quad Blend based on $\gamma$ and proximity to interface
\State Perform particle flow map reinitialization: \hfill $\triangleright$ Section~\ref{subsec-pfmreinit}
\State \quad Redistribute particles in narrow band (e.g., $4 \times 4$ per cell)
\State \quad Reset Jacobians: $\mathcal{T}_{[a,b]} \leftarrow \mathbf{I}$, $\mathcal{T}_{[b,c]} \leftarrow \mathbf{I}$
\State \quad Update time references: $a \leftarrow c$
\State \quad Interpolate $\varphi_p$, $\nabla\varphi_p$, $\nabla^2\varphi_p$ to new particles using APIC-style interpolation
\end{algorithmic}
\end{algorithm}
\subsection{Hybrid Particle--Grid Blending}\label{subsec:hybrid-blend}

Following quasi-Newton redistancing, accurate level-set values are available on narrow-band grid nodes and must be transferred to redistributed particles.
Relying exclusively on either grid interpolation or particle values degrades accuracy, particularly for fine-scale features: purely grid-based transfer behaves like the Hermite-interpolation redistancing baselines of Section~\ref{subsec:experiment-re}, whose errors are an order of magnitude larger, while pure particle values drift after several advection steps.
We therefore employ a hybrid blending scheme that combines the stability of grid interpolation with the sub-grid fidelity of particle data.
The insight is that grid-based Hermite interpolation performs well in well-resolved regions, whereas particles better capture small droplets and thin filaments below grid resolution.
However, after several advection steps, particle values can drift, with distortion increasing away from the interface.
To address this, we blend grid and particle contributions based on both proximity and gradient consistency. For particles within $0.25\Delta x$ of the interface, we compute the gradient consistency measure
$\gamma = \frac{\nabla\varphi_{\text{p}}\cdot\nabla\varphi_{\text{g}}}{|\nabla\varphi_{\text{p}}||\nabla\varphi_{\text{g}}|}$,
where subscripts $\text{p}$ and $\text{g}$ denote particle- and grid-interpolated gradients.
Blending weights are determined as follows:
\begin{itemize}
    \item $\gamma > e_1 = 0.9$: adopt grid values;
    \item $e_2 = 0.7 < \gamma \leq e_1$: linearly blend grid and particle values;
    \item $\gamma \leq e_2$: retain particle values.
\end{itemize}

This mixing of particle-carried and grid-interpolated values through a blend factor is reminiscent of PIC/FLIP-style blending in hybrid particle--grid fluid solvers~\cite{brackbill1986flip,zhu2005animating}, with the blend factor driven here by gradient consistency rather than fixed a priori.

This adaptive blending preserves fine-scale interface structures while maintaining global smoothness, outperforming both purely grid- and particle-based transfers, especially for thin filaments and small droplets.
The complete redistancing and blending procedure is summarized in Algorithm~\ref{alg:reinit}.

\noindent\textbf{Topology changes.}
As with all level-set methods, topological transitions (coalescence and breakup) are handled implicitly by the reconstructed $\varphi$ field. At a topological event the medial axis of the shape approaches the interface, so the particle-carried $\nabla\varphi$ and $\nabla^2\varphi$ become discontinuous there; Section~\ref{subsec:topology} quantifies the resulting behavior on merging and pinch-off benchmarks.
\begin{algorithm}[!htb]
\caption{Particle Flow Map Level Set Method}
\label{alg:nbpfm}
\begin{algorithmic}[1]
\State Initialize: $\varphi_0$, $\nabla\varphi_0$, $\nabla^2\varphi_0$ to initial level set data; $\mathcal{T}_{[a,b]}$, $\mathcal{T}_{[b,c]}$ to $\mathbf{I}$
\For{$k$ in total steps}
    \State $j \leftarrow k \bmod n^L$; \hfill $\triangleright$ Long flow map counter
    \State $l \leftarrow k \bmod n^S$; \hfill $\triangleright$ Short flow map counter
    \If{$j = 0$}
        \State Redistribute particles in narrow band;
        \State Reinitialize $\varphi_p$, $\nabla\varphi_p$, $\nabla^2\varphi_p$; \hfill $\triangleright$ Section~\ref{subsec-recon}
        \State Reinitialize $\mathcal{T}_{[a,b]}$ to $\mathbf{I}$; \hfill $\triangleright$ Section~\ref{subsec-pfmreinit}
    \EndIf
    \If{$l = 0$}
        \State Reinitialize $\nabla^2\varphi_p$; \hfill $\triangleright$ Section~\ref{subsec-advection}
        \State Reinitialize $\mathcal{T}_{[b,c]}$ to $\mathbf{I}$; \hfill $\triangleright$ Section~\ref{subsec-pfmreinit}
    \EndIf
    \State Compute velocity $\mathbf{u}$ at particle positions;
    \State Advect particles with RK4;
    \State Update $\mathcal{T}_{[b,c]}$ with velocity gradient; \hfill $\triangleright$ Eq.~\eqref{eq:evolution}
    \State Compute $\mathcal{T}_{[a,c]}$ with $\mathcal{T}_{[a,b]}$ and $\mathcal{T}_{[b,c]}$; \hfill $\triangleright$ Section~\ref{subsec-advection}
    \State Update $\nabla\varphi_p$ with $\mathcal{T}_{[a,c]}^T\nabla\varphi_0$; \hfill $\triangleright$ Section~\ref{subsec-advection}
    \State Update $\nabla^2\varphi_p$ with $\mathcal{T}_{[b,c]}^T\nabla^2\varphi_0\mathcal{T}_{[b,c]}$; \hfill $\triangleright$ Section~\ref{subsec-advection}
    \State Check particle density in narrow band cells;
    \State Add new particles where density is below threshold; \hfill $\triangleright$ Section~\ref{subsec-adaptive}
    \State Reconstruct $\varphi$ on grid using APIC-style interpolation (if needed); \hfill $\triangleright$ Section~\ref{subsec-recon}
    \If{timeForRedistancing}
        \State Perform particle Newton redistancing; \hfill $\triangleright$ Section~\ref{subsec-renewton}
    \EndIf
\EndFor
\end{algorithmic}
\end{algorithm}
\subsection{Time Integration}\label{subsec-timeintegration}

We now summarize the complete time-stepping procedure that integrates all components of our particle flow map level set method. Each simulation step proceeds through four stages, as illustrated in Algorithm~\ref{alg:nbpfm}.

\noindent\textbf{Stage 1: Flow Map Reinitialization}
At the beginning of each step, we check whether reinitialization is required. Long-range reinitialization (every $n^L$ steps, typically 50) redistributes particles uniformly within the narrow band and resets $\mathcal{T}_{[a,b]}$ to identity. Short-range reinitialization (every $n^S$ steps, typically 5) resets $\mathcal{T}_{[b,c]}$ and recomputes the Hessian $\nabla^2\varphi$ to control curvature error accumulation.

\noindent\textbf{Stage 2: Advection and Flow Map Update}
Particles are advected using fourth-order Runge-Kutta integration. Simultaneously, the short-range Jacobian $\mathcal{T}_{[b,c]}$ is updated by integrating Eq.~\eqref{eq:evolution}. The long-range Jacobian is then computed as $\mathcal{T}_{[a,c]} = \mathcal{T}_{[a,b]}\mathcal{T}_{[b,c]}$, and the gradient $\nabla\varphi$ is updated via $\boldsymbol{g}_p = \mathcal{T}_{[a,c]}^\top \boldsymbol{g}_p(0)$.

\noindent\textbf{Stage 3: Particle Management}
Adaptive particle control (Section~\ref{subsec-adaptive}) monitors particle density in narrow-band cells and generates new particles where coverage falls below the threshold $n_{\min}$. The level set field is then reconstructed on the grid using APIC-style interpolation (Section~\ref{subsec-recon}) when needed for visualization or external coupling.

\noindent\textbf{Stage 4: Redistancing}
When the signed-distance property degrades beyond acceptable tolerance, particle Newton redistancing (Section~\ref{subsec-renewton}) is performed, followed by hybrid particle-grid blending (Section~\ref{subsec:hybrid-blend}) to restore $|\nabla\varphi| = 1$ while preserving fine-scale features.

\section{Numerical Tests}\label{sec:numerical}
In this section, we evaluate the performance of our method through various numerical experiments. We begin with advection tests in both 2D and 3D settings to assess the accuracy of our approach compared to existing methods. Throughout these tests, we measure the accuracy by tracking the area (2D) or volume (3D) of the level-set regions. For 2D area calculations, we measure the area enclosed by the marching-squares isocontour on a grid that is four times finer than the original grid in each dimension. In 3D, we evaluate the liquid volume using the fast marching cubes-style scheme proposed by \citet{takahashi2022fast}.
\begin{table}[!htb]
\centering
\caption{Summary of methods compared in our experiments.}
\label{tab:methods}
\begin{tabular}{lp{10cm}}
\toprule
\textbf{Method} & \textbf{Description} \\
\midrule
Linear & First-order semi-Lagrangian advection \citep{stam1999stable} \\
Linear+RM & First-order semi-Lagrangian with reference map   \\
Quadratic & Second-order semi-Lagrangian advection  \\
Quadratic+RM & Second-order semi-Lagrangian with reference map   \\
Cubic & Third-order semi-Lagrangian advection \citep{fedkiw2001visual} \\
Cubic+RM & Third-order semi-Lagrangian with reference map  \\
BFECC & Back and Forth Error Compensation and Correction \citep{selle2008unconditionally} \\
BFECC+RM & BFECC with reference map augmentation \\
GALS & Gradient-Augmented Level-Set method \citep{nave2010gradient} \\
GARM-LS & Gradient-Augmented Reference Map Level-Set \citep{li2023garmls} \\
PLS(\ensuremath{\times}16) & Particle Level-Set with 16 particles per near-interface cell \citep{enright2002hybrid} \\
PLS(\ensuremath{\times}36) & Particle Level-Set with 36 particles per near-interface cell \citep{enright2002hybrid} \\
NBFLIP & Narrow Band FLIP method (x16 particles in 2D, x64 in 3D)\citep{ferstl2016narrow} \\
Ours(\ensuremath{\times}16) & Our method with 16 particles per near-interface cell \\
Ours(\ensuremath{\times}36) & Our method with 36 particles per near-interface cell \\
\bottomrule
\end{tabular}
\end{table}
Table~\ref{tab:methods} summarizes the methods compared in our experiments. We include several semi-Lagrangian advection schemes with varying interpolation orders: Linear (first-order), Quadratic, and Cubic. For each of these, we also test reference map-augmented versions (denoted with "+RM" suffix). Additionally, we compare against BFECC \citep{selle2008unconditionally}, Gradient-Augmented Level-Set (GALS) \citep{nave2010gradient}, Particle Level-Set (PLS) \citep{enright2002hybrid}, GARM-LS \citep{li2023garmls}, which combines GALS with reference maps, and Narrow Band FLIP (NBFLIP) \citep{ferstl2016narrow}. For particle-based methods such as PLS(\ensuremath{\times}16) and our approach, the notation in parentheses indicates the number of particles initially seeded within each near-interface grid cell. For NBFLIP, we use 16 particles per cell in 2D and 64 particles per cell in 3D (with 4 particles aligned along each axis).

For pure advection tests, we focus solely on the advection capabilities. As such, we disable redistancing for all methods except PLS and NBFLIP, which require redistancing as an intrinsic part of their methodology. Our method performs regular flow map reinitialization as described in Section~\ref{subsec-pfmreinit}. In later tests, we also evaluate the effectiveness of our redistancing approach and compare the combined advection-redistancing performance against other methods.

All 2D tests are conducted within a domain of $[0, 1]^2$, while 3D tests are performed in a domain of $[0, 1]^3$. To ensure temporal accuracy across all methods, we employ a consistent time step size of $\Delta t = 0.02$ s. For grid-based methods, we utilize a fourth-order Runge--Kutta integration scheme to guarantee fair comparison of advection performance.

Table~\ref{tab:protocols} summarizes, for every experiment in this section, the flow-map reinitialization and redistancing schedules together with the particle densities, making the evaluation protocol of each reported number explicit.
\begin{table}[!htb]
\centering
\caption{Per-experiment configuration. ``Advection-only'' disables redistancing (and long-range flow-map reinitialization) to isolate transport accuracy, as stated at the beginning of this section; ``full method'' runs long/short flow-map reinitialization every $f_L/f_S$ steps with redistancing executed at each long-range reinitialization.}
\label{tab:protocols}
\begin{tabular}{lllll}
\toprule
Experiment & Protocol & $f_L/f_S$ & Redistancing & Particles per cell \\
\midrule
2D advection benchmarks (\S\ref{subsec:2DZalesak}--\S\ref{subsec:2DDistorted}) & advection-only & --- & off & 16 / 36 \\
3D advection benchmarks (\S\ref{subsec:3DLeveque}--\S\ref{subsec:3DDistorted}) & advection-only & --- & off & 64 / 216 \\
Redistancing study (\S\ref{subsec:experiment-re}) & full method & 12 / 6 & every \{6, 12, 25, 50\} steps & 16 \\
Convergence study (\S\ref{sec:convergence}) & advection-only & --- & off & 16 (2D), 64 (3D) \\
Topology change (\S\ref{subsec:topology}) & full method & 12 / 6 & at long-range reinit. & 16 \\
Shape-error metrics (\S\ref{subsec:shapemetrics}) & as in source experiment & --- & --- & --- \\
Free-surface coupling (\S\ref{subsec:freesurface}) & full method & 12 / 6 & at long-range reinit. & 16 \\
\bottomrule
\end{tabular}
\end{table}

\subsection{2D Zalesak's disk}\label{subsec:2DZalesak}

Zalesak's disk test \cite{zalesak1979fully} is a widely recognized benchmark for evaluating advection schemes in preserving sharp interfaces at different orientations. This challenging test involves rotating a slotted disk in a prescribed velocity field, with the goal of returning to its initial configuration after one or more complete revolutions.
\begin{figure}[!htb]
    \centering
    \includegraphics[width=0.99\linewidth]{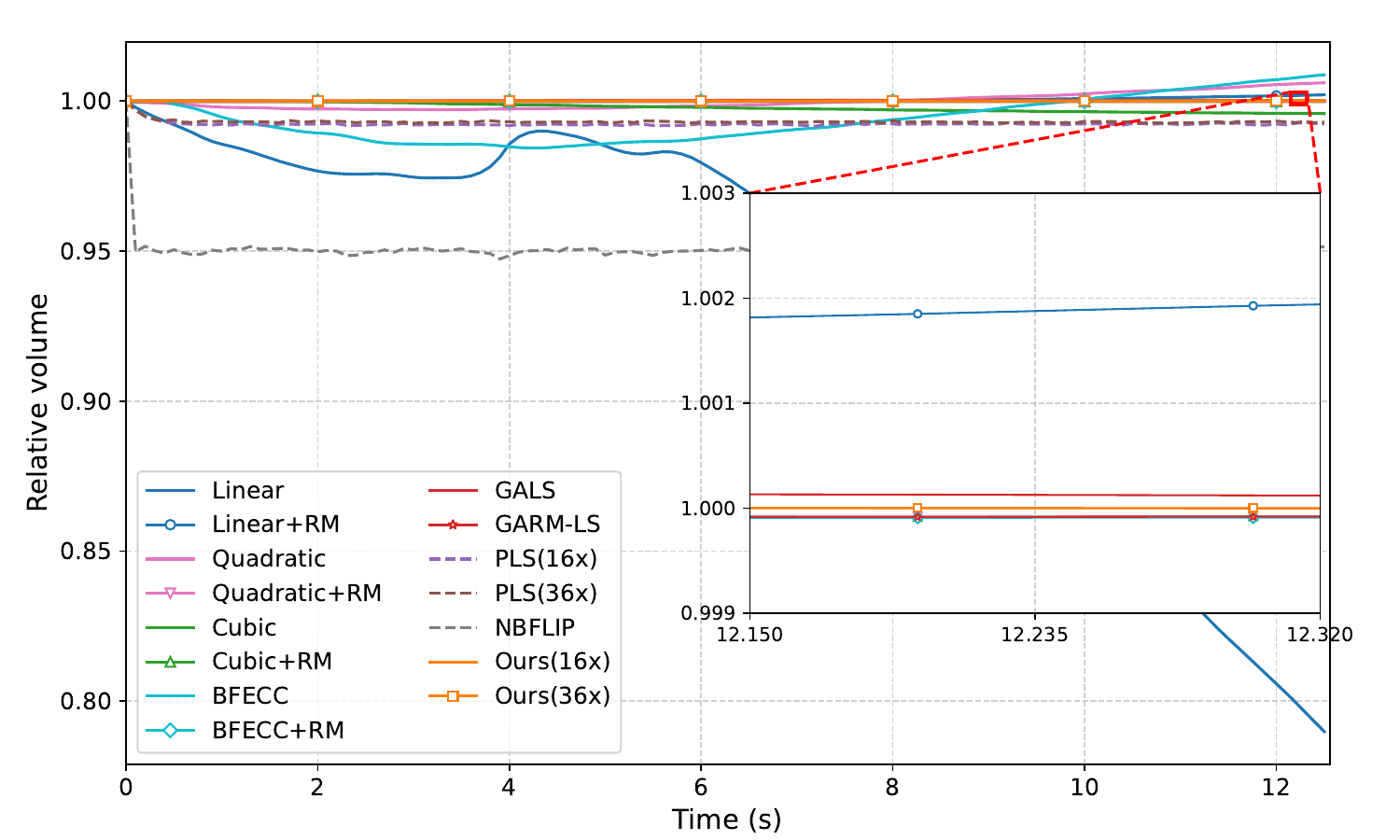}
    \caption{Volume preservation comparison during $200\times200$ \textbf{2D Zalesak's disk rotation test}. Our method (orange lines) maintains nearly perfect volume conservation throughout the entire revolution. The inset shows detailed performance during the final moments, where our method preserves over 99.9999\% of the original volume, significantly outperforming all other methods. Most reference map-augmented methods (RM suffix) achieve good preservation, while traditional methods like Linear show substantial volume loss. Note that NBFLIP struggles in this test despite its particle-based nature.}
    \label{fig:zelasak_curve}
\end{figure}
\begin{figure}[!htb]
    \centering
    \includegraphics[width=0.99\linewidth]{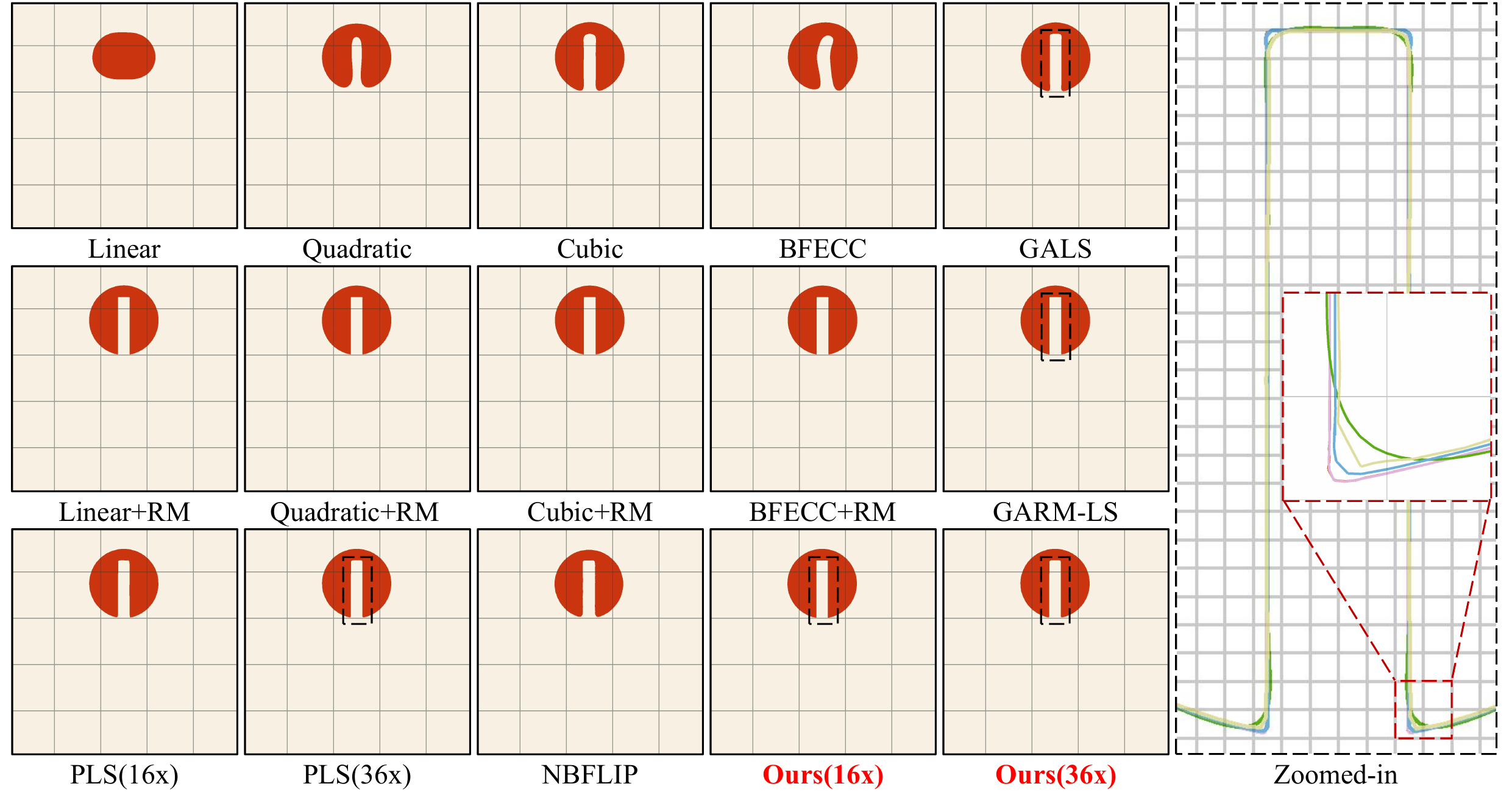}
    \caption{Visual comparison of $200\times200$ \textbf{2D Zalesak's disk} interface shapes after one complete revolution across different methods. The zoomed-in view (bottom right) highlights the preservation of sharp corners and edges. First-order methods (Linear) show significant smoothing of corners, while higher-order methods (Quadratic, Cubic) better maintain the overall shape but still round sharp features. Reference map augmentation (bottom row) consistently improves performance. Our method (bottom row, in red) demonstrates superior preservation of geometric features, maintaining sharp corners and straight edges that other methods fail to capture accurately. Zoomed-in view showing GALS (\colorlabel{green}), PLS (36\ensuremath{\times}) (\colorlabel{yellow}), Ours (16\ensuremath{\times}) (\colorlabel{cyan}), and Ours (36\ensuremath{\times}) (\colorlabel{purple}).}
    \label{fig:zelasak_img}
\end{figure}
\begin{table}[!htb]
\centering
\caption{%
Relative volume errors for the $200\times200$ \textbf{2D Zalesak's disk test} after one full revolution ($T=6.28$\,s).
All metrics are measured at the final frame.
}
\label{tab:zalesak_results}
\begin{tabular}{lcc}
\hline
Method & Volume Preservation (\%) & Relative Error \\
\hline
Linear & 78.79 & 2.12 $\times 10^{-1}$ \\
Linear+RM & 100.21 & 2.10 $\times 10^{-3}$ \\
Quadratic & 100.612 & 6.13 $\times 10^{-3}$ \\
Quadratic+RM & 99.99 & 7.70 $\times 10^{-5}$ \\
Cubic & 99.58 & 4.22 $\times 10^{-3}$ \\
Cubic+RM & 99.99 & 8.09 $\times 10^{-5}$ \\
GALS & 100.01 & 1.10 $\times 10^{-4}$ \\
GARM-LS & 99.99 & 8.18 $\times 10^{-5}$ \\
PLS(16\ensuremath{\times}) & 99.24 & 7.59 $\times 10^{-3}$ \\
PLS(36\ensuremath{\times}) & 99.29 & 7.13 $\times 10^{-3}$ \\
BFECC & 100.89 & 8.87 $\times 10^{-3}$ \\
BFECC+RM & 99.99 & 8.95 $\times 10^{-5}$ \\
NBFLIP & 95.29 & 4.71 $\times 10^{-2}$ \\
\hline
\textbf{Ours(16\ensuremath{\times})} & \textbf{100.00} & \textbf{1.26 $\times 10^{-7}$} \\
\textbf{Ours(36\ensuremath{\times})} & \textbf{100.00} & \textbf{1.17 $\times 10^{-7}$} \\
\hline
\end{tabular}
\end{table}

We adapt classical Zalesak's disk setup to our domain of $[0, 1]^2$. The slotted disk has a radius of $0.15$ and is initially centered at $(0.5, 0.75)$, rotating about the domain center $(0.5, 0.5)$. The slot has dimensions of $0.05$ by $0.25$. To create a rotational velocity field, we define:
\begin{equation}
u = -\omega(y - 0.5), \quad v = \omega(x - 0.5)
\end{equation}
where $\omega = \frac{\pi}{157 \cdot \Delta t}$ is the angular velocity. This configuration completes a full revolution at $T = 6.28$ seconds. All simulations were performed on a $200 \times 200$ grid.

Figure~\ref{fig:zelasak_curve} compares various interface tracking methods in Zalesak's disk rotation test. Our proposed narrow band particle flow map method demonstrates superior performance in both volume conservation and geometric preservation.

Table~\ref{tab:zalesak_results} quantifies the relative volume error for each method. The first-order linear scheme exhibits substantial volume loss with a relative error of $2.12 \times 10^{-1}$. Higher-order methods show improvement, with quadratic, cubic, and BFECC methods achieving errors of $6.7 \times 10^{-3}$, $4.2 \times 10^{-3}$, and $8.9 \times 10^{-3}$, respectively. Reference map augmentation consistently reduces these errors, with Linear+RM showing dramatic improvement to $2.1 \times 10^{-3}$. Advanced level set methods (GALS and GARM-LS) perform well with errors of approximately $1.1 \times 10^{-4}$ and $1.8 \times 10^{-5}$. The Particle Level-Set (PLS) method shows moderate volume loss despite its marker particle correction, with PLS(16\ensuremath{\times}) and PLS(36\ensuremath{\times}) configurations yielding errors of $7.59 \times 10^{-3}$ and $7.13 \times 10^{-3}$ respectively, with minimal improvement at higher particle density.

As shown in Table~\ref{tab:zalesak_results}, our method achieves excellent volume conservation, with Ours(16x) configuration demonstrating 100.00001\% (error of $1.26 \times 10^{-7}$) and Ours(36x) configuration achieving 99.99999\% (error of $1.17 \times 10^{-8}$). The exceptional accuracy of our method in this rotation test can be attributed to the particle flow map's perfect preservation of geometric properties. Since Zalesak's disk undergoes rigid rotation, our particles maintain their original level set values, gradients, and Hessians without distortion, exactly preserving the interface shape, including sharp corners. This demonstrates the fundamental advantage of our approach: under rigid transformations like rotation, the particle flow map precisely captures the motion without the numerical diffusion inherent in grid-based advection schemes.

Visual inspection of Figure~\ref{fig:zelasak_img} further confirms that our approach preserves the geometric features of the slotted disk with high fidelity, maintaining sharp corners and straight edges that other methods fail to capture accurately. These results demonstrate the effectiveness of our PFM-LS method for applications requiring both precise volume conservation and accurate geometric feature preservation.

To assess long-term stability, we extended the Zalesak's disk test to 5 full revolutions ($T = 31.4$\,s). Our method maintains volume preservation of $100.00\%$ at every revolution endpoint ($t = 6.28, 12.56, 18.84, 25.12, 31.40$\,s), with instantaneous volume fluctuating within $[99.56\%, 100.91\%]$ during advection. No long-term drift is observed, confirming the method's stability over extended time horizons.

\subsection{2D LeVeque's Circle}\label{subsec:2DLeVeque}
\begin{figure}[!htb]
    \centering
    \includegraphics[width=0.99\linewidth]{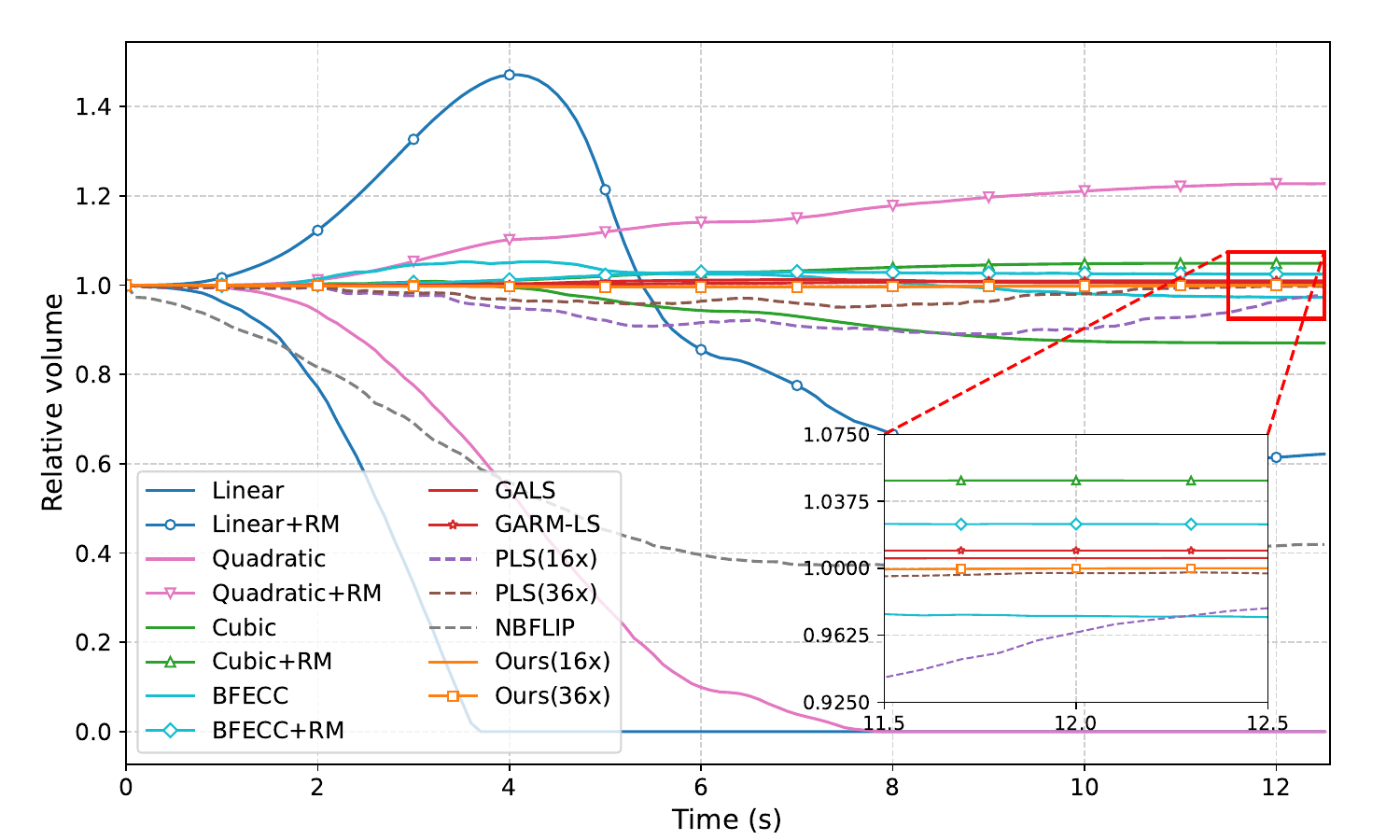}
    \caption{Volume preservation during $200\times200$ \textbf{2D LeVeque's circle test}. Our method (orange lines) maintains exceptional volume conservation during both maximum deformation ($t=6.28$\,s) and recovery phase ($t=12.56$\,s). The inset shows detailed performance at the final state, where our method achieves near-perfect recovery ($\approx 100.0\%$), while most other methods exhibit significant volume errors. Lower-order methods like Linear and Quadratic completely lose the interface during extreme deformation.}
    \label{fig:leveque_curve}
\end{figure}
The LeVeque's circle test provides a challenging scenario for evaluating interface tracking methods under significant deformation. Originally proposed by LeVeque \cite{leveque1996high}, this test has been adapted with modified velocity fields in subsequent studies \cite{nave2010gradient} to create more complex deformations.

In our implementation, we consider a circle with radius $r = 0.15$ centered at $(0.5, 0.75)$ in our $[0, 1]^2$ domain. This circle is deformed by a divergence-free velocity field defined as:
\begin{equation}
\boldsymbol{u} = u_0 \cos \frac{\pi t}{T} \begin{pmatrix}
\cos^2[\pi(x-0.5)] \sin[2\pi(y-0.5)] \\
-\sin[2\pi(x-0.5)] \cos^2[\pi(y-0.5)]
\end{pmatrix},
\end{equation}
where $u_0 = 1.0$ and $T = 12.56$\,s. At $t = T/2 = 6.28$\,s, the circle reaches its maximum deformation, forming a thin filament structure that challenges most interface tracking methods. As the velocity reverses in the second half of the time interval, the shape gradually returns to its initial circular configuration at $t = T = 12.56$\,s.

This test is particularly demanding because the circle undergoes severe stretching, creating thin filamentary structures that are difficult to resolve on standard grids. The ability to capture these fine features and recover the original shape upon velocity reversal provides a stringent test of both advection accuracy and feature preservation. We conduct this experiment on a $200 \times 200$ simulation grid with various advection algorithms.

\begin{figure}[!htb]
    \centering
    \includegraphics[width=0.99\linewidth]{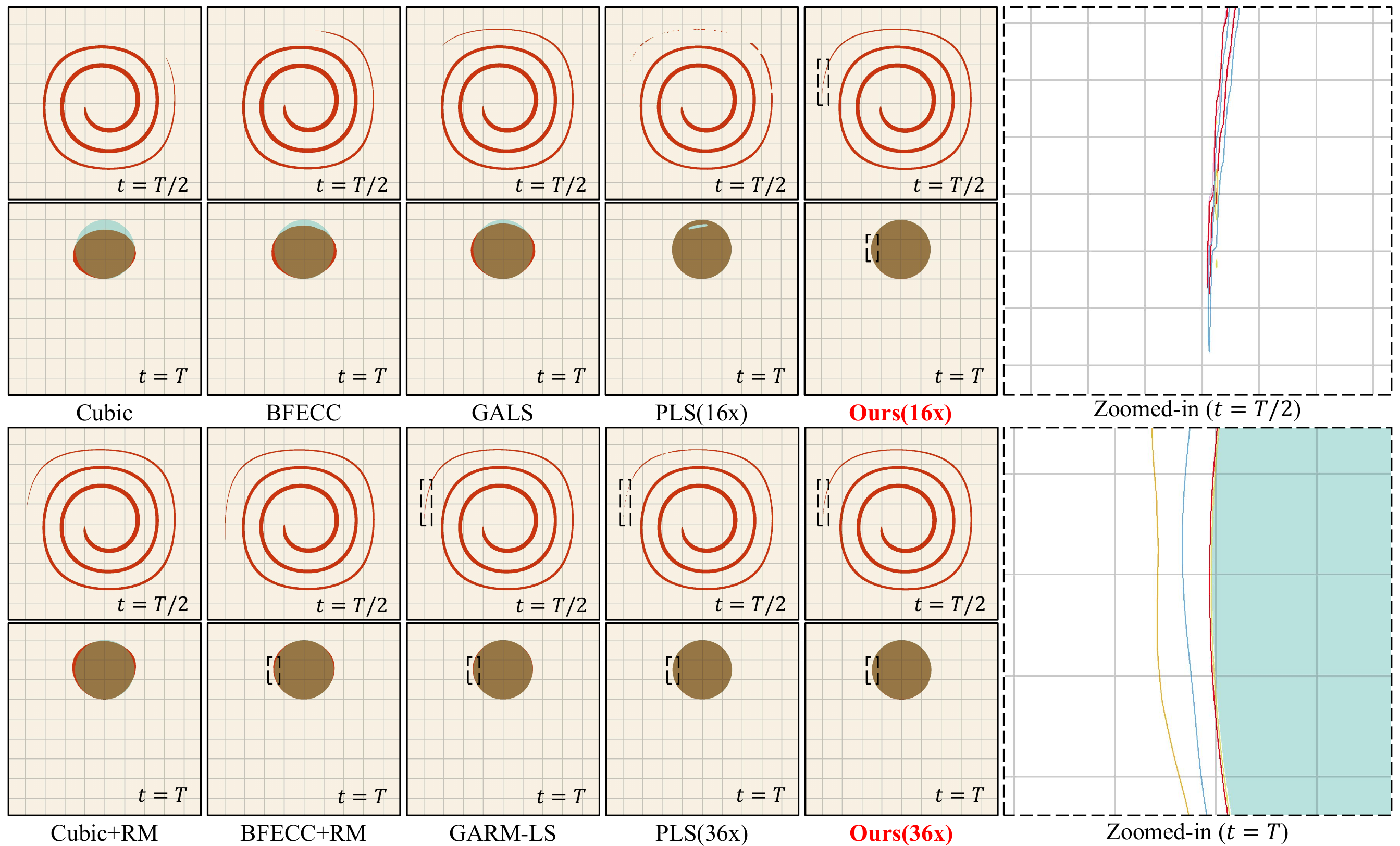}
        \caption{Visual comparison of $200\times200$ \textbf{2D LeVeque's circle test} across different interface-tracking methods.
    Each panel shows the maximum deformation state (top, $t = T/2$) and the final recovered state (bottom, $t = T$), with the blue outline indicating the ground-truth circle.
    The zoomed-in views on the right highlight the thin filament structure and recovery accuracy.
    Zoom-in color legend: \colorlabel{yellow} PLS (36\ensuremath{\times}), \colorlabel{cyan} GARM-LS, \colorlabel{pink} Ours (16\ensuremath{\times}), \colorlabel{red} Ours (36\ensuremath{\times}).}
    \label{fig:leveque_results}
\end{figure}

The visual comparison in Figure~\ref{fig:leveque_results} complements the volume histories by showing the interface at maximum deformation and after recovery, highlighting thin-structure preservation and final shape accuracy across methods. Figure~\ref{fig:leveque_curve} presents volume preservation results for LeVeque's circle test, which poses a more severe challenge than Zalesak's disk due to extreme deformation creating thin filaments at the midpoint ($t=6.28s$) before reverting to its original shape. Our proposed narrow-band particle flow map method demonstrates exceptional performance throughout the simulation, maintaining superior volume conservation at both the maximum deformation point and the final state.

Table~\ref{tab:leveque_mid_results} quantifies the volume preservation at maximum deformation ($t=6.28$\,s). The first-order linear scheme struggles significantly, capturing only {0.00\%} of the initial volume, essentially losing the entire interface. Higher-order methods show progressive improvement, with quadratic and cubic schemes preserving {8.75\%} and {94.14\%} respectively. The BFECC method slightly overestimates volume at {102.45\%}. Reference map augmentation generally improves performance, with Quadratic+RM achieving {114.19\%} and Cubic+RM yielding {102.91\%}. Advanced level set methods (GALS and GARM-LS) perform well with GALS at {101.15\%} and GARM-LS at {100.43\%}. Particle-based methods show varied results, with PLS(x16) preserving {92.01\%} and PLS(x36) achieving {96.98\%}. NBFLIP underperforms at {38.60\%}.

\begin{table}[!htb]
\centering
\caption{Volume preservation at maximum deformation ($t=6.28s$) for the $200\times200$ \textbf{2D LeVeque's circle test}}
\label{tab:leveque_mid_results}
\begin{tabular}{lcc}
\hline
Method & Volume Preservation (\%) & Relative Error \\
\hline
Linear & 0.00 & 1.00 \\
Quadratic & 8.75 & 9.13 $\times 10^{-1}$ \\
Cubic & 94.14 & 5.86 $\times 10^{-2}$ \\
BFECC & 102.45 & 2.45 $\times 10^{-2}$ \\
Linear+RM & 83.49 & 1.65 $\times 10^{-1}$ \\
Quadratic+RM & 114.20 & 1.42 $\times 10^{-1}$ \\
Cubic+RM & 102.91 & 2.91 $\times 10^{-2}$ \\
BFECC+RM & 102.93 & 2.92 $\times 10^{-2}$ \\
Hermite(GALS) & 101.15 & 1.15 $\times 10^{-2}$ \\
Hermite+RM(GARM-LS) & 100.43 & 4.30 $\times 10^{-3}$ \\
PLS(x16) & 92.01 & 7.99 $\times 10^{-2}$ \\
PLS(x36) & 96.98 & 3.02 $\times 10^{-2}$ \\
NBFLIP & 38.60 & 6.14 $\times 10^{-1}$ \\
\hline
\textbf{Ours(16\ensuremath{\times})} & \textbf{99.62} & \textbf{3.80 $\times 10^{-3}$} \\
\textbf{Ours(36\ensuremath{\times})} & \textbf{99.63} & \textbf{3.75 $\times 10^{-3}$} \\
\hline
\end{tabular}
\end{table}
Our method demonstrates remarkable consistency, with Ours(x16) and Ours(x36) configurations preserving {99.62\%} and {99.63\%} respectively at maximum deformation. This near-perfect volume conservation at mid-time is comparable to GARM-LS ({100.43\%}), with both methods achieving near-ideal preservation despite extreme deformation. Our approach's strong performance can be attributed to the particle flow map's ability to accurately track the interface throughout extreme deformation. Unlike other methods that struggle with the thin filamentary structures formed at maximum deformation, our approach maintains precise level set values, gradients, and Hessians even in these challenging regions.

At the final time ($t=12.56s$), when the circle should theoretically return to its initial state, our method demonstrates exceptional recovery capabilities. While most methods show some recovery, with Linear reaching {0.00\%}, quadratic achieving {87.04\%}, and cubic at {97.26\%}, our approach maintains nearly perfect volume preservation with Ours(16\ensuremath{\times}) at {100.000022\%} ($2.2 \times 10^{-5}$ error) and Ours(36\ensuremath{\times}) at {99.99998\%} ($2.0 \times 10^{-5}$ error). Notably, GARM-LS overestimates the volume at {100.99\%} and performs worse than both GALS ({100.57\%}) and even PLS(x36) ({99.69\%}) in final state recovery. This demonstrates that our method not only handles severe deformation well but also accurately tracks the reversing flow to recover the original shape with unprecedented precision.

The superior performance of our method in this test stems from its ability to maintain high-order interface information through particle trajectories, enabling accurate representation of thin structures during extreme deformation and precise recovery when the flow reverses. This capability sets our approach apart from both traditional grid-based methods and other advanced techniques such as GARM-LS.

\subsection{2D Distorted circle}\label{subsec:2DDistorted}

\begin{figure}[!htb]
    \centering
    \includegraphics[width=0.99\linewidth]{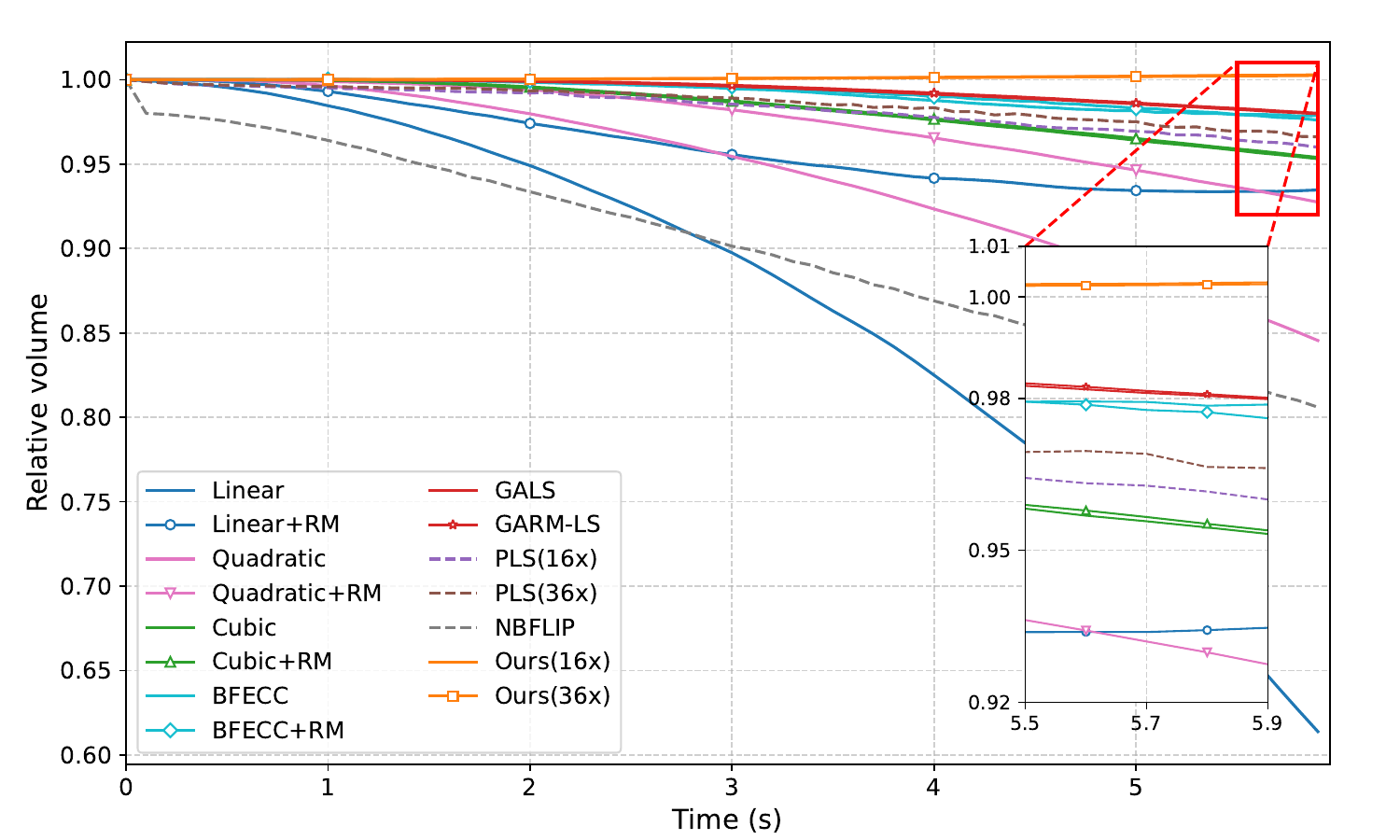}
    \caption{Volume preservation during the $200\times200$ \textbf{2D distorted circle test}. Our method (orange lines) maintains superior volume conservation as the spiral structure develops, achieving 100.3\% preservation at final time compared to significant volume loss in other methods. The inset highlights the final stages, where even advanced methods like GALS (97.9\%) and GARM-LS (97.9\%) lose approximately 2\% volume, while traditional methods show more substantial degradation.}
    \label{fig:distorted_curve}
\end{figure}
\begin{figure}[!htb]
    \centering
    \includegraphics[width=0.99\linewidth]{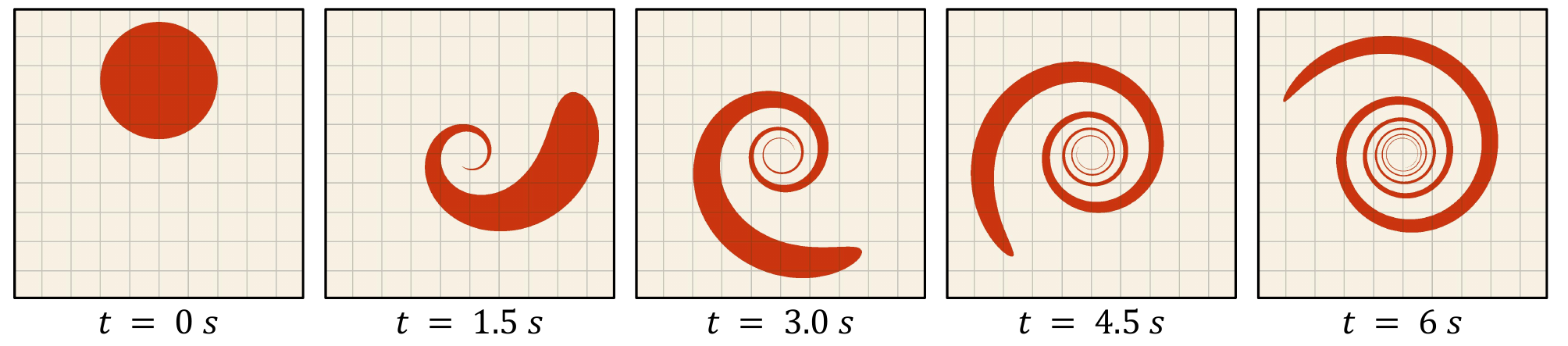}
    \caption{Evolution of the $200\times200$ \textbf{2D distorted circle test} using our method over time. The initially circular interface progressively transforms into a spiral structure due to the differential rotational velocity field, creating increasingly thin filamentary structures that challenge interface tracking methods. Our method preserves these thin structures with high fidelity compared to other approaches.}
    \label{fig:distorted_half_img}
\end{figure}
The distorted circle test presents an extremely challenging case for interface tracking methods, as it creates spiral structures with progressively thinner features. In this test, a circle with radius $r = 0.2$ centered at $(0.5, 0.75)$ is subjected to a divergence-free rotational velocity field (centered at the domain center) defined as:

\begin{equation}
\boldsymbol{u} = \frac{u_0}{\sqrt{(x-0.5)^2 + (y-0.5)^2 + r_0}} ((y-0.5), -(x-0.5))
\end{equation} where $u_0 = 0.4$ and $r_0 = 0$ (the regularization is unnecessary since the interface never reaches the rotation center). This velocity field creates a differential rotation that is stronger near the center and weaker farther away, causing the circle to spiral inward as time progresses.

\begin{figure}[!htbp]
    \centering
    \includegraphics[width=0.99\linewidth]{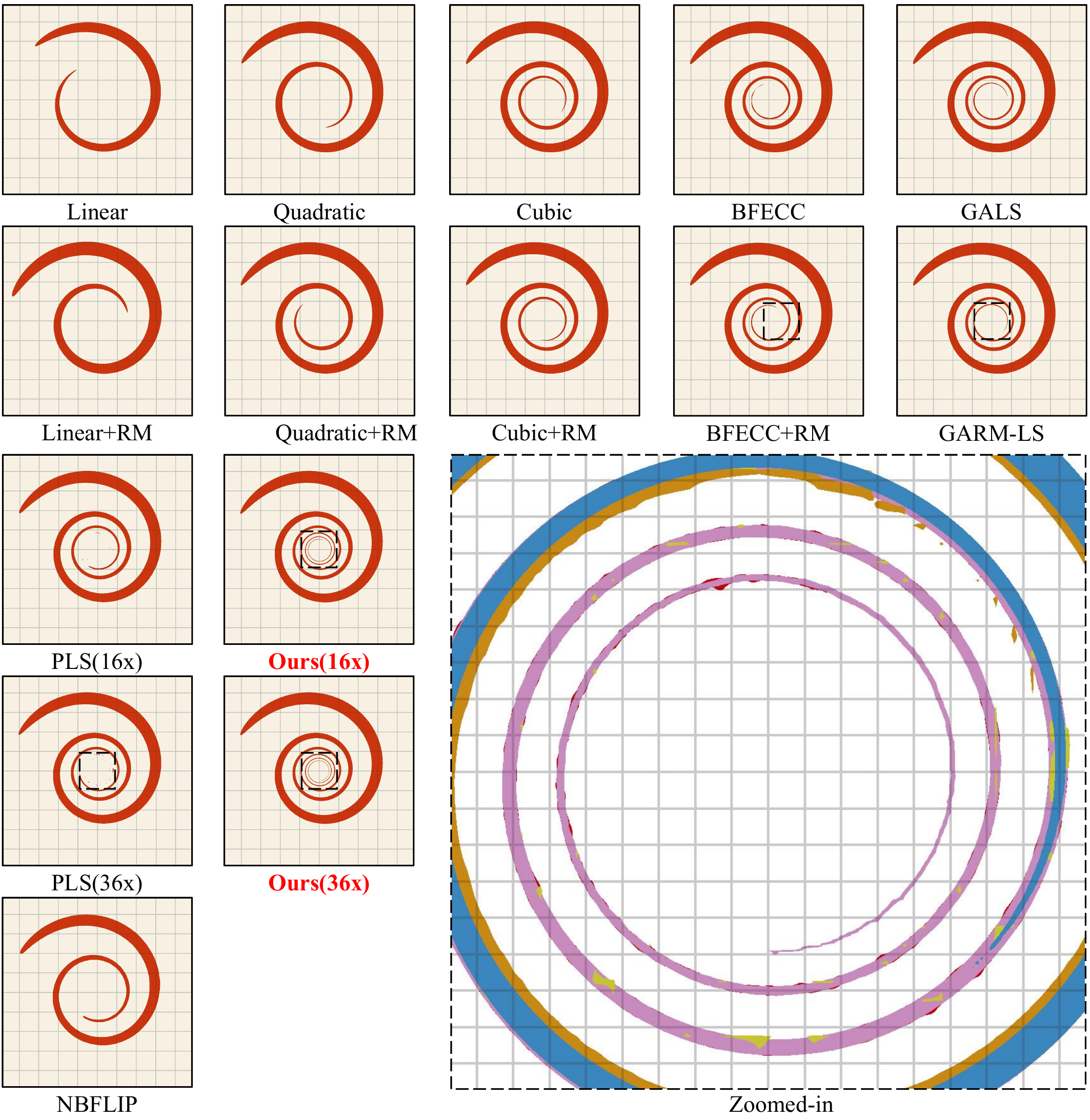}
    \caption{Visual comparison of the $200\times200$ \textbf{2D distorted circle test} across different interface tracking methods. Lower-order methods (top row) lose significant portions of the spiral structure, while higher-order and reference map methods (middle rows) preserve more but still lose inner details. Our method (highlighted in red) preserves the complete spiral structure with exceptional fidelity, maintaining continuous thin features even where they become extremely thin. The zoomed-in view (right) reveals our method's superior ability to capture sub-grid features that other methods fail to represent.
Zoom-in method color:
\colorlabel{darkBrown} BFECC+RM,
\colorlabel{darkYellow} PLS (36\ensuremath{\times}),
\colorlabel{gtBlue} GARM-LS,
\colorlabel{darkRed} Ours (16\ensuremath{\times}),
\colorlabel{deepPink} Ours (36\ensuremath{\times}), }
    \label{fig:distorted_img}
\end{figure}

After $t = 6.0$ seconds, the circle has undergone severe distortion and transforms into a tightly wound spiral. The inner end of the spiral becomes extremely thin, creating sub-grid features that challenge even high-resolution simulations. This test specifically evaluates a method's ability to preserve thin structures and small-scale features that are smaller than the grid resolution. Figure~\ref{fig:distorted_half_img} shows the progressive deformation of the interface during the distorted circle test using our method, illustrating how the initially circular shape develops into a thin spiral structure that challenges conventional interface tracking approaches. Even as the spiral becomes increasingly thin, our method maintains excellent interface definition throughout the simulation.

Figure~\ref{fig:distorted_curve} presents volume preservation results across different methods. The distorted circle test reveals dramatic differences in performance, with most methods struggling to maintain even basic volume conservation as the spiral structures become increasingly thin.

Table~\ref{tab:distorted_results} quantifies the final volume preservation after severe distortion. Traditional methods show significant volume loss, with the linear scheme preserving only {60.01\%} of the initial volume. Even higher-order methods like quadratic and cubic achieve just {84.05\%} and {95.20\%} respectively. Reference map augmentation helps but still falls short, with Linear+RM, Quadratic+RM, and Cubic+RM reaching {93.51\%}, {92.53\%}, and {95.27\%}. Advanced level set methods (GALS and GARM-LS) perform better with {97.91\%} and {97.94\%}, but still lose approximately {2\%} of volume. Particle Level-Set approaches show moderate results, with PLS(x16) and PLS(x36) preserving {95.83\%} and {96.62\%}, while NBFLIP maintains only {80.36\%}.

Most remarkably, our method significantly outperforms all competitors, with Ours(16\ensuremath{\times}) and Ours(36\ensuremath{\times}) configurations achieving {100.30\%} and {100.26\%} volume preservation. While this represents a slight volume gain rather than loss, the magnitude of error remains exceptionally small at approximately {0.3\%}, dramatically better than all other methods tested. This outstanding performance can be attributed to our particle flow map's ability to maintain high-order information even in regions where spiral structures become extremely thin and fall below grid resolution.

\begin{table}[!htb]
\centering
\caption{Volume preservation for the $200\times200$ \textbf{2D distorted circle test} at final time}
\label{tab:distorted_results}
\begin{tabular}{lcc}
\hline
Method & Volume Preservation (\%) & Relative Error \\
\hline
Linear & 60.01 & 4.00 $\times 10^{-1}$ \\
Quadratic & 84.05 & 1.60 $\times 10^{-1}$ \\
Cubic & 95.20 & 4.80 $\times 10^{-2}$ \\
BFECC & 97.81 & 2.19 $\times 10^{-2}$ \\
Linear+RM & 93.51 & 6.49 $\times 10^{-2}$ \\
Quadratic+RM & 92.53 & 7.47 $\times 10^{-2}$ \\
Cubic+RM & 95.27 & 4.73 $\times 10^{-2}$ \\
BFECC+RM & 97.54 & 2.46 $\times 10^{-2}$ \\
Hermite(GALS) & 97.91 & 2.09 $\times 10^{-2}$ \\
Hermite+RM(GARM-LS) & 97.94 & 2.06 $\times 10^{-2}$ \\
PLS(x16) & 95.83 & 4.17 $\times 10^{-2}$ \\
PLS(x36) & 96.62 & 3.38 $\times 10^{-2}$ \\
NBFLIP & 80.36 & 1.96 $\times 10^{-1}$ \\
\hline
\textbf{Ours(x16)} & \textbf{100.30} & \textbf{3.00 $\times 10^{-3}$} \\
\textbf{Ours(x36)} & \textbf{100.26} & \textbf{2.60 $\times 10^{-3}$} \\
\hline
\end{tabular}
\end{table}

Visual inspection of Figure~\ref{fig:distorted_img} further confirms the superiority of our approach. While other methods lose the thin spiral tails once they become sub-grid features, our method preserves these fine structures with high fidelity. The particle flow map accurately tracks the interface as it evolves into complex geometries, maintaining both the global shape and intricate details that competing methods fail to capture.

This challenging test highlights the strength of the narrow-band particle flow map framework for simulations involving extreme deformation and sub-grid features. By leveraging the high-order information carried by particles and the precise motion tracking of flow maps, our method faithfully represents thin structures even under conditions that defeat conventional grid-based and particle schemes.

\subsection{3D LeVeque's Sphere}\label{subsec:3DLeveque}

Following our 2D experiments, we extend our evaluation to challenging 3D scenarios to assess how our method handles volume preservation in higher dimensions. Our first 3D test implements a spherical version of the LeVeque's circle, subjecting a sphere to severe stretching and compression before returning it to its original shape, following the methodology in \citep{nave2010gradient}.

In our implementation, a sphere with radius $r = 0.15$ is initially centered at $(0.35, 0.35, 0.35)$ within our $[0,1]^3$ domain. This sphere is deformed by a divergence-free velocity field $\boldsymbol{u} = (u, v, w)$ defined as:

\begin{equation}
\begin{cases}
u = 2u_0 \cos \frac{\pi t}{T} \cos^2 [2\pi(x-0.5)] \sin [2\pi(y-0.5)] \sin [2\pi(z-0.5)], \\
v = -u_0 \cos \frac{\pi t}{T} \sin [2\pi(x-0.5)] \cos^2 [2\pi(y-0.5)] \sin [2\pi(z-0.5)], \\
w = -u_0 \cos \frac{\pi t}{T} \sin [2\pi(x-0.5)] \sin [2\pi(y-0.5)] \cos^2 [2\pi(z-0.5)],
\end{cases}
\end{equation} where $u_0 = 0.25$ m/s and $T = 10.0$ s. Under this velocity field, the sphere undergoes maximum deformation at $t = T/2 = 5.0$ seconds, developing into a complex folded structure with thin features. The velocity then reverses direction, theoretically returning the sphere to its original shape at $t = T = 10.0$ seconds.
\begin{figure}[!htbp]
    \centering
    \includegraphics[width=0.99\linewidth]{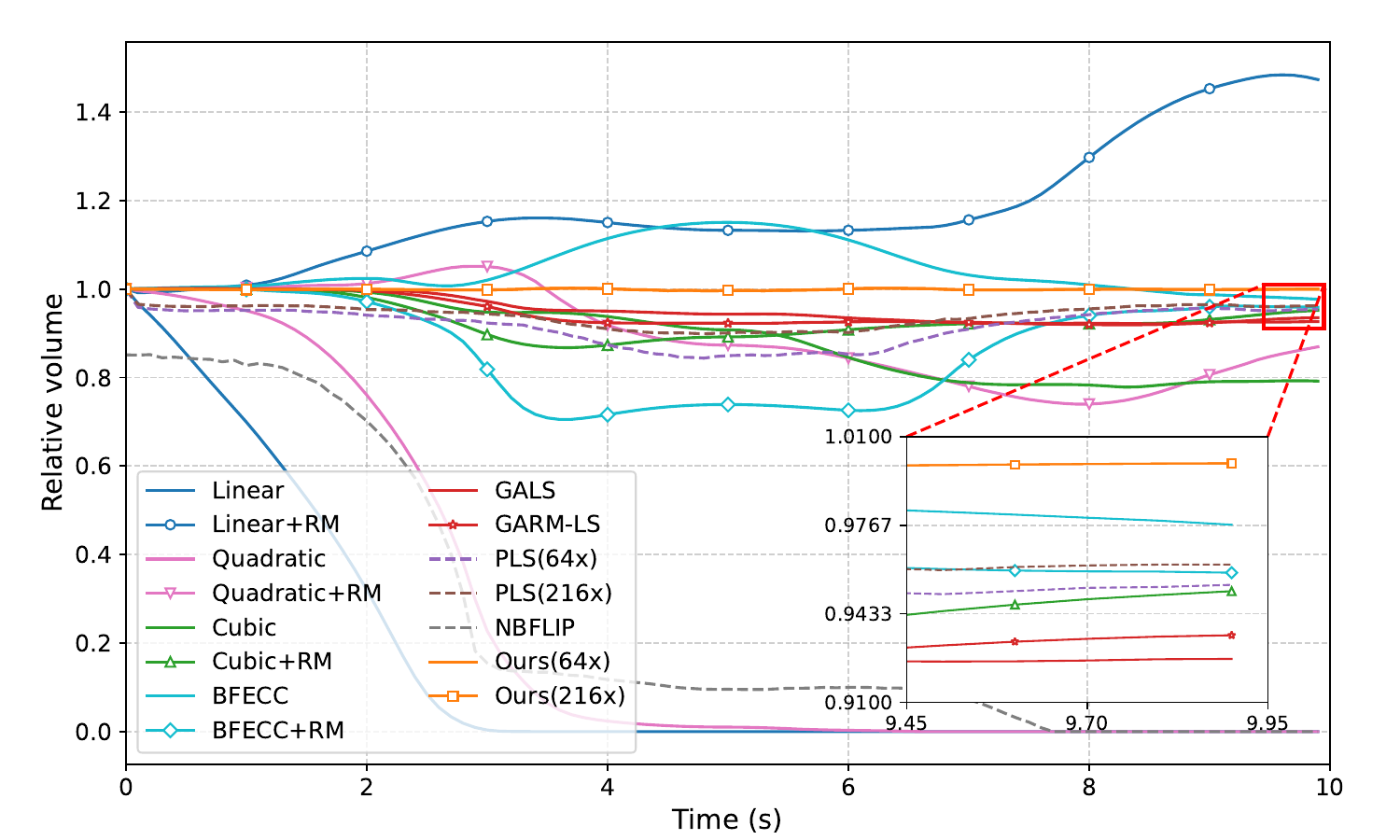}
    \caption{Volume preservation comparison during the $50^3$ \textbf{3D LeVeque sphere test}. Our method (orange lines) maintains nearly perfect volume conservation throughout the simulation, including both the deformation phase (0-5s) and the recovery phase (5-10s). The inset shows detailed performance during the critical recovery phase, where our method achieves over 99.9\% volume preservation while other methods exhibit significant volume loss or inaccurate volume gain. Linear and quadratic methods completely lose the interface, while most advanced methods struggle to maintain accurate volume during the complex 3D deformation and reversal.}
    \label{fig:leveque3d_curve}
\end{figure}
We conduct these experiments on a $50^3$ grid and compare the volume preservation capabilities of various methods throughout the deformation and recovery process. For particle-based methods, we use $4\times4\times4=64$ particles (4 particles along each axis) for the lower density configurations (PLS(x64), ours(x64)) and $6\times6\times6=216$ particles for the higher density configurations (PLS(x216), ours(x216)). The results demonstrate how different approaches handle the challenges of extreme 3D deformation.

Figure~\ref{fig:leveque3d_curve} presents volume preservation results for the 3D LeVeque's sphere test. The three-dimensional case introduces substantially greater challenges than its 2D counterpart, with most methods struggling to maintain volume under extreme deformation. Table~\ref{tab:leveque3D} quantifies volume preservation at maximum deformation ($t = 5.0$ s). The results reveal dramatic performance differences across methods. Lower-order methods show significant volume loss: the first-order linear scheme completely fails (0.00\% preservation), while the quadratic scheme preserves just 1.01\%. Even the cubic scheme preserves only 90.78\% of the initial volume. Grid-based methods with reference map augmentation show inconsistent results---Linear+RM overestimates at 113.27\%, while Quadratic+RM and Cubic+RM preserve 87.31\% and 89.18\%, respectively. Advanced level set approaches GALS and GARM-LS perform better with 94.31\% and 92.28\% preservation. Particle-based methods show moderate success, with PLS(x64) and PLS(x216) preserving 84.86\% and 90.14\%, while NBFLIP struggles severely at just 9.55\%.

Our narrow band PFM-LS method substantially outperforms all competitors at maximum deformation, with both Ours(64\ensuremath{\times}) and Ours(216\ensuremath{\times}) configurations achieving 99.68\% volume preservation. This near-perfect conservation under extreme 3D deformation demonstrates our approach's exceptional capability to track complex interfaces in three dimensions.

Performance differences become even more pronounced at the final time ($t = 10.0$ s), when the sphere should theoretically return to its original shape. The linear and quadratic schemes completely lose the interface (0.00\% preservation), while the cubic scheme recovers only 79.09\%. Reference map methods show highly variable results---Linear+RM significantly overestimates at 146.65\%, while Quadratic+RM and Cubic+RM achieve 87.25\% and 95.32\%. GALS and GARM-LS perform moderately with 92.61\% and 93.56\%, demonstrating that even advanced level set methods struggle with flow reversal in 3D.

\begin{figure}[!htbp]
    \centering
    \includegraphics[width=0.99\linewidth]{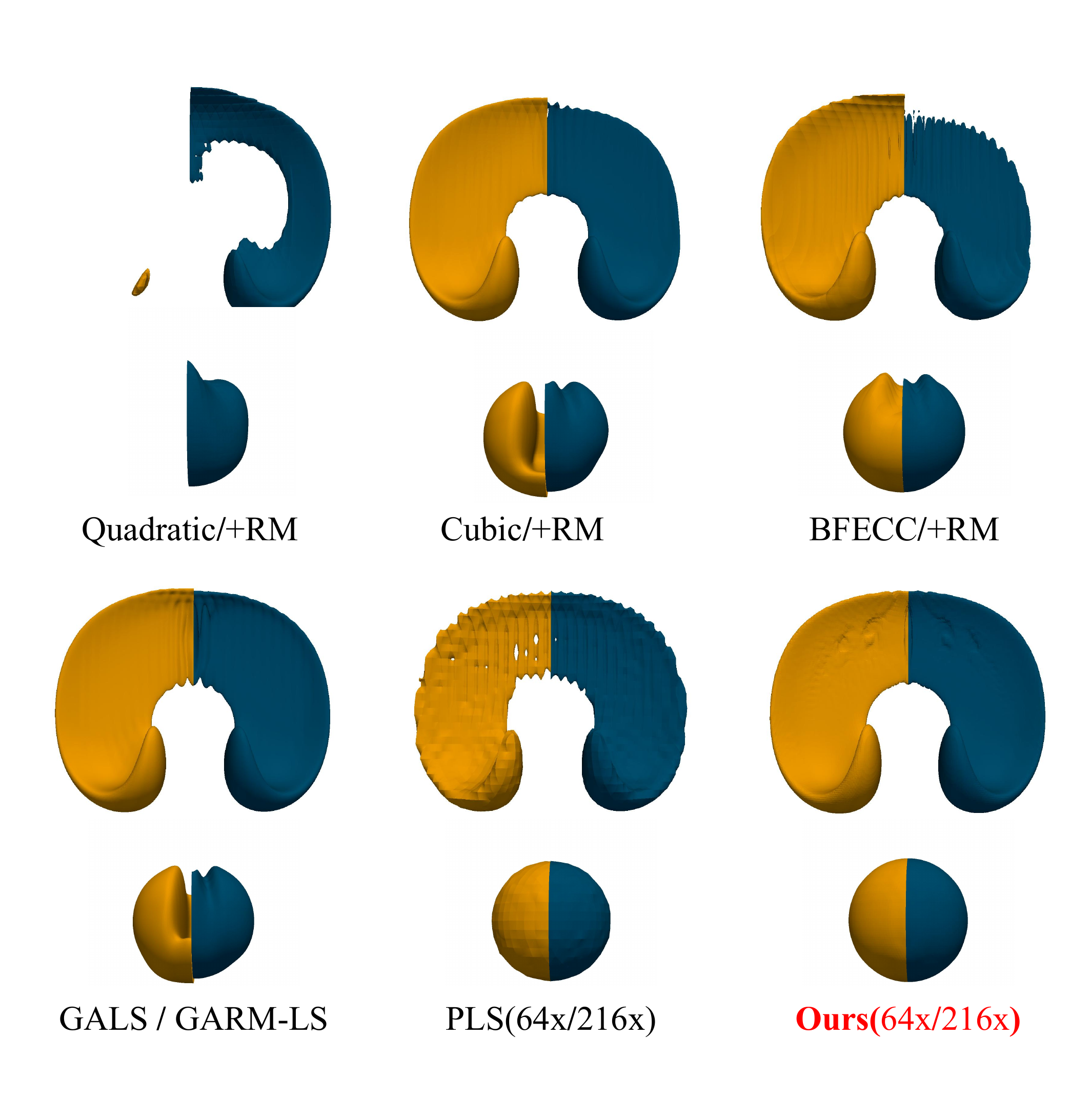}
    \caption{Visual comparison of $50^3$ \textbf{3D LeVeque sphere test} results across different interface tracking methods. Each panel shows maximum deformation state (top) and final recovered state (bottom). The two-color visualization (gold/blue) in each panel compares method pairs: Quadratic without/with Reference Map, Cubic without/with Reference Map, BFECC without/with Reference Map, GALS/GARM-LS, and PLS with different particle densities (64x/216x). Our method (shown in red) is displayed with both 64x and 216x particle configurations. This comparative visualization reveals that while Reference Map augmentation generally improves performance for traditional methods, our approach delivers superior shape preservation and nearly perfect recovery regardless of particle density, outperforming all competing methods in this challenging 3D deformation test.}
    \label{fig:leveque_3d_results}
\end{figure}

Particle-based methods generally show advantages in this reversal scenario. PLS(x64) and PLS(x216) recover 95.44\% and 96.33\% of the original volume, indicating that particle methods typically capture reversing flows better than pure grid-based approaches. NBFLIP, however, fails completely with 0.00\% recovery. The corresponding visual comparison is shown in Figure~\ref{fig:leveque_3d_results}, which illustrates the deformation and recovery behavior of each method in three dimensions.
Most remarkably, our method achieves nearly perfect recovery, with Ours(64\ensuremath{\times}) and Ours(216\ensuremath{\times}) configurations preserving 99.97\% and 99.99\% of the original volume, respectively. This extraordinary accuracy in both deformation and reversal phases highlights the fundamental advantage of our particle flow map approach: by tracking interface points with high-order information through accurate trajectories, we faithfully capture even the most complex 3D deformations and recoveries. We note that all 3D interfaces are visualized using marching cubes on a $250^3$ grid. In 3D, redistancing is performed on the background grid with the same schedule as in 2D; the pipeline requires no dimensional modification.

The 3D LeVeque test results clearly demonstrate that three-dimensional interface tracking presents substantially greater challenges than 2D scenarios. Many methods that perform adequately in 2D struggle in 3D, particularly during the reversal phase. Our method's consistent excellence across both 2D and 3D tests, and in both deformation and reversal phases, confirms its robustness and superiority for complex interface tracking applications.

\begin{table}[!htb]
\centering
\caption{Volume preservation at maximum deformation ($t=5.0$ s) for the $50^3$ \textbf{3D LeVeque's sphere test}}
\label{tab:leveque3D}
\begin{tabular}{lcc}
\hline
Method & Volume Preservation (\%) & Relative Error \\
\hline
Linear & 0.00\% & 1.00000 \\
Quadratic & 1.01\% & 9.90 $\times 10^{-1}$ \\
Cubic & 90.78\% & 9.22 $\times 10^{-2}$ \\
BFECC & 115.06\% & 1.51 $\times 10^{-1}$ \\
Linear+RM & 113.27\% & 1.33 $\times 10^{-1}$ \\
Quadratic+RM & 87.31\% & 1.27 $\times 10^{-1}$ \\
Cubic+RM & 89.18\% & 1.08 $\times 10^{-1}$ \\
BFECC+RM & 73.90\% & 2.61 $\times 10^{-1}$ \\
Hermite(GALS) & 94.31\% & 5.69 $\times 10^{-2}$ \\
Hermite+RM(GARM-LS) & 92.28\% & 7.72 $\times 10^{-2}$ \\
PLS(x64) & 84.86\% & 1.51 $\times 10^{-1}$ \\
PLS(x216) & 90.14\% & 9.86 $\times 10^{-2}$ \\
NBFLIP & 9.55\% & 9.04 $\times 10^{-1}$ \\
\hline
\textbf{Ours(64\ensuremath{\times})} & \textbf{99.68\%} & \textbf{3.16 $\times 10^{-3}$} \\
\textbf{Ours(216\ensuremath{\times})} & \textbf{99.68\%} & \textbf{3.16 $\times 10^{-3}$} \\
\hline
\end{tabular}
\end{table}

\subsection{3D Distorted Sphere}\label{subsec:3DDistorted}

Building upon our previous 3D evaluation, we now examine another demanding scenario by extending the 2D distorted circle test into three dimensions. This 3D distorted sphere test represents one of the most challenging benchmarks for interface tracking methods, as it generates complex spiral structures with progressively thinner features that evolve in three-dimensional space.
\begin{figure}[!htbp]
    \centering
    \includegraphics[width=0.99\linewidth]{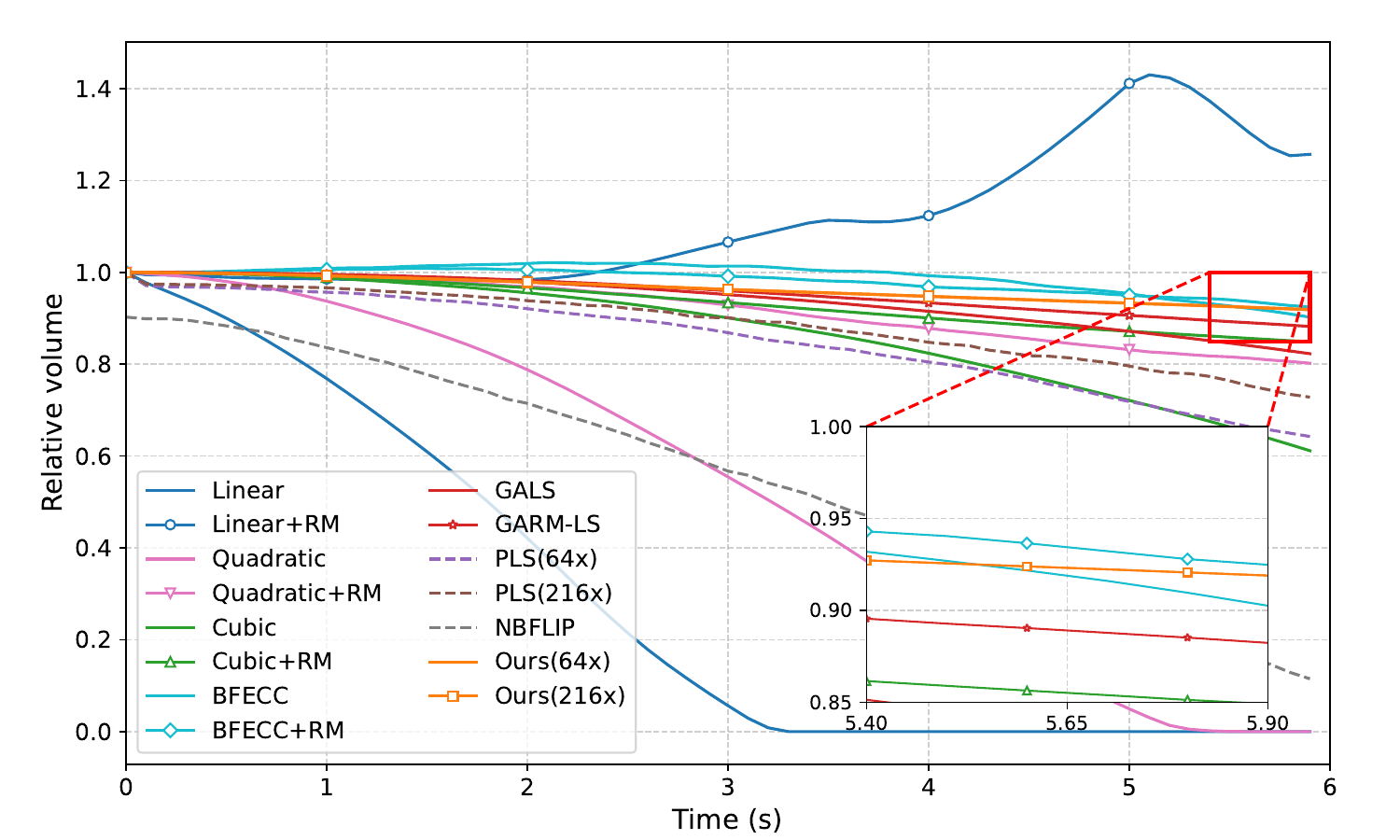}
    \caption{Volume preservation comparison during the $50^3$ \textbf{3D distorted sphere test}. Our method (orange lines) maintains good volume preservation throughout the simulation. Although BFECC+RM shows slightly better final volume preservation, it erroneously increases volume around 2.5s (visible in the inset), while our method exhibits only physically consistent volume loss. Most other methods suffer from significant volume deterioration as the spiral structure develops.}
    \label{fig:distorted3d_curve}
\end{figure}
In our implementation, a sphere with radius $r = 0.2$ is initially centered at $(0.5, 0.75, 0.5)$ within our $[0, 1]^3$ domain. The sphere is subjected to a rotational velocity field defined as:

\begin{equation}
\boldsymbol{u} = \frac{u_0}{\sqrt{(x-0.5)^2 + (y-0.5)^2 + r_0}} ((y-0.5), -(x-0.5), 0)
\end{equation} where $u_0 = 0.4$ m/s and $r_0 = 0.01$ m. This velocity field creates a differential rotation in the xy-plane that is stronger near the center axis and weaker farther away, causing the sphere to spiral inward as time progresses.

From $t = 0$ to $t = 6.0$ seconds, the sphere undergoes severe distortion and transforms into a complex spiral structure. The inner regions of the spiral become extremely thin, creating sub-grid features that challenge even high-resolution simulations. This test specifically evaluates a method's ability to preserve thin structures and small-scale features in three dimensions that may be smaller than the grid resolution.

We conduct these experiments on a $50^3$ grid, consistent with our 3D LeVeque's sphere test setup. As in the previous experiment, we use $4\times4\times4=64$ particles per cell for the lower density configurations (PLS(x64), Ours(64\ensuremath{\times})) and $6\times6\times6=216$ particles per cell for the higher density configurations (PLS(x216), Ours(216\ensuremath{\times})).

Figure~\ref{fig:distorted3d_curve} illustrates the performance of various methods in this highly challenging test, with quantitative results presented in Table~\ref{tab:distorted3d}. Table~\ref{tab:distorted3d} quantifies the volume preservation at the final simulation time. The results reveal significant performance differences across methods. Lower-order schemes struggle severely, with both linear and quadratic methods completely failing to preserve any volume (0.00\%). Even the cubic scheme maintains only 60.08\% of the original volume. Grid-based methods with reference map augmentation show mixed behavior---Linear+RM significantly overestimates at 126.89\%, while Quadratic+RM and Cubic+RM preserve 79.96\% and 84.76\%, respectively. The advanced level set methods GALS and GARM-LS achieve moderate preservation with 81.79\% and 88.01\%, but still lose substantial volume under this extreme deformation.

Particle-based methods demonstrate varied performance. Traditional particle level set approaches struggle considerably, with PLS(x64) and PLS(x216) configurations preserving only 63.67\% and 72.70\% of the original volume. NBFLIP shows poor results with just 10.27\% preservation, confirming its limitations in handling complex 3D structures.

Our PFM-LS approach delivers significantly better performance, with both Ours(64\ensuremath{\times}) and Ours(216\ensuremath{\times}) configurations achieving approximately 91.72\% volume preservation. While BFECC+RM slightly outperforms our method at 92.28\%, analysis of the volume preservation curve in Figure~\ref{fig:distorted3d_curve} reveals an important distinction: BFECC+RM erroneously increases the volume during advection, reaching a peak of 100.89\% before decreasing to its final value. In contrast, our method exhibits a monotone volume evolution without spurious oscillation; whether volume loss or gain is preferable is application-dependent, and we report both signs of error without ranking them.

\begin{figure}[!htbp]
    \centering
    \includegraphics[width=0.99\linewidth]{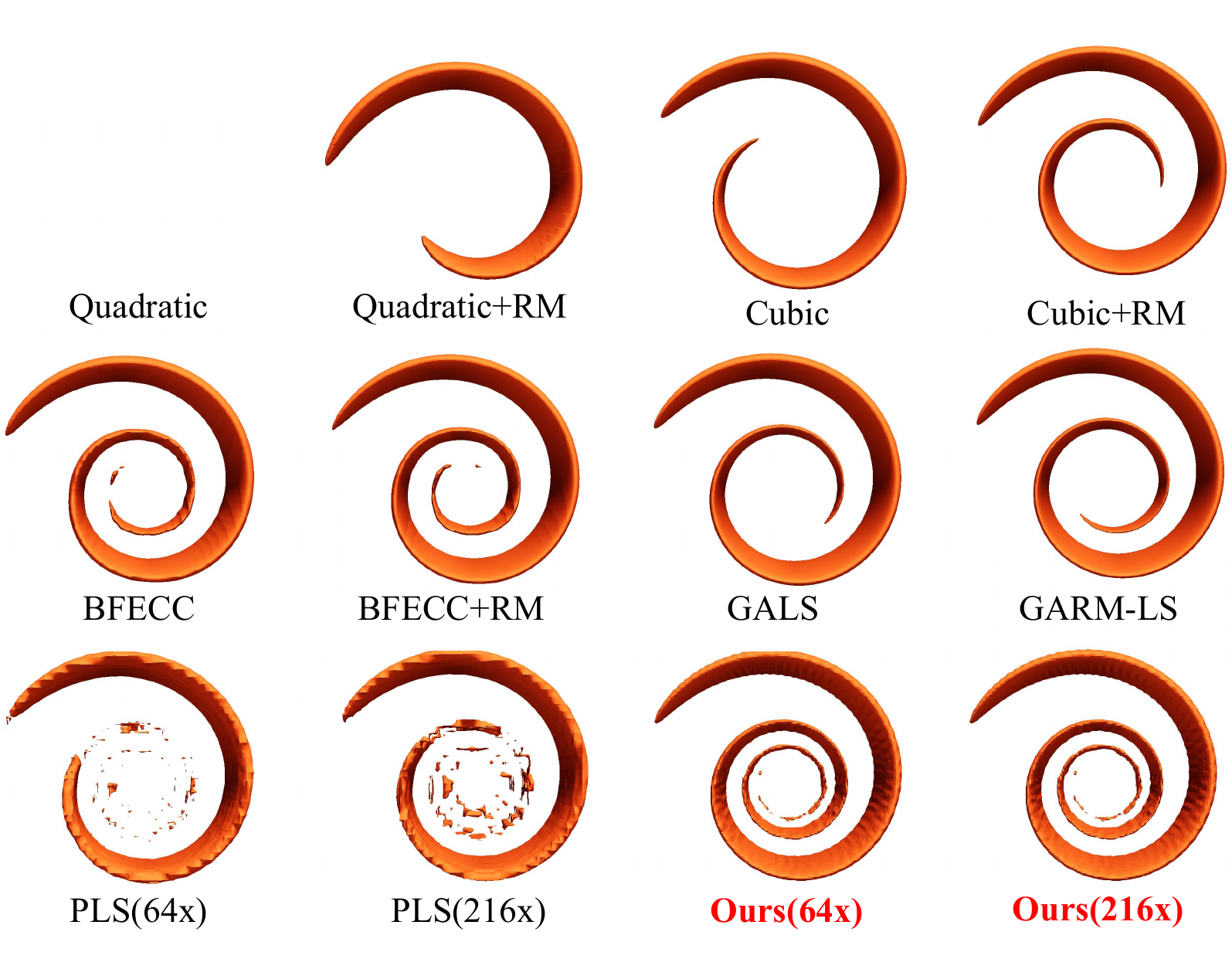}
    \caption{Visual comparison of the $50^3$ \textbf{3D distorted sphere test} results across different interface tracking methods after 6.0 seconds of differential rotation. Each method demonstrates varying abilities to preserve the evolving spiral structure. Lower-order methods (Quadratic, Quadratic+RM) capture only outer portions of the spiral. Higher-order methods (Cubic, Cubic+RM, BFECC, BFECC+RM) preserve more of the structure but still lose inner details. Advanced level set methods (GALS, GARM-LS) maintain better continuity but exhibit some distortion. Traditional particle-based methods (PLS with 64\ensuremath{\times} and 216\ensuremath{\times} particles) suffer from significant discontinuities in the inner spiral. In contrast, our method (shown in red) preserves the complete spiral structure with superior fidelity, maintaining continuous thin features and clear boundaries even in regions where the spiral becomes extremely thin, demonstrating the effectiveness of our narrow band particle flow map approach for capturing sub-grid features.}
    \label{fig:distroted3d_results}
\end{figure}
\begin{table}[!htb]
\centering
\caption{Volume preservation for the $50^3$ \textbf{3D distorted sphere test} at final time}
\label{tab:distorted3d}
\begin{tabular}{lcc}
\hline
Method & Volume Preservation (\%) & Relative Error \\
\hline
Linear & 0.00\% & 1.00 \\
Quadratic & 0.00\% & 1.00 \\
Cubic & 60.08\% & 3.99 $\times 10^{-1}$ \\
BFECC & 89.66\% & 1.03 $\times 10^{-1}$ \\
Linear+RM & 126.89\% & 2.70 $\times 10^{-1}$ \\
Quadratic+RM & 79.96\% & 2.00 $\times 10^{-1}$ \\
Cubic+RM & 84.76\% & 1.52 $\times 10^{-1}$ \\
\textbf{BFECC+RM} & \textbf{92.28\%} & \textbf{7.72 $\times 10^{-2}$} \\
Hermite(GALS) & 81.79\% & 1.82 $\times 10^{-1}$ \\
Hermite+RM(GARM-LS) & 88.01\% & 1.20 $\times 10^{-1}$ \\
PLS(x64) & 63.67\% & 3.63 $\times 10^{-1}$ \\
PLS(x216) & 72.70\% & 2.73 $\times 10^{-1}$ \\
NBFLIP & 10.27\% & 8.97 $\times 10^{-1}$ \\
\hline
\textbf{Ours(64\ensuremath{\times})} & \textbf{91.72\%} & \textbf{8.28 $\times 10^{-2}$} \\
\textbf{Ours(216\ensuremath{\times})} & \textbf{91.72\%} & \textbf{8.28 $\times 10^{-2}$} \\
\hline
\end{tabular}
\end{table}
Visual comparison in Figure~\ref{fig:distroted3d_results} further validates our method's advantages. Despite the similar final volume preservation percentages between our approach and BFECC+RM, our method captures the fine spiral tail structures with significantly greater fidelity. These thin features, which fall below grid resolution in the most challenging regions, are preserved with remarkable clarity in our simulations while appearing diffused or disconnected in other methods.

The 3D distorted sphere test demonstrates that tracking complex spiral structures in three dimensions presents extreme challenges for interface methods. The fact that our approach maintains over 91\% volume preservation while accurately representing sub-grid features underscores the effectiveness of our narrow band particle flow map for handling the most demanding interface tracking scenarios.

\subsection{Redistance}\label{subsec:experiment-re}

In this section, we evaluate the performance of our particle Newton redistance method and compare it with two established approaches---the Hermite Fast Marching Method (HFMM) and Hermite Newton (HNewton) from \citet{li2023garmls}.
Redistancing frequency plays a critical role in level-set methods: more frequent redistancing (corresponding to smaller "steps" values in our tests) better preserves the signed-distance property but can introduce interface distortion.
We conduct our evaluation using the $200\times200$ LeVeque's circle test, which provides a standard benchmark for assessing redistancing accuracy and interface stability.

\begin{figure}[!htb]
    \centering
    \begin{minipage}[t]{0.48\linewidth}
        \centering
        \includegraphics[width=\linewidth]{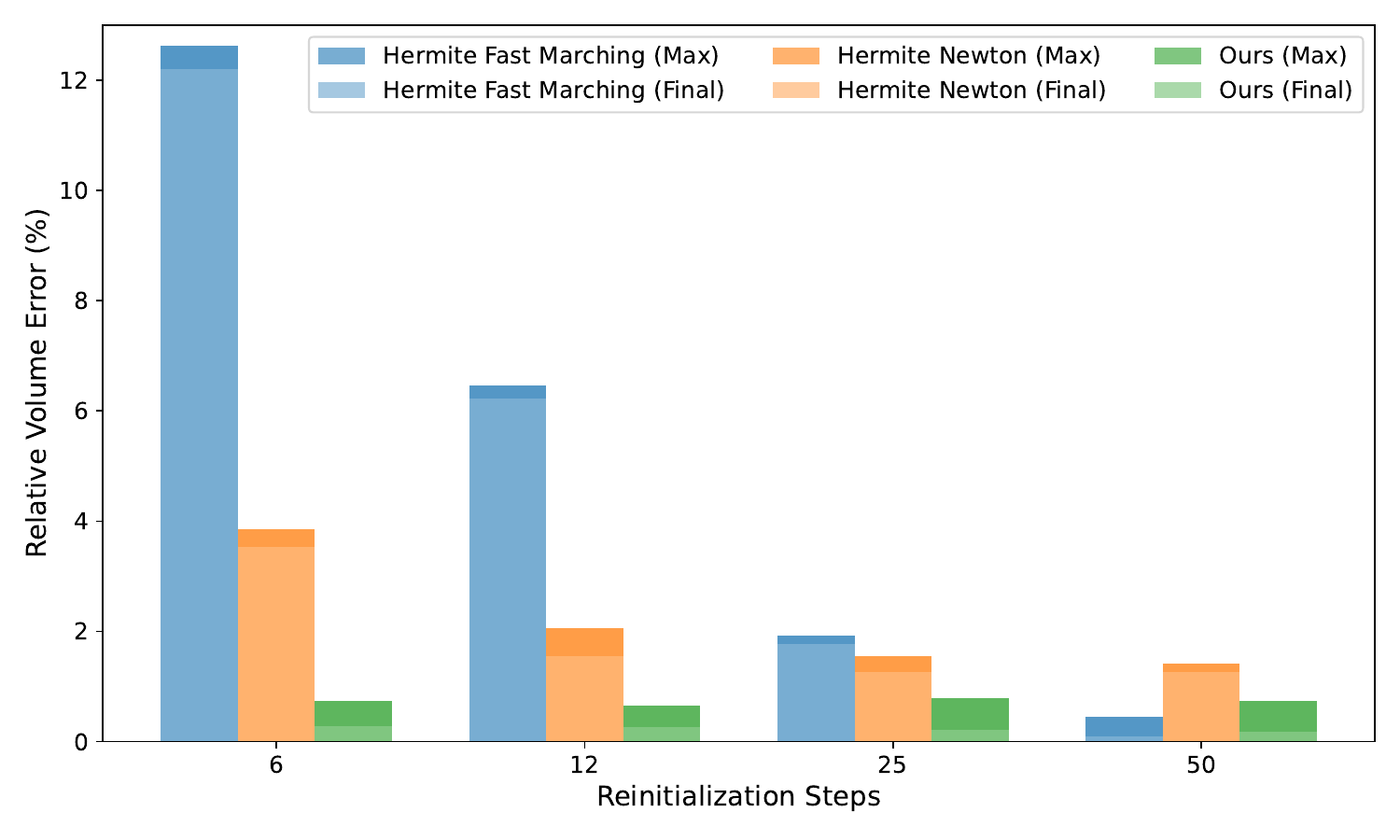}
        \caption{Relative volume error comparison of different \textbf{redistancing methods} in the $200\times200$ 2D LeVeque's circle test across varying redistancing frequencies (6, 12, 25, and 50 steps). Lower step values indicate more frequent redistancing. Our method (green) maintains consistently low error across all frequencies, whereas Hermite Fast Marching (blue) and Hermite Newton (orange) exhibit higher sensitivity to redistancing frequency.}
        \label{fig:redistance_volume_comparison}
    \end{minipage}
    \hfill
    \begin{minipage}[t]{0.48\linewidth}
        \centering
        \includegraphics[width=\linewidth]{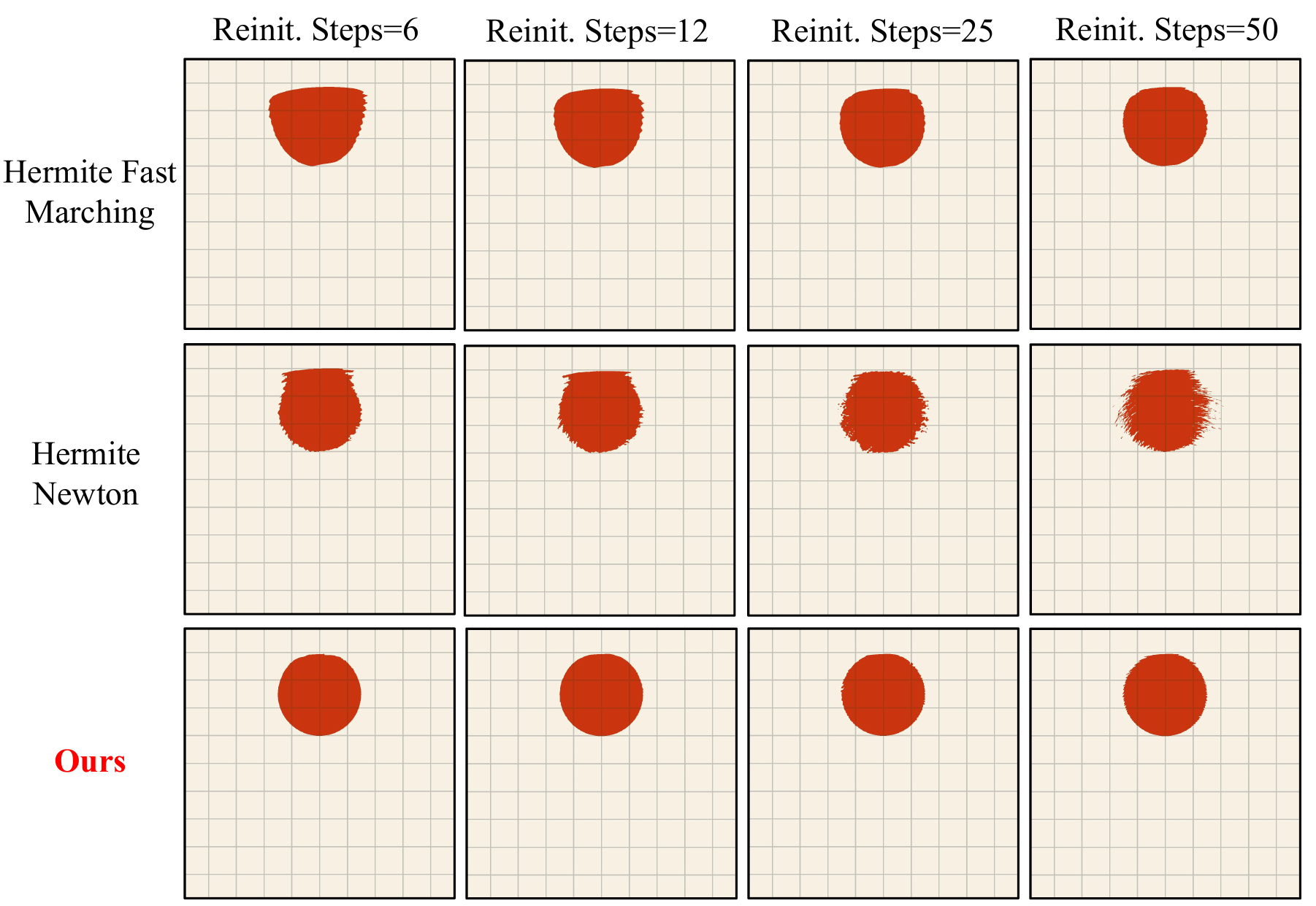}
        \caption{\textbf{Redistancing test}. Visual comparison of interface shapes after redistancing using different methods and frequencies. Top, middle, and bottom rows correspond to Hermite Fast Marching, Hermite Newton, and our method. Columns represent redistancing frequencies of 6, 12, 25, and 50 steps. Our method preserves the circular interface across all frequencies, whereas others exhibit noticeable distortions. For the least frequent schedules, close inspection reveals mild high-frequency perturbations along our interface as well; connected-component analysis bounds the associated spurious fragments below $0.05\%$ of the enclosed volume.}
        \label{fig:redistance_results}
    \end{minipage}
\end{figure}

Figure~\ref{fig:redistance_volume_comparison} presents a comparative analysis of relative volume errors for each method across different redistancing frequencies. Lower steps indicate more frequent redistancing, which is particularly relevant for applications like liquid simulation where redistancing occurs often to maintain numerical stability.

At high redistancing frequency (6 steps), HFMM shows significant volume distortion with errors exceeding 12\% at maximum deformation and approximately 4\% in the final state. Hermite Newton performs better but still exhibits errors of approximately 3.5\% at maximum deformation. In contrast, our method maintains excellent volume preservation with errors below 1\% regardless of redistancing frequency, demonstrating remarkable insensitivity to this parameter.

As redistancing becomes less frequent (increasing to 50 steps), HFMM and HNewton show improved volume preservation, though still not matching our method's consistency. This insensitivity to redistancing frequency represents a significant advantage of our approach, as it allows for flexible redistancing schedules without compromising accuracy.

The quantitative data reveals that our method preserves between 99.21\% and 99.82\% of the original volume across all tested frequencies. Even with the most aggressive redistancing (6 steps), our approach achieves 99.29\% volume preservation at maximum deformation and 99.72\% at the final state, demonstrating exceptional stability.

Figure~\ref{fig:redistance_results} provides visual confirmation of these numerical findings. The ideal outcome of redistancing should maintain the original circular shape without introducing artifacts. Our method consistently preserves the circle's shape across all frequencies, with virtually no visible difference between 6 steps and 50 steps. This indicates that our approach maintains geometric fidelity regardless of how frequently redistancing occurs. In contrast, HFMM exhibits significant shape distortion at high redistancing frequencies, with the interface appearing flattened and irregular, progressively improving as redistancing becomes less frequent. Hermite Newton shows better shape preservation than HFMM but still develops noticeable artifacts at various frequencies, particularly evident in the 50 steps case where small perturbations appear along the interface.

The visual results clearly demonstrate our method's ability to maintain the original circular interface shape regardless of redistancing frequency. This consistency is crucial for applications like liquid simulation where frequent redistancing is necessary, as our approach allows for aggressive redistancing schedules without compromising interface fidelity. This superior redistancing performance stems from our hybrid particle-grid blending approach described in Section~\ref{subsec:hybrid-blend}. By intelligently combining particle information near the interface with redistributed grid values elsewhere, our method effectively maintains both the signed distance property and high-order geometric information captured by particles.

These results confirm that our particle Newton redistance method offers the best combination of shape preservation and insensitivity to redistancing frequency among the tested approaches, making it particularly well-suited for applications requiring frequent redistancing while maintaining precise interface geometries.

\subsection{Convergence Analysis}\label{sec:convergence}
To evaluate the numerical convergence of our method compared to existing approaches, we conducted \textbf{convergence test} on $200\times200$ 2D distorted circle benchmark cases introduced in Sections~\ref{subsec:2DDistorted}, respectively. For these tests, we performed short-duration simulations (30 time steps, equivalent to 0.6s) to focus specifically on the numerical order of convergence rather than long-term behavior.

We analyzed convergence by computing the error between numerical solutions at different grid resolutions and fitting these errors to a power law using least squares regression. The quality of the fit is measured by the coefficient of determination ($R^2$), with values closer to 1.0 indicating more consistent convergence behavior.

\begin{figure}[!htbp]
    \centering
    \includegraphics[width=0.99\linewidth]{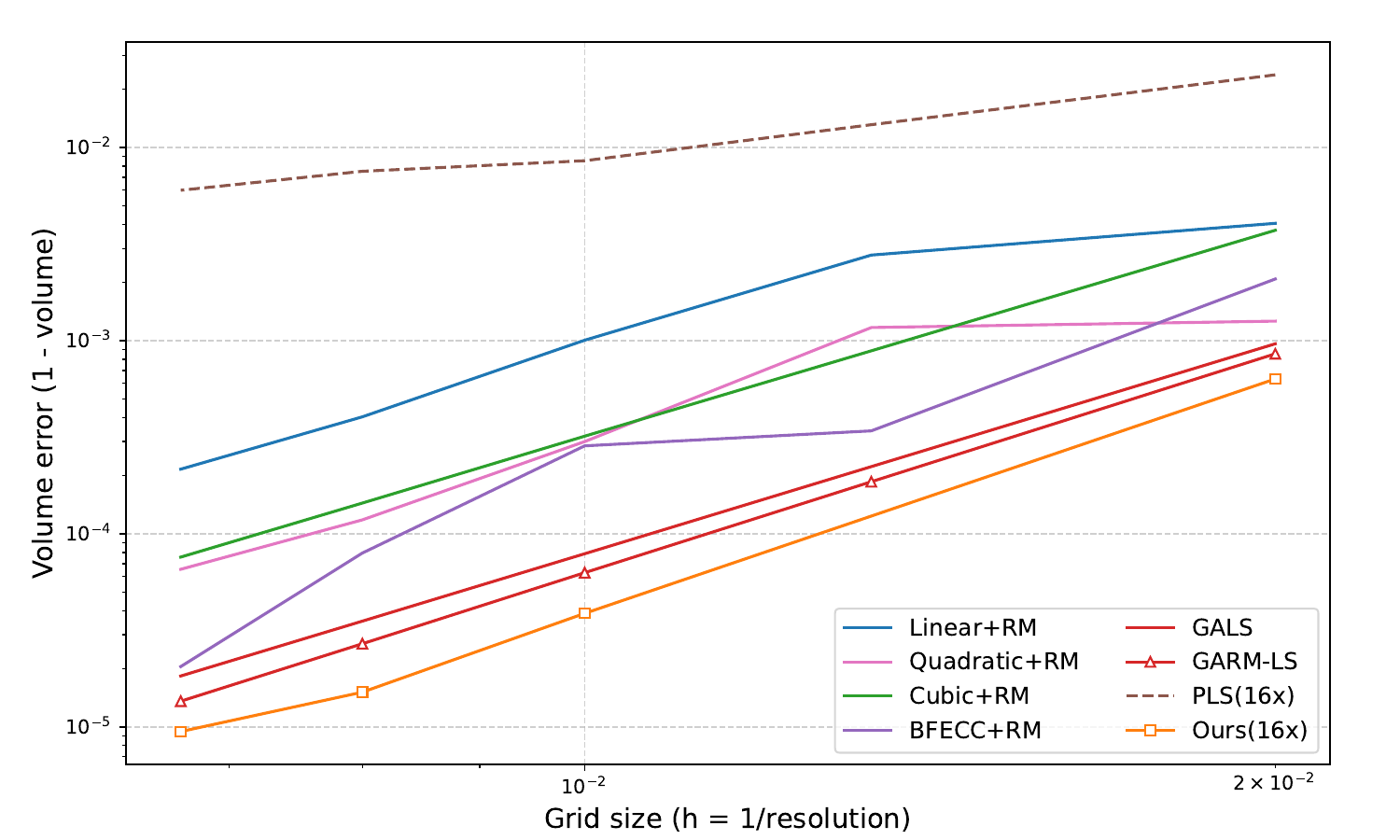}
    \caption{\textbf{Convergence analysis} for  the $200\times200$ 2D distorted circle test, comparing error reduction across grid refinements for different interface tracking methods. Our method achieves nearly fourth-order convergence.}
    \label{fig:distorted_convergence}
\end{figure}

In $200\times200$ 2D distorted circle test, as illustrated in Figure~\ref{fig:distorted_convergence}, our method maintained superior performance with a convergence rate of 3.9084 ($R^2$ = 0.9969), again approaching fourth-order accuracy. Other high-performing methods included BFECC+RM with 3.8600 ($R^2$ = 0.9391), GARM-LS with 3.7715 ($R^2$ = 1.0000), and GALS with 3.6058 ($R^2$ = 1.0000). Cubic+RM achieved 3.5474 ($R^2$ = 1.0000), while lower-order methods showed improved but still inferior rates: Quadratic+RM at 2.9074 ($R^2$ = 0.8990) and Linear+RM at 2.7744 ($R^2$ = 0.9390). PLS(x16) exhibited first-order convergence at 1.2430 ($R^2$ = 0.9833).

Our method achieves nearly fourth-order convergence in the test case due to its integration of particle flow maps with high-order information. Combined with the volume preservation results from previous sections, these findings demonstrate our approach's suitability for precise interface tracking applications.

\begin{figure}[!htb]
    \centering
    \includegraphics[width=0.5\linewidth]{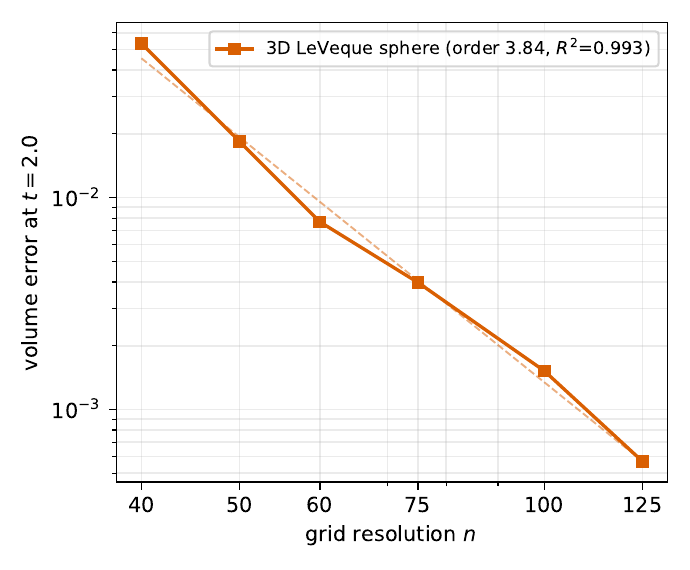}
    \caption{3D convergence: volume error after 100 steps ($t=2.0$) versus grid resolution, measured on a fixed $512^3$ reconstruction, with the least-squares power-law fit.}
    \label{fig:conv3d}
\end{figure}

To extend the convergence assessment to three dimensions, we additionally report a convergence study on the 3D LeVeque sphere with 64 particles per cell: the volume error against the analytic solution is measured after 100 time steps ($t=2.0$) at grid resolutions $\{40,50,60,75,100,125\}^3$, with the volume evaluated by marching cubes on a fixed $512^3$ reconstruction for every resolution, so that the measurement resolution is decoupled from the simulation grid (Figure~\ref{fig:conv3d}). The fitted order is $3.84$ ($R^2=0.993$), consistent with the near fourth-order behavior observed in 2D. We note that shorter horizons are unsuitable for this purpose: after only 30 steps the fine-grid errors fall below the resolution of the volume measurement itself, leaving no asymptotic range to fit. Beyond the asymptotic rate, the error constants of our method are one to four orders of magnitude smaller than those of competing schemes at equal resolution (e.g., $280\times$ smaller than the strongest baseline on the Zalesak benchmark, Table~\ref{tab:zalesak_results}).

\subsection{Topology change}\label{subsec:topology}
Level set methods are routinely adopted for their implicit handling of topology change, yet quantitative evaluations of this capability are rare. Because the medial axis of the shape approaches the interface at a topological event, the particle-carried $\nabla\varphi$ and $\nabla^2\varphi$ are discontinuous there, and it is natural to ask whether this degrades the solution. We therefore evaluate both directions of topology change under the full method (Table~\ref{tab:protocols}). In these benchmarks the normal-flow velocity is discretized on the background grid: at every step the field $\boldsymbol{u}=V\mathbf{n}$ is rebuilt with the normal $\mathbf{n}=\nabla\varphi/|\nabla\varphi|$ evaluated from the current grid level set, stored on the staggered grid faces, and interpolated to the particle positions for the trajectory and Jacobian integration, exactly as for an external velocity field; the particle-carried gradients are not used to define the driving velocity.

\begin{figure}[!htb]
    \centering
    \includegraphics[width=0.92\linewidth]{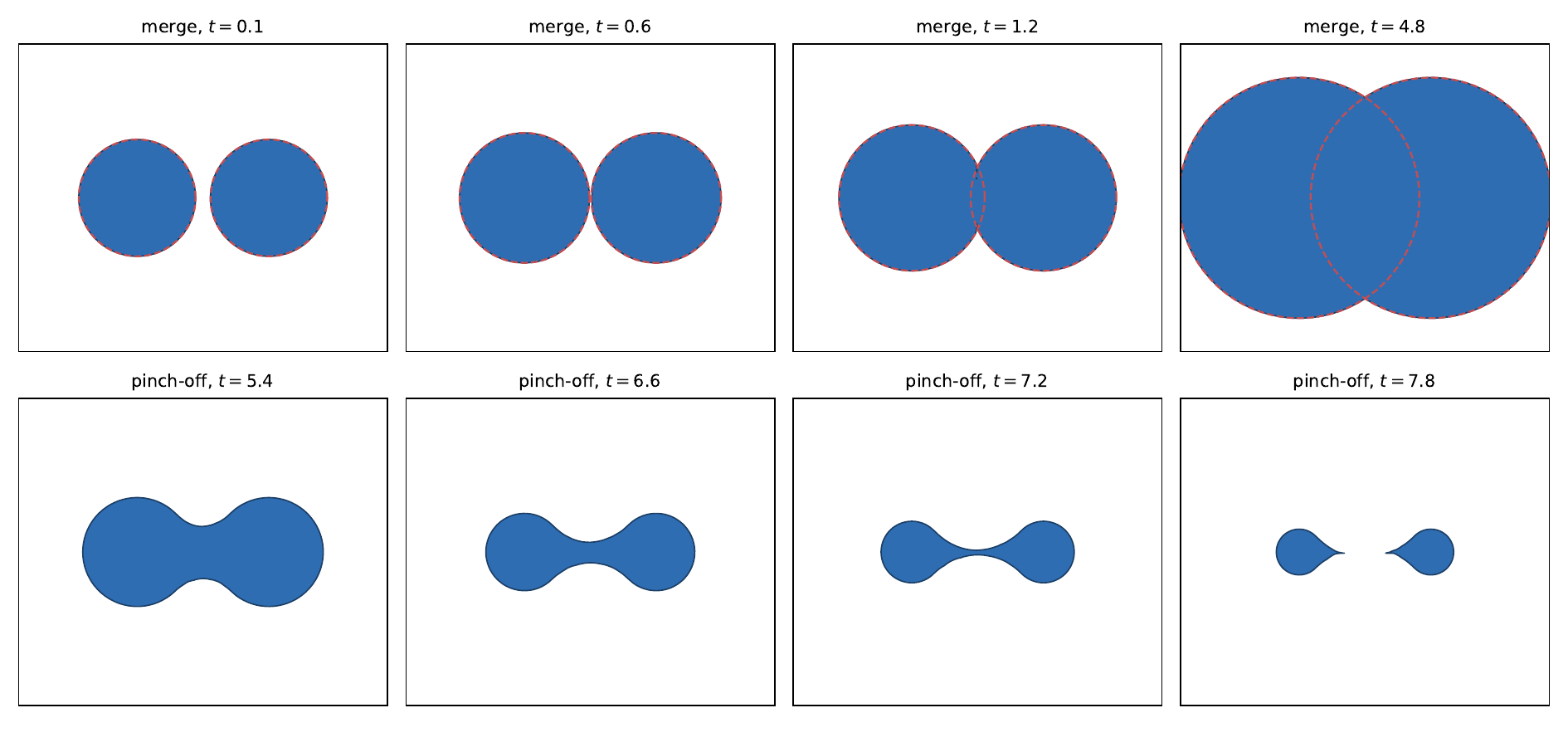}
    \caption{Topology-change benchmarks under normal flow. Top: two expanding circles merge; red dashed curves show the analytic solution. Bottom: reversing the flow contracts the dumbbell to pinch-off, which proceeds through a smoothly thinning filament and completes cleanly into two drops, delayed relative to the sharp-interface limit (see text).}
    \label{fig:topology}
\end{figure}

\noindent\textbf{Coalescence.} Two circles of radius $0.13$ centered at $(0.35,0.5)$ and $(0.65,0.5)$ expand under the normal flow $\boldsymbol{u}=V\mathbf{n}$ with $V=0.03$, for which the exact solution is the union of two circles of radius $r(t)=0.13+Vt$ and the analytic merge time is $t=0.67$ ($33$ steps). Figure~\ref{fig:topology} (top) overlays the computed interface on the analytic solution: the merge occurs at the predicted time, the enclosed-area error remains at $3\times10^{-4}$ through the topological event (peak $6.2\times10^{-4}$ at the merge instant), and the final-state shape error is $0.034\Delta x$ (chamfer) and $0.305\Delta x$ (Hausdorff). Consistent with the expectation above, the largest deviation localizes exactly at the coalescence neck where the medial axis meets the interface; however, it remains below a third of a grid cell, with no spurious artifacts.

\noindent\textbf{Pinch-off.} Reversing the normal velocity ($V=-0.03$) after merging contracts the dumbbell until the neck breaks (Figure~\ref{fig:topology}, bottom). Breakup occurs later than in the sharp-interface limit and completes cleanly, through a smoothly thinning filament, into two symmetric drops without spurious fragments. The delay reflects a trade-off inherent in sub-grid feature preservation: the same mechanism that preserves thin filaments in the distorted-circle test resists their breakup during topological splitting.

\subsection{Shape-error metrics}\label{subsec:shapemetrics}
Volume preservation alone does not capture errors in interface position. Following standard practice in geometry processing, we therefore measure symmetric chamfer and Hausdorff distances between the computed zero isocontour and the analytic solution, in units of the grid spacing $\Delta x$. For the comparison methods we extract the final interfaces from the reference implementation of \citet{li2023garmls} on the same $200\times200$ grids and evaluate them against the same analytic references with the same metric code; all horizons match those of Section~\ref{sec:numerical}. Table~\ref{tab:shape_errors} reports the results.

\begin{table}[!htb]
\centering
\caption{Shape errors (chamfer / Hausdorff, in units of $\Delta x$) versus the analytic solution at the final state of each 2D benchmark, $200\times200$. Baselines are measured from the reference implementation of \citet{li2023garmls} with identical metrics and horizons. For the LeVeque test, linear and quadratic (without reference maps) lose the interface entirely and are marked ``lost''; our LeVeque result uses the full method (redistancing active), a strictly harder protocol than the advection-only baselines.}
\label{tab:shape_errors}
\begin{tabular}{lcccccc}
\toprule
 & \multicolumn{2}{c}{Zalesak (2 rev.)} & \multicolumn{2}{c}{Distorted ($t{=}6$)} & \multicolumn{2}{c}{LeVeque (final)} \\
Method & chamfer & Hausdorff & chamfer & Hausdorff & chamfer & Hausdorff \\
\midrule
Linear & 5.31 & 20.2 & 5.86 & 42.8 & \multicolumn{2}{c}{lost} \\
Linear+RM & 0.052 & 0.213 & 3.02 & 32.7 & 25.4 & 59.0 \\
Quadratic & 0.784 & 5.11 & 1.77 & 28.3 & \multicolumn{2}{c}{lost} \\
Quadratic+RM & 0.039 & 0.209 & 1.16 & 21.4 & 5.29 & 20.3 \\
Cubic & 0.158 & 1.99 & 0.668 & 15.2 & 6.48 & 10.9 \\
Cubic+RM & 0.039 & 0.209 & 0.567 & 13.9 & 4.85 & 8.73 \\
BFECC & 1.21 & 6.28 & 0.457 & 10.2 & 5.25 & 10.4 \\
BFECC+RM & 0.039 & 0.209 & 0.374 & 9.59 & 4.51 & 7.72 \\
GALS & 0.059 & 0.925 & 0.316 & 10.0 & 4.82 & 9.61 \\
GARM-LS & 0.039 & 0.209 & 0.258 & 7.48 & 4.49 & 7.57 \\
PLS($\times$16) & 0.114 & 0.748 & 0.223 & 4.95 & 4.63 & 9.57 \\
PLS($\times$36) & 0.101 & 0.615 & 0.170 & 3.82 & 4.45 & 7.64 \\
\midrule
\textbf{Ours($\times$16)} & \textbf{0.025} & \textbf{0.134} & \textbf{0.131} & \textbf{1.53} & \textbf{2.26} & \textbf{4.27} \\
\bottomrule
\end{tabular}
\end{table}

Our method attains the lowest chamfer and Hausdorff distances on all three benchmarks. The chamfer distances are uniformly sub-cell for Zalesak and Distorted, confirming that the body of the interface is captured to sub-grid accuracy; the larger values on the LeVeque test are dominated, for every method, by the sub-grid tail filament that lies below the representable resolution of the grid; these are precisely the regions visible in the zoom-in figures. In 3D (LeVeque sphere, $50^3$), our final-state errors are $0.168\Delta x$ (chamfer) and $0.484\Delta x$ (Hausdorff) with spurious fragments totaling $0.02\%$ of the volume.

Since our particles carry $\nabla\varphi$ explicitly, interface-normal accuracy is available at no extra cost: on the severely shearing distorted-circle test at $t=6$ the particle-carried normals deviate from the analytic normals by $0.98^\circ$ on average (median $0.67^\circ$, excluding the sub-$\Delta x$ wound core, consistently with the chamfer protocol), and grid normals on Zalesak after a full revolution deviate by $1.15^\circ$ on average.

\subsection{Hessian transport ablation}\label{subsec:hessian-ablation}
To justify the approximate Hessian transport of Section~\ref{subsec-advection}, we compare three variants under identical settings ($200\times200$, 16 particles per cell): the approximate transport used throughout the paper, a variant including the correction term, and a variant that does not transport the Hessian at all (recomputing it at reinitialization only). Table~\ref{tab:hessian_ablation} shows that all three agree to within $0.1$ percentage points of volume preservation and a few hundredths of $\Delta x$ in chamfer distance; on these pure-advection metrics the no-transport variant is in fact marginally better (by at most $0.07$ percentage points), i.e., the Hessian contribution lies within measurement noise here. We retain Hessian transport not for these metrics but for the roles they do not exercise: it supplies the second-order term of the particle-to-grid reconstruction (Section~\ref{subsec-recon}), which carries sub-grid features, and it provides direct curvature access for coupled physics (Section~\ref{subsec:freesurface}). The purpose of the ablation is therefore to confirm that the inexpensive approximate transport is safe, not to claim that Hessian transport improves advection volume metrics.

\begin{table}[!htb]
\centering
\caption{Hessian-transport ablation ($200\times200$, 16 ppc). Volume preservation and shape errors agree within $0.1$ percentage points across variants (the no-transport variant marginally best); see text for why Hessian transport is nonetheless retained.}
\label{tab:hessian_ablation}
\begin{tabular}{lcccc}
\toprule
Variant & Distorted vol.\ ($t{=}6$) & chamfer & Hausdorff & LeVeque vol.\ (final) \\
\midrule
Approximate transport (default) & 99.630\% & 0.084$\Delta x$ & 3.70$\Delta x$ & 99.655\% \\
With correction term & 99.651\% & 0.072$\Delta x$ & 3.60$\Delta x$ & 99.663\% \\
No Hessian transport & 99.697\% & 0.071$\Delta x$ & 3.50$\Delta x$ & 99.730\% \\
\bottomrule
\end{tabular}
\end{table}

\subsection{Computational cost}\label{subsec:cost}
Table~\ref{tab:cost} profiles the wall-clock compute cost of the method (visualization and file export excluded), measured over full benchmark runs on a single consumer GPU. The flow-map-specific components (particle advection with Jacobian integration, particle-to-grid reconstruction, and the $\mathcal{T}$ compositions) account for less than $3\%$ of the total; the remainder is standard level-set infrastructure (redistancing and narrow-band maintenance) shared by any level-set implementation. Regarding 3D scalability, both the particle count and the redistancing work scale with the narrow-band \emph{area}, $O(N^2)$ for an $N^3$ grid, rather than with the domain volume. Note that redistancing is not executed every step: it runs once every $f_L$ steps ($357$\,ms per invocation at the default $f_L=12$, i.e., roughly $30$\,ms per step amortized), so a step without redistancing costs about $6$\,ms; the background Eikonal propagation is the largest component of each invocation.

\begin{table}[!htb]
\centering
\caption{Per-step compute cost ($200\times200$, 16 ppc; visualization and I/O excluded).}
\label{tab:cost}
\begin{tabular}{lc}
\toprule
Configuration & Cost per step \\
\midrule
Full method (advection + reinitialization + redistancing) & 36\,ms \\
\quad of which: redistancing ($357$\,ms per invocation, every $f_L{=}12$ steps, amortized) & 30\,ms \\
\quad of which: Hessian/gradient transport ($\mathrm{d}\mathcal{T}/\mathrm{d}x$) & 2.5\,ms \\
\quad of which: flow-map core (advect + P2G + composition) & $<$4\,ms \\
Advection-only & $\sim$6\,ms \\
\bottomrule
\end{tabular}
\end{table}

\subsection{Parameter sensitivity and defaults}\label{subsec:params}
\begin{figure}[!htb]
    \centering
    \includegraphics[width=0.99\linewidth]{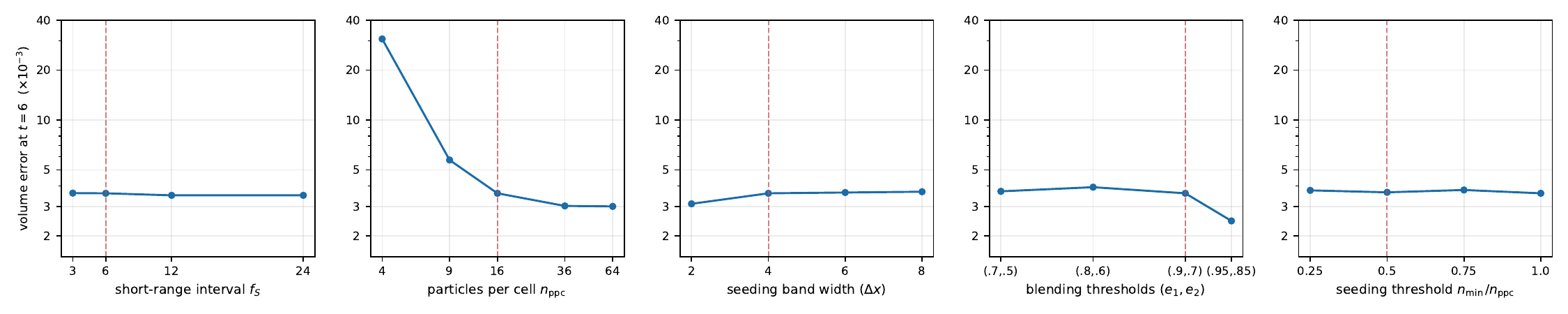}
    \caption{One-at-a-time parameter sensitivity on the $200\times200$ distorted-circle benchmark (full method, $300$ steps). Each panel varies one parameter with all others at their defaults (red dashed lines); the vertical axis is the relative volume error at $t=6$. Only the particle density produces a non-negligible response; the redistancing/reinitialization interval is studied separately in Figure~\ref{fig:redistance_volume_comparison}.}
    \label{fig:sensitivity}
\end{figure}

Table~\ref{tab:params} collects the parameters of the method with their default values, the ranges tested, and the measured effect of each variation. The method exhibits a wide insensitivity plateau: volume errors vary by less than $0.1$ percentage points across redistancing intervals of $6$--$25$ steps (Section~\ref{subsec:experiment-re}) and by less than $0.3$ percentage points between $16$ and $36$ particles per cell. For rigid or reversible flows, longer flow-map horizons further improve accuracy (down to $1.3\times10^{-7}$ relative volume error on the Zalesak benchmark). To cover the remaining parameters, we performed a one-at-a-time sweep on the distorted-circle benchmark (full method, $300$ steps; Figure~\ref{fig:sensitivity}), varying each parameter around its default with all others fixed. Only the particle density produces a non-negligible response, and it is converged from $16$ particles per cell onward ($\leq 0.06$ percentage points between $16$ and $64$). The short-range interval, seeding band width, blending thresholds, and adaptive seeding threshold all lie on plateaus, with at most $0.12$ percentage points of variation across the tested ranges.

\begin{table}[!htb]
\centering
\caption{Parameter summary: defaults, tested ranges, and measured effect on volume preservation. Effects are given in percentage points (pp) of preserved volume: e.g., a change from 99.6\% to 99.5\% preservation is 0.1\,pp.}
\label{tab:params}
\begin{tabularx}{\linewidth}{@{}llll>{\raggedright\arraybackslash}X@{}}
\toprule
Parameter & Symbol & Default & Tested & Measured effect on volume error \\
\midrule
Reinit.\ / redistancing interval & $f_L$ & 12 & $\{6,12,25\}$ & $<0.1$\,pp (Fig.~\ref{fig:redistance_volume_comparison}) \\
Short-range reinit.\ interval & $f_S$ & 5--6 & $\{3,6,12,24\}$ & $\leq 0.01$\,pp (Fig.~\ref{fig:sensitivity}) \\
Particles per cell & $n_{\mathrm{ppc}}$ & 16 & $\{4,9,16,36,64\}$ & converged from 16 ($\leq 0.06$\,pp to 64); 16 vs.\ 36: $\leq 0.3$\,pp on all benchmarks \\
Seeding band width & $n\Delta x$ & $4\Delta x$ & $\{2,4,6,8\}\Delta x$ & $\leq 0.05$\,pp (Fig.~\ref{fig:sensitivity}) \\
Blending thresholds & $e_1, e_2$ & 0.9, 0.7 & $(0.7,0.5)$ to $(0.95,0.85)$ & $\leq 0.12$\,pp (Fig.~\ref{fig:sensitivity}) \\
Adaptive seeding threshold & $n_{\min}$ & $n_{\mathrm{ppc}}/2$ & $\{0.25,0.5,0.75,1\}\,n_{\mathrm{ppc}}$ & $\leq 0.02$\,pp (Fig.~\ref{fig:sensitivity}) \\
\bottomrule
\end{tabularx}
\end{table}

\subsection{Coupling to free-surface flow}\label{subsec:freesurface}
\begin{figure}[!htb]
    \centering
    \includegraphics[width=0.92\linewidth]{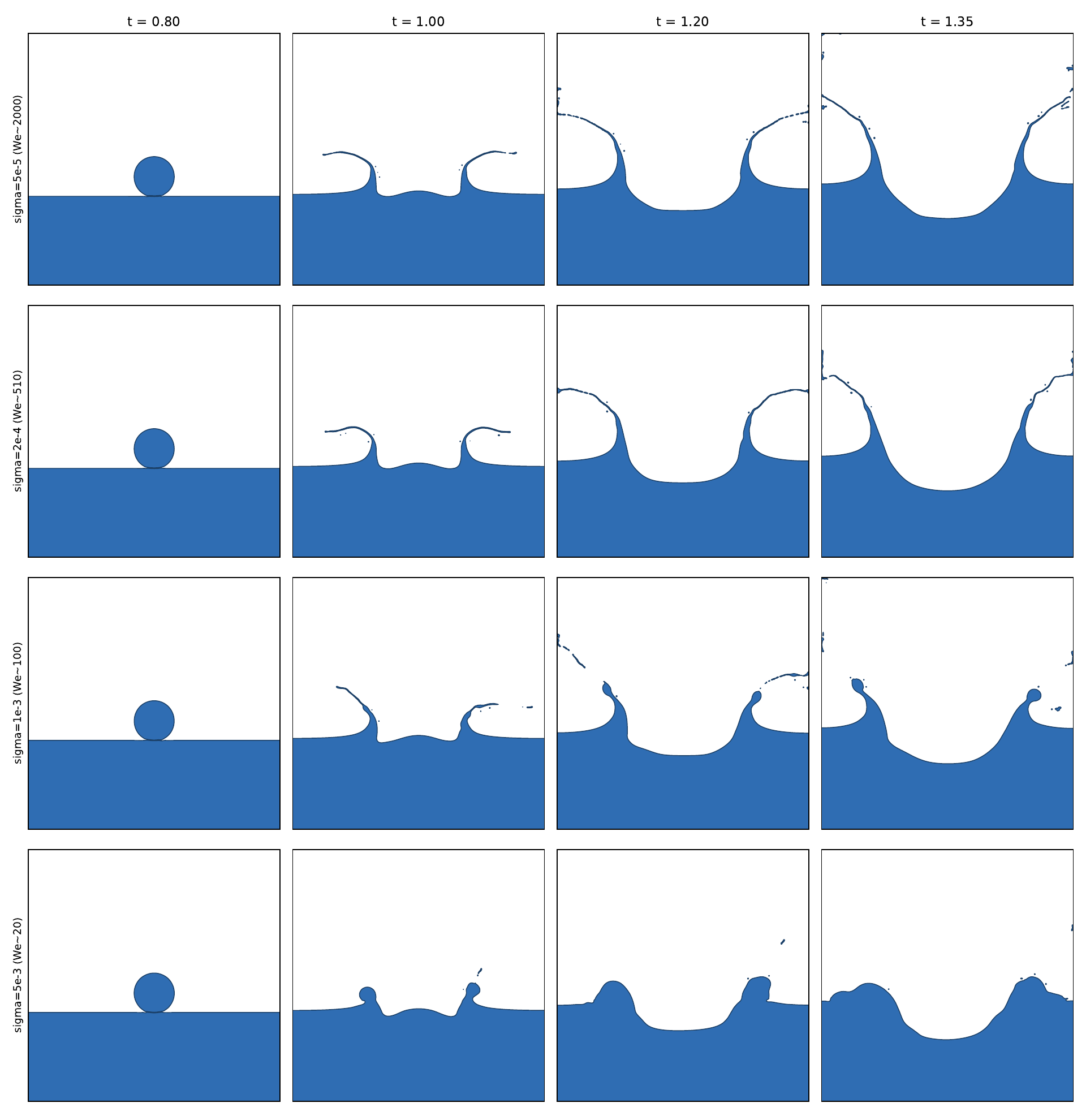}
    \caption{Surface-tension sweep of the droplet--pool impact ($200\times200$; rows: $\sigma=5\times10^{-5}$, $2\times10^{-4}$, $1\times10^{-3}$, $5\times10^{-3}$, i.e., $\mathrm{We}\approx2000$, $510$, $100$, $20$; columns: $t=0.8$, $1.0$, $1.2$, $1.35$). Sheet thickness, breakup scale, and the presence of a crown vary monotonically with $\sigma$; below the splashing threshold ($\mathrm{We}\approx20$) no crown forms.}
    \label{fig:sigmasweep}
\end{figure}

\begin{figure}[!htb]
    \centering
    \includegraphics[width=0.99\linewidth]{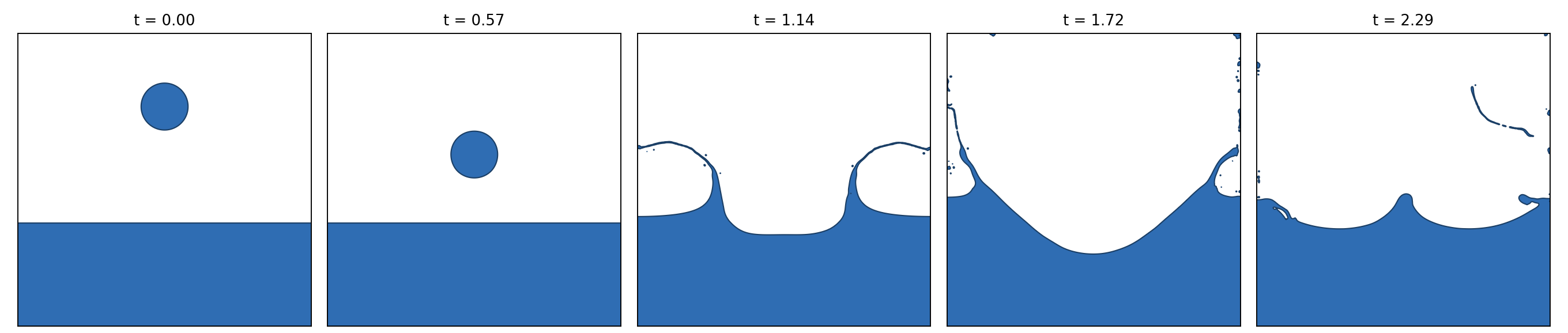}
    \caption{Free-surface coupling: a droplet falls into a pool ($200\times200$, $\sigma=2\times10^{-4}$). Uniformly sampled frames show the initial configuration, free fall, crown splash, maximum cavity, and central rebound jet.}
    \label{fig:freesurface}
\end{figure}

Finally, we demonstrate the method in the liquid-simulation setting targeted by our conclusion by coupling PFM-LS to an incompressible free-surface solver. Velocity is transported with the impulse-based particle flow map scheme of \citet{zhou2024eulerian}, sharing the flow-map machinery of the interface transport; pressure is solved on the liquid domain with a sharp (ghost-fluid) interfacial boundary condition $p=\sigma\kappa$, with curvature evaluated from the reconstructed level-set field, and velocity is extended smoothly into the air region. A standing-wave benchmark verifies the coupling quantitatively: the oscillation period matches linear dispersion theory to $-0.3\%$, with an amplitude decay of $0.6\%$ per period over three periods. Figure~\ref{fig:freesurface} shows a droplet falling into a pool, simulated over $10{,}000$ steps: the impact ejects a thin crown sheet, the cavity collapses into a central rebound jet, secondary droplets detach, and the surface returns toward rest while conserving $98.5\%$ of the liquid volume; these are the qualitative regimes expected of a high-Weber-number impact. To demonstrate the effect of the surface-tension coupling, Figure~\ref{fig:sigmasweep} repeats the experiment from $\mathrm{We}\approx20$ to $2000$, varying the surface-tension coefficient, $\sigma\in\{5\times10^{-5},\,2\times10^{-4},\,1\times10^{-3},\,5\times10^{-3}\}$ (impact Weber numbers $\mathrm{We}\approx2000$, $510$, $100$, and $20$). The morphology varies monotonically as expected: thin, long crown sheets that fragment into fine spray at the lowest $\sigma$; the curling crown of Figure~\ref{fig:freesurface} at the baseline; short, stubby jets with bulbous Rayleigh--Plateau tips at $\mathrm{We}\approx100$; and no crown at all below the splashing threshold ($\mathrm{We}\approx20$), where the rim merely curls over a shallow cavity. Over the interval shown ($t\leq1.35$) all four cases conserve the liquid volume to better than $99.5\%$.

\section{Conclusion}\label{conclusion}

In this work, we presented PFM-LS, a particle flow map method for high-fidelity level-set interface tracking. The method incorporates bidirectional flow maps to advect particle-embedded level set values, gradients, and Hessians. By integrating redistancing and resampling within a narrow band structure, the method achieves accurate interface reconstruction while maintaining computational efficiency.

Our method bridges several lines of prior work. Compared to Particle Level Set~\cite{enright2002hybrid}, which uses particles primarily for error correction, PFM-LS employs particles as primary carriers of geometric information, achieving significantly better volume preservation ($10^{-7}$ vs.\ $10^{-3}$ relative error). Compared to GALS~\cite{nave2010gradient} and GARM-LS~\cite{li2023garmls}, our particle-based formulation naturally captures sub-grid features without grid-based interpolation errors. Among particle flow map methods~\cite{zhou2024eulerian, nabizadeh2022covector, deng2023neural}, PFM-LS is the first to apply this framework to level set interface tracking, extending the differential forms perspective from impulse and vorticity to interface geometry. The results from 2D and 3D benchmark tests demonstrate that PFM-LS delivers competitive performance compared to established methods in terms of geometric fidelity, volume conservation, and the preservation of fine-scale structures. Notably, the solver effectively handles complex interface deformations, including high-curvature features, without excessive particle clustering or loss of resolution.

\noindent\textbf{Limitations and Future Work.}
Several limitations warrant further investigation. First, while adaptive particle control maintains coverage in most scenarios, particles near fine-scale features such as thin filaments or high-curvature regions may disperse during advection, causing the method to fall back to the less accurate background grid representation. More refined adaptive control that accounts for local interface complexity is needed to address this issue. Second, the current topological change detection relies on gradient consistency thresholds; overly aggressive settings may erroneously merge nearby interfaces or smooth out fine details, while conservative settings may fail to detect legitimate topology changes. The topology benchmarks (Section~\ref{subsec:topology}) quantify this trade-off: coalescence is handled cleanly ($0.3\Delta x$ at the merge neck), while pinch-off occurs later than in the sharp-interface limit, which is the price of the thin-feature preservation demonstrated in the distorted-circle tests. Developing robust, geometry-aware detection mechanisms remains an open challenge. Finally, per-particle Jacobian storage incurs higher memory overhead than purely grid-based methods, which may become a concern for large-scale 3D simulations.

In conclusion, the PFM-LS framework provides an efficient and accurate approach for level-set interface tracking.
Future work will extend this framework to applications involving surface tension, multiphase flows, liquid simulations, fire dynamics, and differentiable fluid simulators with free surfaces, where precise and stable interface handling remains essential.

\bmsubsection*{Author Contributions}
\textbf{Jinjin He}: Conceptualization, Methodology, Software, Validation, Writing---original draft.
\textbf{Taiyuan Zhang}: Software, Validation, Visualization.
\textbf{Zhiqi Li}: Methodology.
\textbf{Junwei Zhou}: Methodology.
\textbf{Duowen Chen}: Software.
\textbf{Bo Zhu}: Conceptualization, Supervision, Funding acquisition, Writing---review \& editing.

\bmsubsection*{Acknowledgments}
The Georgia Tech authors acknowledge support from the National Science Foundation (NSF) under grants CAREER \#2420319, IIS \#2433307, OISE \#2433313, IIS \#2433322, ECCS \#2318814, and CNS \#2450401. We thank NVIDIA for providing computational resources through the NVIDIA Academic Grant.

\bmsubsection*{Data Availability Statement}
The source code is publicly available at \url{https://github.com/jinjinhe2001/Particle-Flowmap-Levelset}.

\bmsubsection*{Conflicts of Interest}
The authors declare no conflicts of interest.

\bibliography{references}

\end{document}